\documentclass[11pt, preprint]{aastex}
\usepackage{subfigure}
\let\tablehead\undefined
\let\tabletail\undefined
\usepackage{supertabular}
\usepackage{rotating}
\let\affil\undefined
\usepackage{authblk}
\usepackage{amsmath, amsthm, amssymb}
\def\ha{H$\alpha~$}
\def\hap{H$\alpha$}
\def\iha{$I_{\textrm{H}\alpha}~$}
\def\ihap{$I_{\textrm{H}\alpha}$}
\def\lha{$[L_{\textrm{H}\alpha}/L_{\textrm{bol}}]~$}
\def\lhap{$[L_{\textrm{H}\alpha}/L_{\textrm{bol}}]$}

\title{\bf \ha Activity of Old M Dwarfs: Stellar Cycles and Mean Activity Levels For 93 Low-Mass Stars in the Solar Neighborhood}
\author[]{Paul Robertson}
\author[]{Michael Endl}
\author[]{William D. Cochran}
\author[]{Sarah E. Dodson-Robinson}
\affil[]{Department of Astronomy and McDonald Observatory, University of Texas at Austin, Austin, TX 78712, USA; paul@astro.as.utexas.edu}

\begin{abstract}

Through the McDonald Observatory M Dwarf Planet Search, we have acquired nearly 3,000 high-resolution spectra of 93 late-type (K5-M5) stars over more than a decade using HET/HRS.  This sample provides a unique opportunity to investigate the occurrence of long-term stellar activity cycles for low-mass stars.  In this paper, we examine the stellar activity of our targets as reflected in the \ha feature.  We have identified periodic signals for 6 stars, with periods ranging from days to more than 10 years, and find long-term trends for 7 others.  Stellar cycles with $P \ge 1$ year are present for at least $5 \%$ of our targets.  Additionally, we present an analysis of the time-averaged activity levels of our sample, and search for correlations with other stellar properties.  In particular, we find that more massive, earlier type (M0-M2) stars tend to be more active than later type dwarfs.  Furthermore, high-metallicity stars tend to be more active at a given stellar mass.  We also evaluate \ha variability as a tracer of activity-induced radial velocity (RV) variation.  For the M dwarf GJ 1170, \ha variation reveals stellar activity patterns matching those seen in the RVs, mimicking the signal of a giant planet, and we find evidence that the previously identified stellar activity cycle of GJ 581 may be responsible for the recently retracted planet f (Vogt et al. 2012) in that system.  In general, though, we find that \ha is not frequently correlated with RV at the precision (typically 6-7 m/s) of our measurements.

\end{abstract}

\begin{document}

\section{\bf Introduction}

The study of \ha activity for M dwarfs has provided a wealth of insight into the physics of low-mass stars.  For young M stars with \ha emission lines (so-called ``active" M dwarfs), stellar activity has been shown to be tightly coupled with rotation \citep[e.g.][]{browning10,reiners12}.  Furthermore, as these stars age and spin down, their \ha emission and variability decrease significantly, offering a diagnostic for distinguishing stellar populations of different ages \citep{west08,west09,ls10}.

Absorption line variability in inactive M dwarfs is often caused by periodic stellar activity cycles \citep{kurster03,cincunegui07a,zechmeister09,buccino11,gds12}.  Because such cycles may appear in radial velocity (RV) measurements (as evidenced by correlation between RV and activity tracers such as Ca II H and K emission), potentially mimicking Keplerian planet signals \citep{queloz01,if10,gds12}, they are of particular interest to planet search surveys.  As a somewhat happy coincidence, the long-term, multi-epoch spectral monitoring of stars conducted by RV surveys makes them uniquely sensitive to analogs of the 11-year solar cycle. These cycles may not appear in photometry if they manifest as variable heating of active chromospheric regions and not starspot modulation.  

Identification and characterization of solar-type cycles will be instrumental in better understanding the solar magnetic dynamo \citep{thompson03,wright11}.  Since stellar cycles are reflective of a star's magnetic field strength and variability \citep[e.g.][]{brown11}, internal structure \citep[][and references therein]{ossendrijver03}, and long-term evolution \citep{wright04}, exploring activity across all spectral types is an excellent way to understand how fundamental stellar properties vary, and how stellar magnetic fields are generated and maintained.

In the Sun, the origin of the magnetic activity driving the observed 11-year cycle is believed to be the $\alpha \Omega$ dynamo maintained through differential rotation at the tachocline, the interface between the radiative and convective layers of the Sun.  The prevalence of similar cycles for FGK stars \citep{baliunas95} confirms that such dynamos are common amongst solar-type stars.  However, the tachocline lies at increasing depths at later spectral types, disappearing altogether around $\sim$ M4 \citep{chabrier97}.  Thus, a direct comparison between the stellar activity levels and cyclic behavior of solar-type and lower-mass stars is essential to characterizing the effect of stellar mass on internal architecture and the resulting magnetic activity.

In this paper, we present the first systematic analysis of stellar activity for the McDonald Observatory M Dwarf Planet Survey \citep{endl03,endl06}.  While the long-term surveys at Mount Wilson \citep{baliunas95} and HARPS \citep{lovis11} have already conducted similar studies for over 400 quiet FGK stars, analyses of M stars \citep[e.g.][]{kurster03,cincunegui07a,gds12} are significantly lacking, with less than 50 total stars examined.  With 93 stars included in this paper, it is the largest such study to date for inactive M stars, and represents a substantial addition to the total collection surveyed.  

Here, we will use the flux in the \ha line as a tracer of stellar chromospheric activity.  In areas of the stellar chromosphere where magnetic field lines influence the local convective behavior (so-called ``active regions" or plages), \ha photons are emitted, resulting in an activity-dependent depth of the \ha absorption line.  While this effect is more commonly observed in the calcium H and K lines, \ha is potentially a more suitable line for M stars because of their lack of flux near the calcium lines.  Additionally, because \ha and Ca H and K are emitted from different depths in the chromosphere (and hence different distances from the tachocline), valuable information may be gained by comparing the results of M dwarf activity using both indices.  Since \ha is a standard tracer of M dwarf activity, our measurements can easily be compared with previously published results.  

The paper is organized as follows.  In Section 2, we describe how we acquired and reduced our data.  We have identified a number of periodic \ha signals, which we discuss in Section 3.  Additionally, we have computed the mean \ha flux levels of our targets, and include a detailed examination of how these average fluxes depend on stellar mass and metallicity.  The broader implications of our results are discussed in Section 4, and we summarize our conclusions in Section 5.

\section{\bf Observations and Data Analysis}

The McDonald Observatory Planet Search \citep[e.g.][]{cochran93,robertson12a,robertson12b} has monitored nearby stars for RV variation induced by exoplanets with the Hobby-Eberly Telescope \citep[HET,][]{ramsey98} since 2001.  During this period, we have also surveyed 100 M dwarfs to explore the frequency of planets around low-mass stars.  Our 100 targets are specifically selected to have low activity based on a lack of X-ray emission from the \emph{ROSAT} All-Sky Survey \citep{hunsch99}, and should therefore represent the old, quiet population of nearby M stars.

\begin{sidewaysfigure}
\begin{center}
\subfigure[\label{mass_hist}]{\includegraphics[scale=0.4]{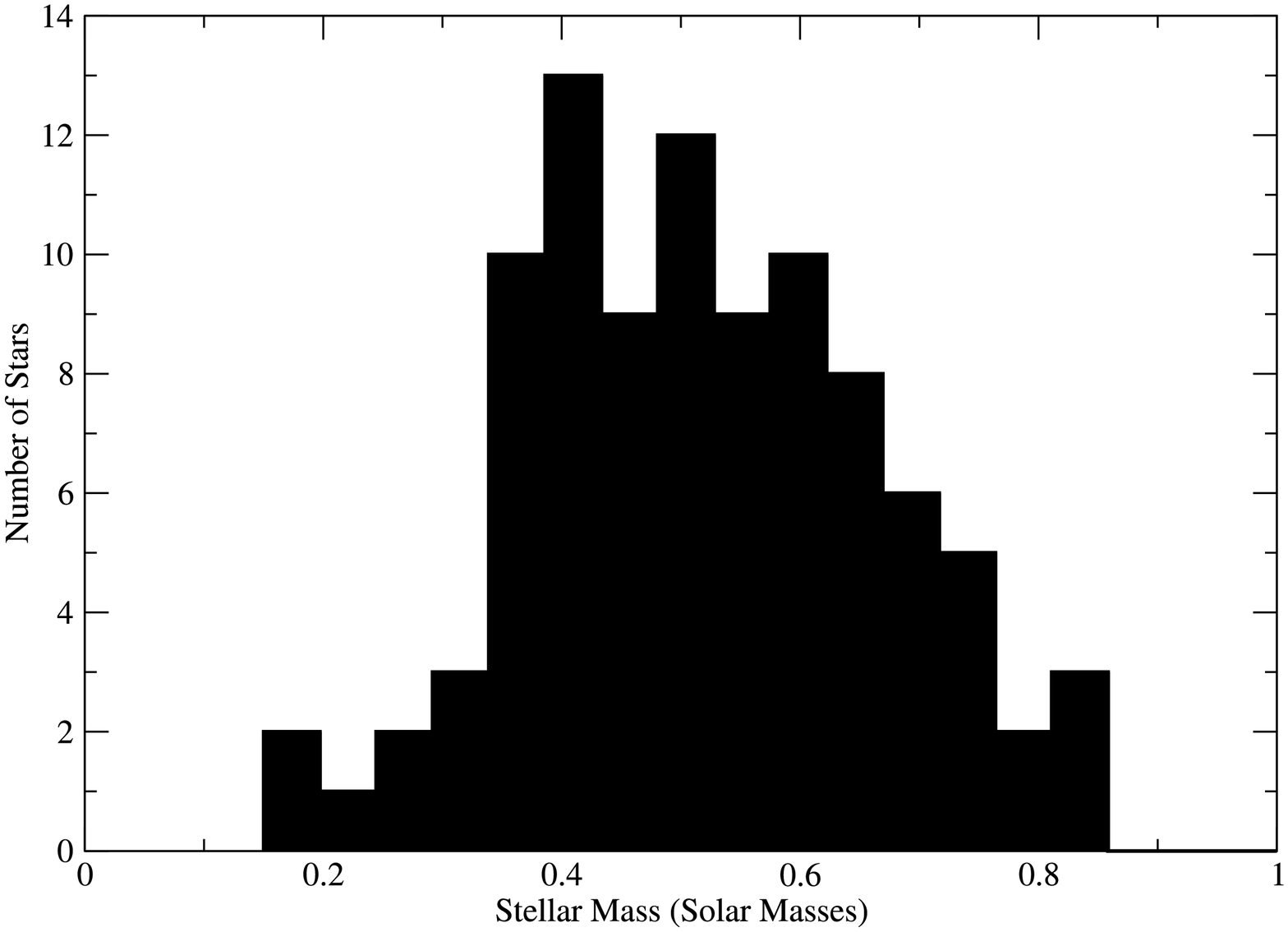}}
\subfigure[\label{Z_hist}]{\includegraphics[scale=0.4]{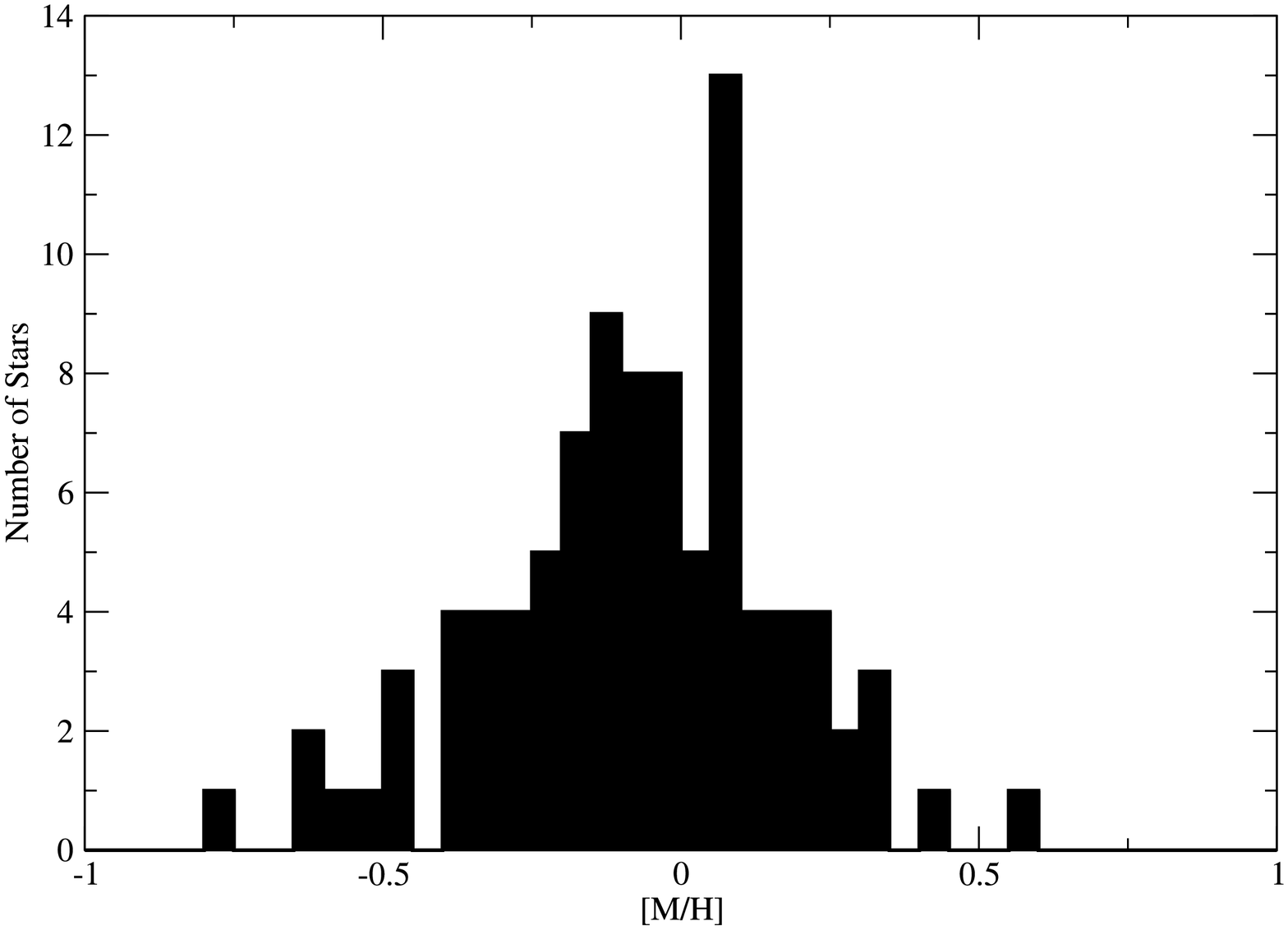}}
\subfigure[\label{type_hist}]{\includegraphics[scale=0.4]{type_hist.eps}}
\caption{The McDonald Observatory M Dwarf Planet Search is a dedicated long-term survey of late-type stars in the Solar neighborhood.  Here, we give histograms of the stellar masses, metallicities, and spectral subtypes of our targets.}
\label{histos}
\end{center}
\end{sidewaysfigure}

In order to better characterize our targets, we computed stellar masses using the \citet{delfosse00} K-band mass-luminosity relation.  Additionally, we calculate stellar metallicity using the \citet{schlaufman10} photometric metallicity calibration.  We note that this calibration is largely consistent with earlier photometric metallicity estimates of \citet{bonfils05} and \citet{johnsonapps09}, but is less susceptible to systematically over- or underestimating stellar [M/H].  Where available, we have also obtained spectral subtypes from the Orion Spiral Arm Catalogue \citep{bobylev06}.  In Figure \ref{histos}, we show the distributions of these stellar properties for our sample.

After removing double-lined spectroscopic binaries (SB2s), the M dwarf survey has amassed 2933 high-resolution spectra of 93 stars as of July 2012.  All of the 93 targets have been observed at high cadence for a brief amount of time (typically $\sim 5$ observations over one week) to explore short-period RV variability, and are observed at least once per season to ensure sensitivity to long-period signals.  Stars with potentially interesting RV signals have been observed several times each season.

Our HET M dwarf spectra are taken with the High Resolution Spectrograph \citep[HRS,][]{tull98}.  Nearly all the spectra are taken at a resolving power $R = 60,000$, with a small number of bright targets for which we observed at $R = 120,000$.  We note that while our observations are taken with an I$_2$ absorption cell in front of the slit for the purpose of obtaining precise RVs, no I$_2$ absorption lines are present in the spectral region around \hap.  Flat-fielding, bias subtraction, cosmic ray removal, and wavelength calibration are performed with standard IRAF\footnote{IRAF is distributed by the National Optical Astronomy Observatories, which are operated by the Association of Universities for Research in Astronomy, Inc., under cooperative agreement with the National Science Foundation.} routines, and we have divided our spectra by the blaze function to normalize the continuum.

\begin{figure}
\begin{center}
\includegraphics[scale=0.5]{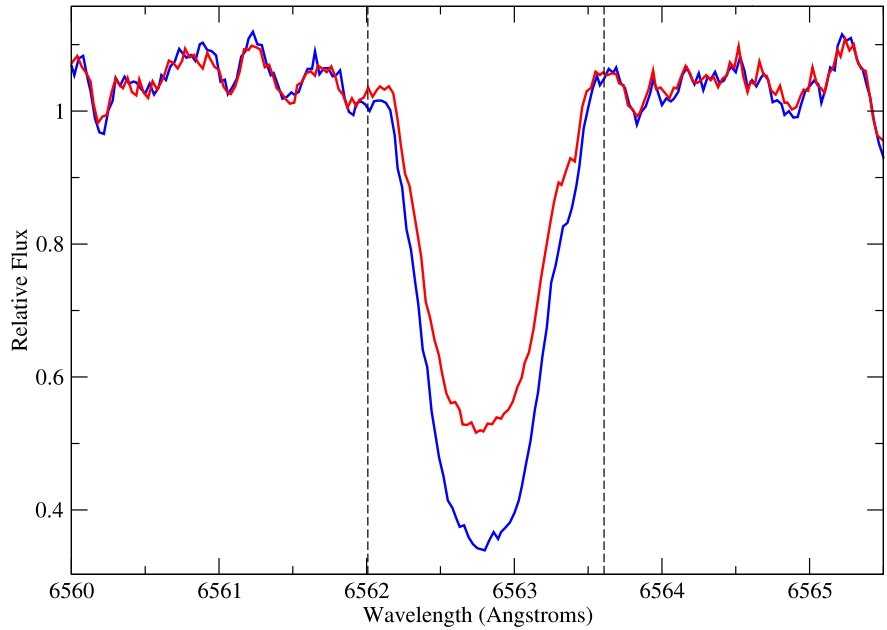}
\caption{For quiet M stars, the \ha line traces stellar magnetic activity through the variable ``filling in" of the absorption line with photons emitted from magnetically active regions of the chromosphere.  Here, we show the \ha absorption line of GJ 270, in states of high (red) and low (blue) chromospheric emission.  The dashed vertical lines indicate the 1.6 \AA~window in which the \ha flux is calculated.}
\label{line}
\end{center}
\end{figure}

Although it is common practice for RV surveys to trace stellar activity with the Ca II H and K lines \citep[$S_{\textrm{HK}}$, e.g.][]{paulson04,if10,robertson12a}, those lines are not accessible to HRS.  Instead, we examine the \ha line, which appears in absorption for our targets, but becomes increasingly filled in as chromospheric activity increases \citep{kurster03,gds11}.  Figure \ref{line} shows the \ha absorption line for GJ 270 at times of high and low emission.  In addition to the issue of availability, analyzing \ha reveals stellar behavior which will not necessarily appear in $S_{\textrm{HK}}$ since \ha activity and Ca H and K activity do not always correlate, as shown in previous comparisons between tracers \citep{cincunegui07b,santos10,gds11}.  We measure this chromospheric \ha ``filling in" with an index \ihap.  Following \citet{kurster03}, we define the index as

\begin{equation}
\label{iha_eq}
I_{\textrm{H}\alpha} = \frac{F_{\textrm{H}\alpha}}{F_1 + F_2}
\end{equation}

\noindent As in \citet{gds11}, we have taken $F_{\textrm{H}\alpha}$ to be the sum of the flux in a band of width 1.6 \AA~centered on the \ha $\lambda 6562.828$ \AA~line.  We find the line center with our fitted wavelength solution, making manual corrections for large stellar RV offsets where necessary.  $F_1$ and $F_2$ are the sum of the fluxes in the reference bands $[-700~\textrm{km s}^{-1}, -300~\textrm{km s}^{-1}]$ (from the \ha line center) for band 1 and $[600~\textrm{km s}^{-1}, 1000~\textrm{km s}^{-1}]$ for band 2, as defined by \citet{kurster03}.  To ensure any variation observed is in fact from stellar chromospheric activity and not instrumental effects, we also measured a similar index for the Ca I $\lambda 6572.795$ \AA~line, where $F_{\textrm{CaI}}$ is the sum of the flux from $[441.5~\textrm{km s}^{-1}, 472.5~\textrm{km s}^{-1}]$ around the Ca I line center, and using the same reference bands.  Because the Ca I line should not be sensitive to stellar activity, the Ca I index serves as a control against which to verify \ha activity.

For each stellar line index, we calculated an error bar using a method analogous to the one employed for the Mount Wilson Ca II $S_{\textrm{HK}}$ index we compute for our 2.7m spectra \citep[e.g.][]{robertson12a}.  Specifically, we multiplied the RMS in the continuum in the 0.5 \AA~adjacent to the \ha or Ca I line by $\sqrt{N}$, where $N$ is the number of pixels in the line (i.e. $\sigma_{F_{\textrm{H}\alpha}} = \textrm{RMS}_{F_{\textrm{H}\alpha \pm 0.5}} \times \sqrt{N}$).  Likewise, we multiplied the RMS scatter of each reference band by the square root of the number of pixels in the band, and added the errors in quadrature for the final line index.  In equation form,

\begin{equation}
\label{iha_err}
(\frac{\sigma_{I_{\textrm{H}\alpha}}}{I_{\textrm{H}\alpha}})^2 = (\frac{\sigma_{F_{\textrm{H}\alpha}}}{F_{\textrm{H}\alpha}})^2 + (\frac{\sqrt{\sigma_{F_1}^2 + \sigma_{F_2}^2}}{F_1 + F_2})^2
\end{equation}

\noindent Rather than take the approach of \citet{gds11} and bin our observations, we assign these error bars so as to accurately assess the quality of each spectrum, and to best make use of our more sparsely-sampled data sets, which nonetheless show significant activity.

\section{\bf Analysis}

We see time-dependent stellar activity for a number of stars from our RV survey.  In addition to detailing this behavior, we have also examined the time-averaged overall activity levels of our targets, and related those levels to other fundamental stellar properties.  We present these results in separate subsections below.

\subsection{Periodic \ha Activity}

We began our analysis by searching the time series \iha measurements of each star for periodic activity.  Since M dwarf stellar activity signals have been observed with periods as short as weeks or months \citep[as is typical of rotation periods, e.g.][]{kurster03,forveille09} and as long as years \citep{cincunegui07a,gds12}, we searched a broad range of frequencies.  We search for periodicity using the fully generalized Lomb-Scargle periodogram \citep{zk09} due to its ability to handle uneven time sampling and individually-weighted data points.  In cases where our periodograms show significant power levels, we have estimated false alarm probabilities (FAPs) using three different methods.  The first two computations, which we consider preliminary estimates, use Equation 24 of \citet{zk09} and the method outlined in \citet{sturrock10}.  For the \citet{zk09} FAP estimate, we normalize the periodogram assuming gaussian noise, so our false alarm probability for a peak of power P is 

$\textrm{FAP} = 1 - [1 - (1 - \frac{2P}{N - 1})^\frac{N - 3}{2}]^M$, 

\noindent where $N$ is the number of data points and $M = \frac{\Delta f}{\delta f}$ is an estimate of the number of independent frequencies sampled.  The FAP estimate of \citet{sturrock10}, which compares the power of a periodogram peak to an expected distribution of power values based on Bayesian statistics, agrees well with the probabilistic calculation for our power spectra.  Our final FAPs for periodic signals presented herein are calculated according to the bootstrap resampling technique of \citet{kurster97}.  Briefly, the bootstrap FAP technique retains the time stamps of the original data set, then creates a series of fake data sets by assigning a random \iha value (with replacement) from the set to each time stamp.  The FAP is then taken as the percentage of resampled data sets which at any frequency give an equal or higher periodogram power value than the highest peak in the original data.  

Because stellar activity is neither strictly periodic nor perfectly sinusoidal, and will experience stochastic behavior (flares, etc.) regardless of any regular cycles, the FAP required to confirm a \ha activity cycle is somewhat higher than for a more rigidly periodic phenomenon such as a Keplerian planet orbit.  Previous studies \citep[e.g.][]{cincunegui07a} have considered FAPs as high as 0.5.  While we see many signals in our sample with a FAP in the 0.5-0.1 range, we do not present them here, pending confirmation either through more dense time sampling (for short-period signals) or a second activity tracer such as the Na I D feature (for low-amplitude signals).  Here, we only claim detection of a periodic signal if it has a bootstrap FAP at or near the 0.01 level, with the exception of GJ 581 (FAP $\sim 0.1$), for which we offer a confirmation of a previously observed cycle (see below).  This criterion is essentially equivalent to the one adopted by \citet{buccino11}.  The periodograms for our confirmed activity cycles are shown in Figure \ref{periodogram}, and the individual FAPs are discussed in later sections.

For each activity cycle, we have fit a sinusoid of the form $I_{\textrm{H}\alpha}(t) = a_0 + a_1 \sin(\omega t + \phi)$, where $\omega = \frac{2\pi}{P}$, $P$ is the period of the cycle, and $a_0,~a_1$, and $\phi$ are free parameters to set the \iha zero point, amplitude, and phase, respectively.  In Figure \ref{cycles}, we show the time-series \ha index of the six stars for which we observe periodic activity cycles and the fitted sine curves.  Included with each plot is the time series of the Ca I index.  We note that we have examined the periodograms of the Ca I index for each star in Figure \ref{cycles}, and do not see periodicity matching the signals seen in \hap.  Furthermore, the level of variation in Ca I is considerably lower than we see in \hap, confirming that the cycles observed are in fact caused by stellar activity.  

While we show sinusoidal fits to the activity cycles shown in Figure \ref{cycles}, it is clear from visual inspection that the cycles are not all perfectly sinusoidal in shape.  Interestingly, the two most obviously non-sinusoidal signals--GJ 552 and GJ 630 (see below)--both exhibit gradual declines and rapid rises in \ihap.  This behavior is reminiscent of the Waldmeier effect seen in the Sun, where higher-amplitude activity cycles are preceded by more rapid rises in activity \citep[see, e.g.][]{cameron08}.  While the details of the Waldmeier effect are the subject of active research, its qualitative properties can be explained by the variable nonlinear feedback of hydrodynamic turbulence and small-scale magnetic fields on the large-scale stellar magnetic field \citep{pipin11}.  If this is the case for our targets, we should expect to see varying shapes and durations of each cycle as we observe successive periods.

We find that in cases of non-sinusoidal \iha series, a better approach is to fit a Fourier series.  In Figure \ref{fourier}, we show the results of fitting three terms of the Fourier series

$I_{\textrm{H}\alpha}(t) = a_0 + \displaystyle\sum_{i=1}^{n} a_{i}\sin(\omega_i t + \phi_i)$

\noindent for two of the cycles presented in Figure \ref{cycles}.  We set the initial guess for $\omega_1$ equal to the frequency identified in the periodogram of each star's \iha index.  For both stars, the best-fit frequency of each successive term falls very close to twice the frequency of the previous term, leading us to conclude that we are likely observing harmonics of the initially-observed activity cycle.  The drawback to fitting Fourier series is that each new term adds three free parameters to the fit.  For targets without a large number of observations (e.g. GJ 552, Figure \ref{gj552_fourier}), we quickly lose the ability to constrain the parameters of higher-order terms, leading to over-fitting as shown in the Figure.

Below, we discuss each observed activity signal in detail.  Additionally, we have included the details of our fits to the data in Table \ref{fits}.

\subsubsection{GJ 270 and GJ 476}

With periods $\ge 3$ years, the \iha signals of GJ 270, GJ 476, and GJ 581 (discussed below) could be considered the ``stellar cycles" in our survey.  Here, we see the advantage of utilizing the data set of an RV survey; the same observational strategy intended to reveal long-period giant planets gives us excellent coverage of very slow activity cycles, the durations of which have never been previously observed for M dwarf stars.

At $M_* = 0.68 M_{\odot}$, GJ 270 (M2) is among the more massive M dwarfs in our survey.  As first reported in \citet{endl06}, we see evidence for a long-period binary companion in the RVs of GJ 270, but at present its period is longer than our observational time baseline.  The \iha time series, shown in Figure \ref{gj270_cycle} shows clear cyclic variation, with a distinct peak in the corresponding periodogram (Figure \ref{gj270_ps}) at 2553 days.  Our bootstrap analysis generated no false alarms in $10^4$ trials, thus giving an upper limit on the FAP of $10^{-4}$.  We fit a sinusoidal model to this signal with a period of 2687 days and amplitude 0.00158 in \ihap, the longest-period cycle for which we have observed a full period.  We include our fit to the data in Figure \ref{gj270_cycle}.

\citet{bobylev06} list GJ 476 as subtype M4, although with a mass estimate of $0.47 M_{\odot}$, it would be among the more massive M4 stars.  In Figure \ref{gj476_cycle}, we show the \iha time series, which again shows long-period variation.  The power spectrum for GJ 476 peaks at 993 days, with a bootstrap FAP estimate again giving an upper limit of $10^{-4}$.  Our fitted sinusiod gives a final period of 1066 days.  The cycle has an \iha amplitude of 0.00193, the largest of the periodic signals presented herein.

\subsubsection{GJ 581}

\begin{sidewaysfigure}
\subfigure[\label{gj270_ps}]{\includegraphics[scale=0.3]{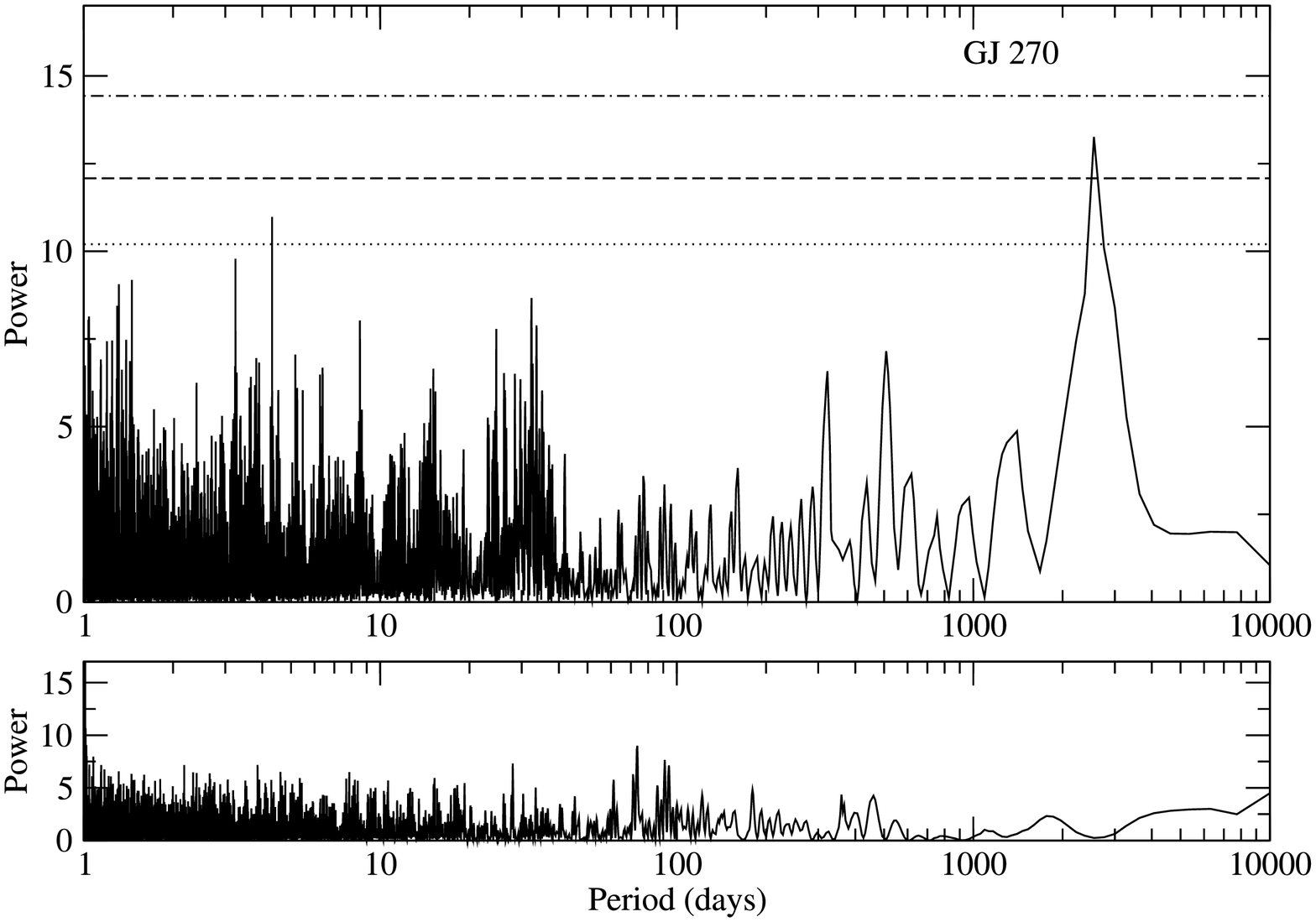}}
\subfigure[\label{gj476_ps}]{\includegraphics[scale=0.3]{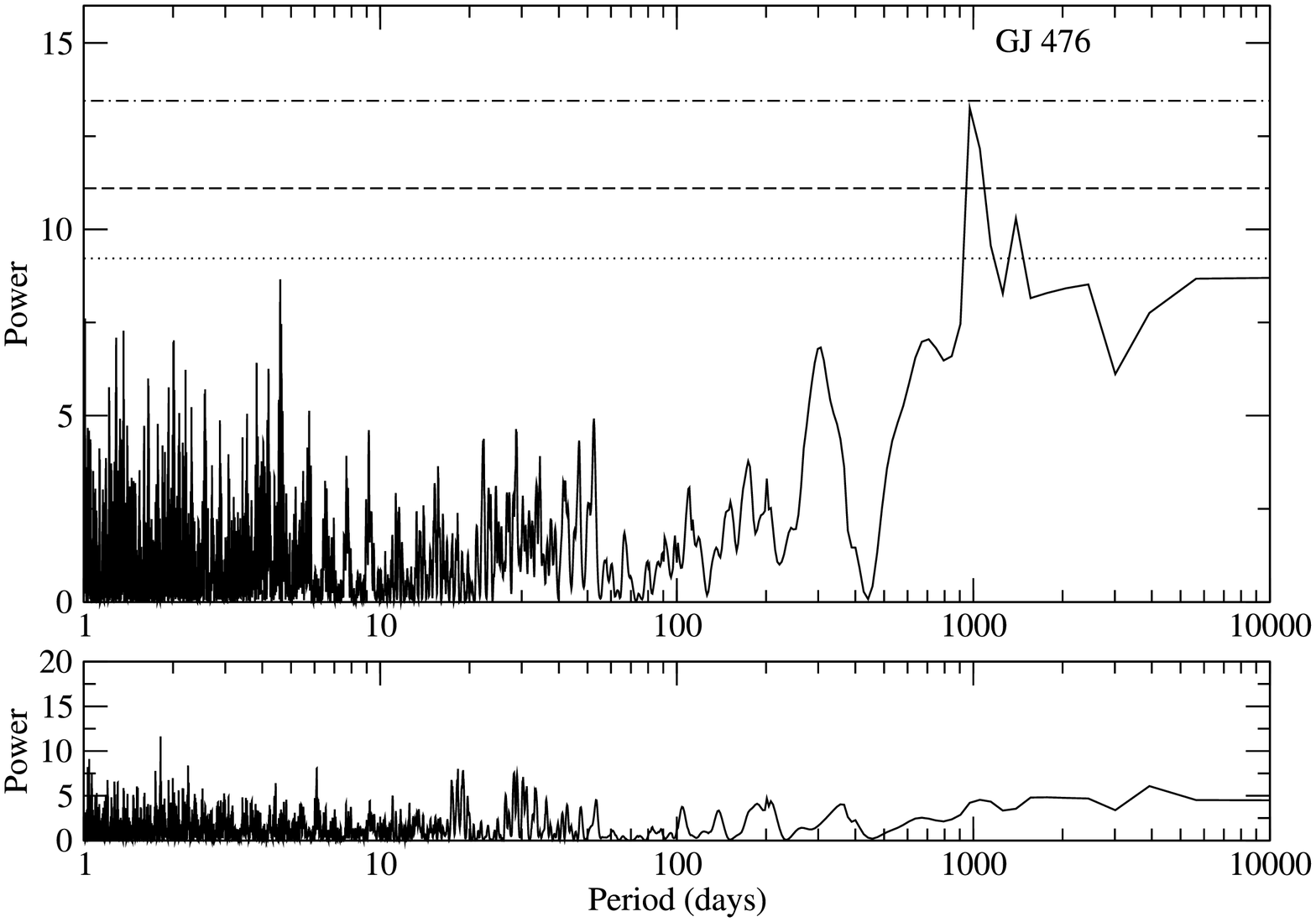}}
\subfigure[\label{gj581_ps}]{\includegraphics[scale=0.3]{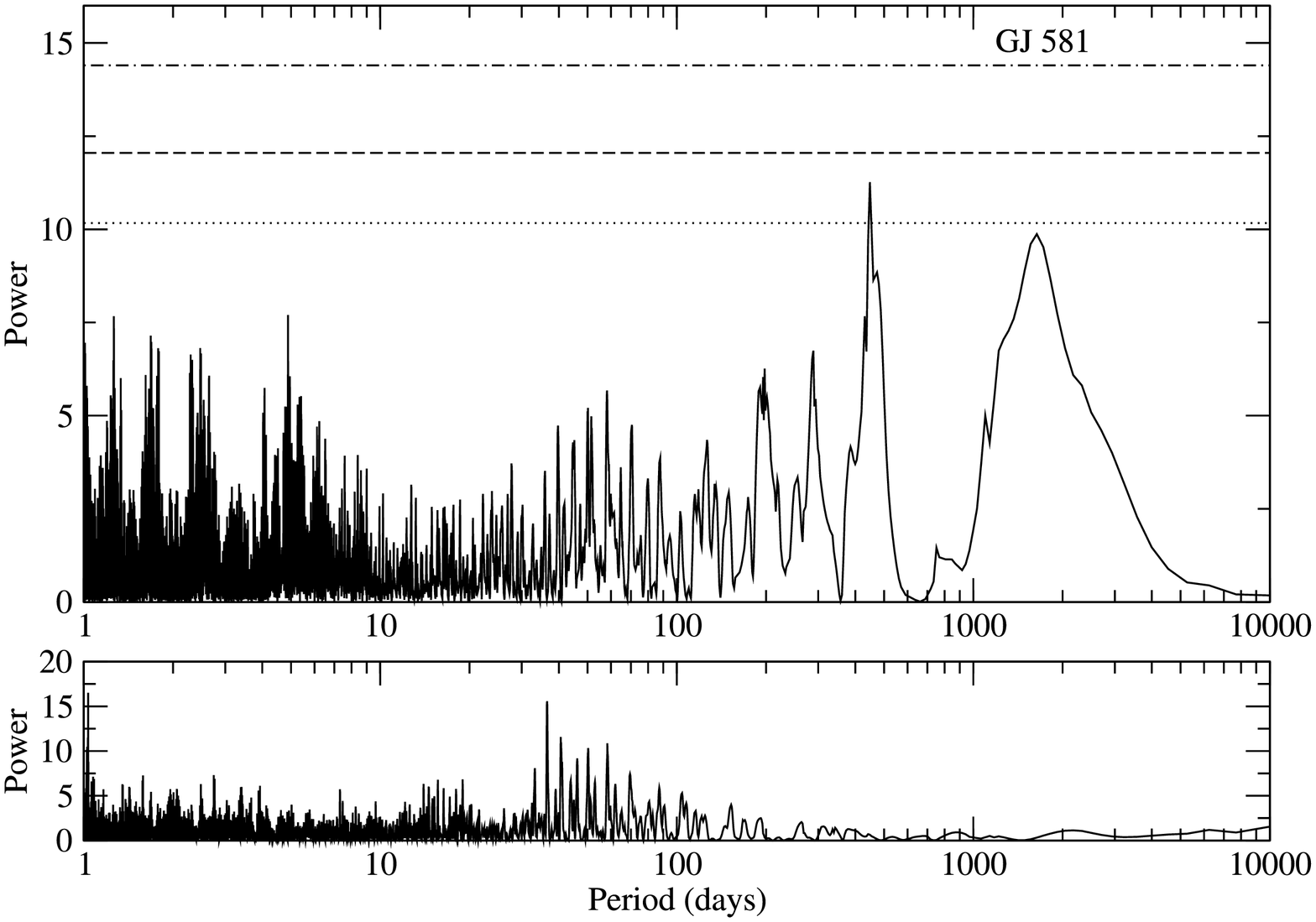}}
\subfigure[\label{gj708_ps}]{\includegraphics[scale=0.3]{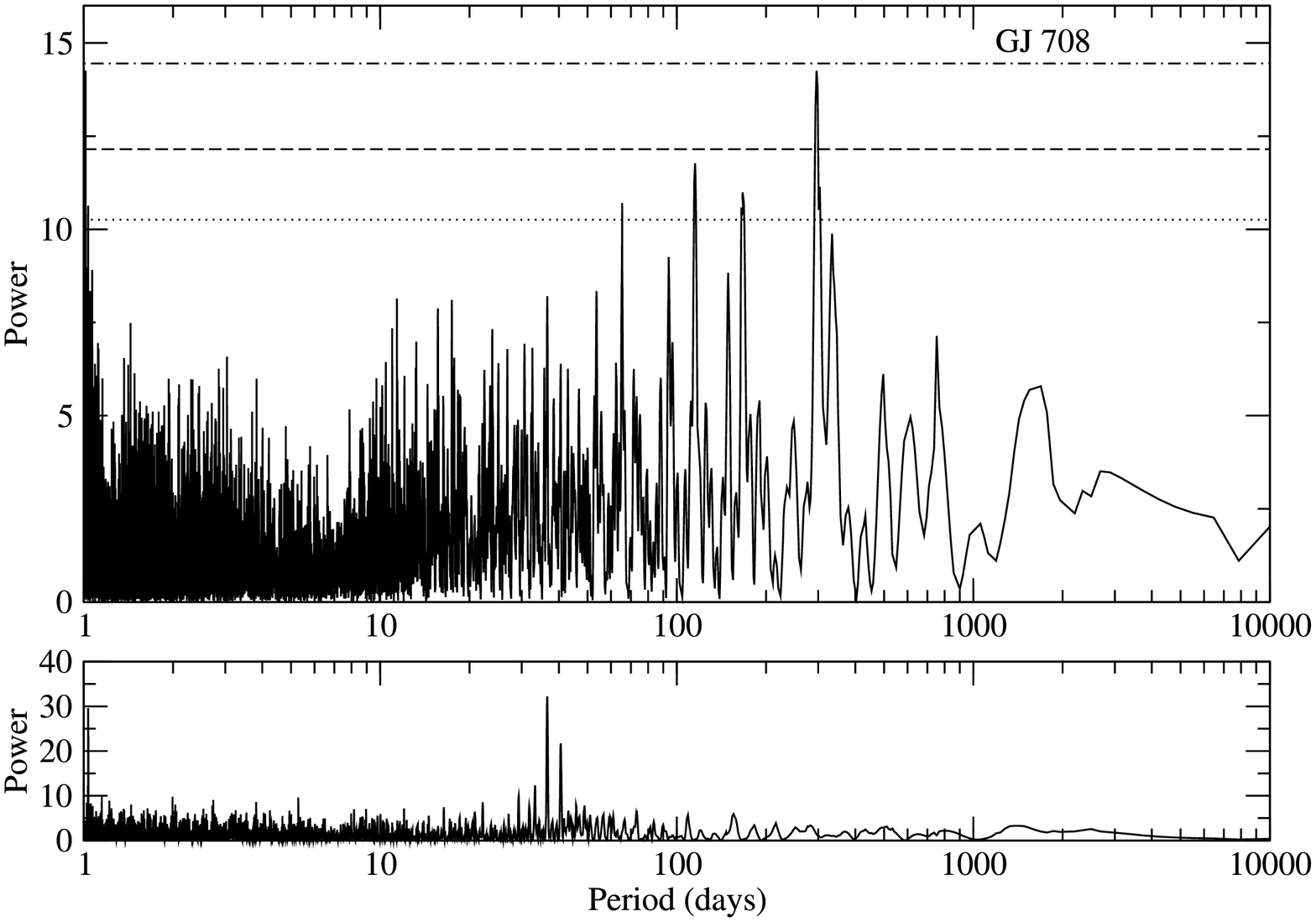}}
\subfigure[\label{gj730_ps}]{\includegraphics[scale=0.3]{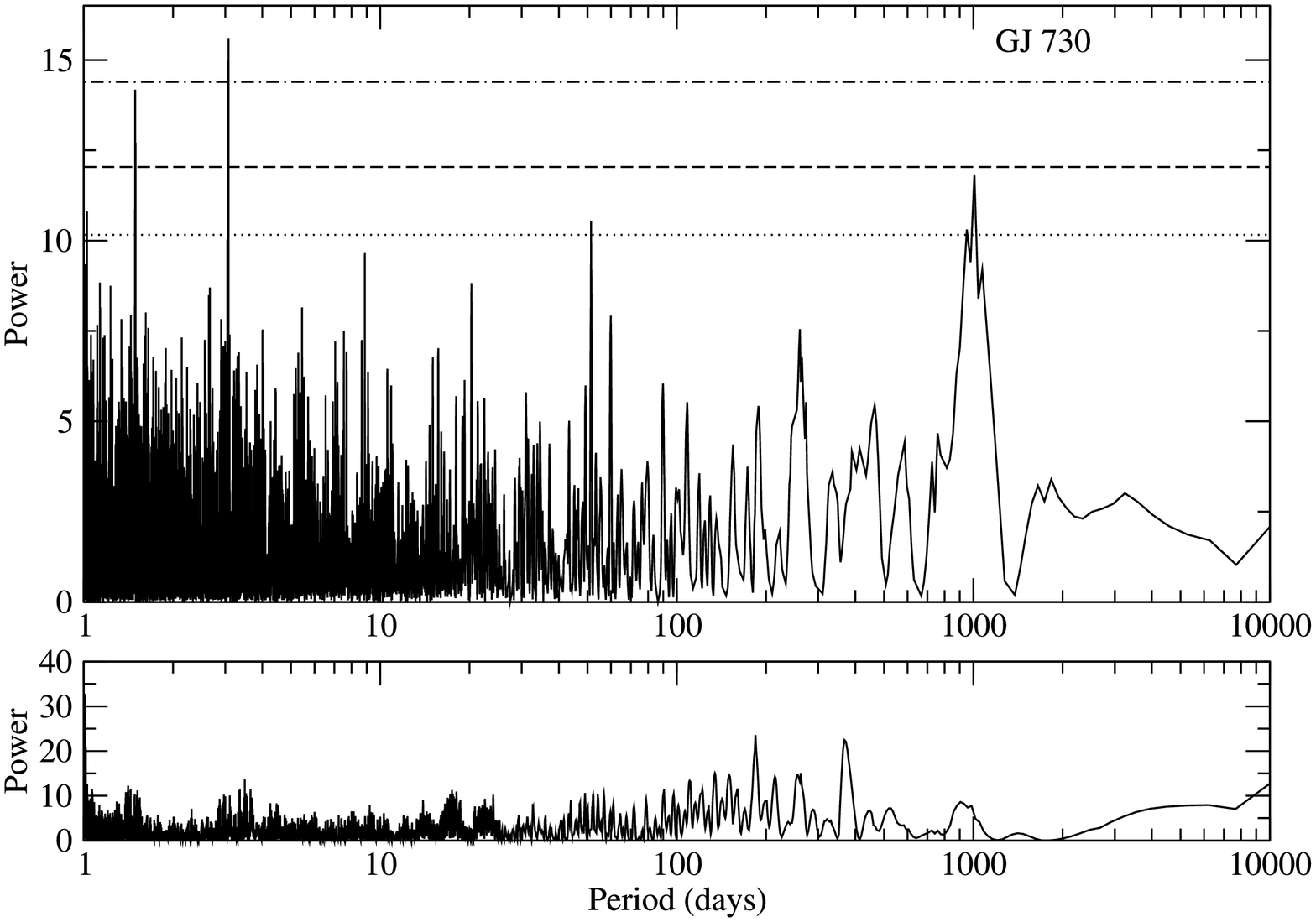}}
\subfigure[\label{gj552_ps}]{\includegraphics[scale=0.3]{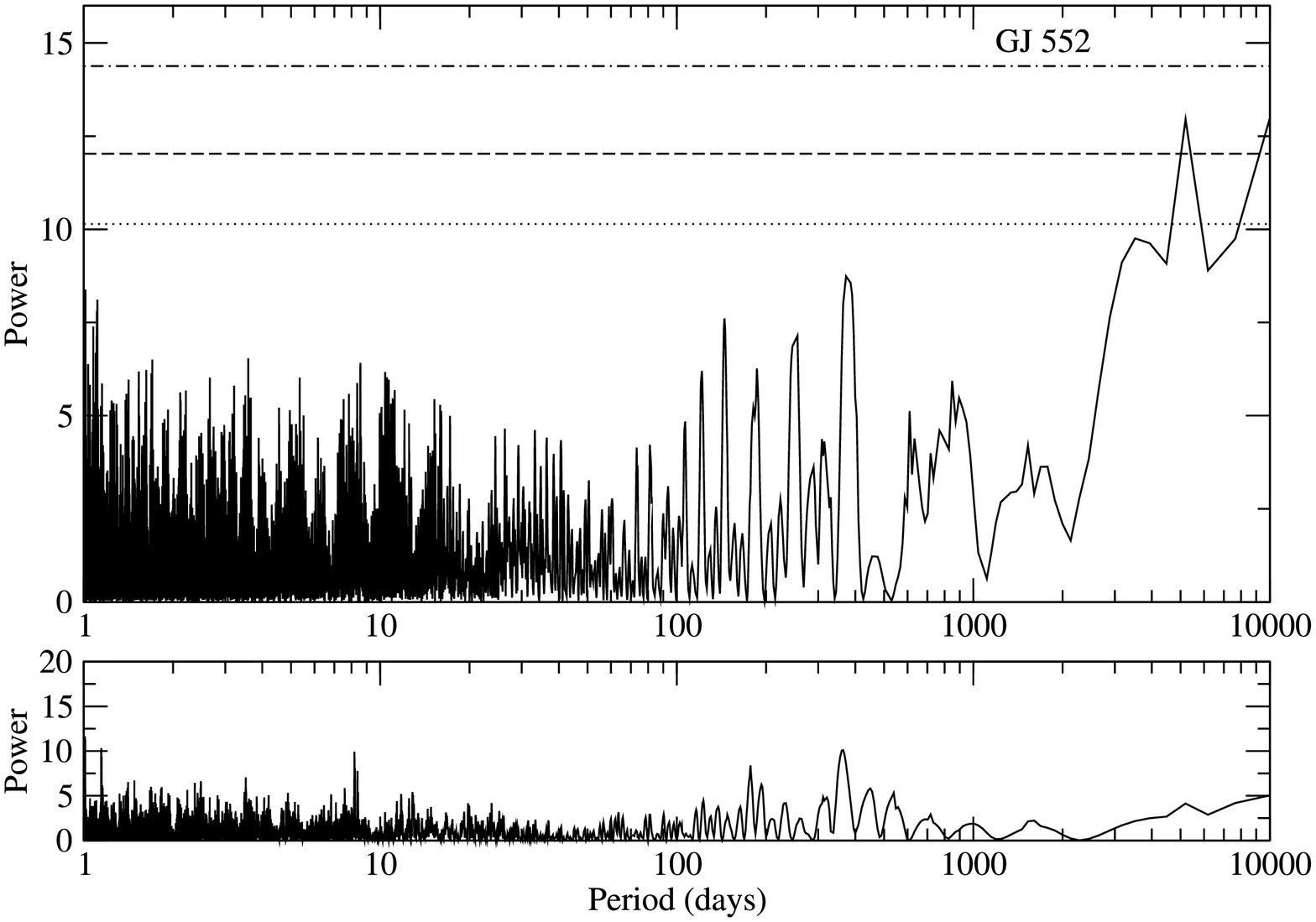}}
\caption{Fully generalized Lomb-Scargle periodograms for the six stellar activity cycles detected in our M dwarf sample.  Below each periodogram is the power spectrum of our time sampling (the window function).  The horizontal lines represent the power levels corresponding to a false alarm probability (FAP) of 0.5 (dotted line), 0.1 (dashed line), and 0.01(dash-dotted line), as calculated from Equation 24 of \citet{zk09}.  We note that these FAP levels represent a preliminary estimate, and our formal FAP values are obtained through a bootstrap analysis, which we give in Section 3.}
\label{periodogram}
\end{sidewaysfigure}

We take particular note of GJ 581, both because of the considerable interest in its planetary system \citep[e.g.][]{bonfils05b,mayor09,vogt12}, and because our observed periodicity confirms the discovery of \citet{gds12}.  It is important to note that our periodogram analysis (Figure \ref{gj581_ps}) shows a peak at 448 days with higher power than the longer-period signal claimed in \citet{gds12}.  However, since the longer-period peak leads to a marginally better fit--yielding an RMS of 0.00118 in \iha versus 0.00123--we adopt the 1633-day peak as the true period.  The 448-day peak disappears in the residuals to a 1633-day fit, indicating it is likely an alias of the fitted signal.

As seen in Figure \ref{periodogram}, the 1633-day signal of GJ 581 appears at somewhat lower power in our periodogram analysis.  In our bootstrap FAP analysis, our randomly-resampled \iha indices produced higher power spectrum values in 1196 of $10^4$ trials, for a FAP of 0.12.  However, since our observed period is similar to the 1407-day period derived from the HARPS data, and because our observed \iha maximum in December 2007/January 2008 matches the \ha maximum found by \citet{gds11}, we consider our detection valid despite having a FAP of $\sim 0.1$.  It is especially remarkable that we observe the cycle in \hap, while \citet{gds12} use the Na I D feature.  Furthermore, the reversal of the minima/maxima between our \ha index and the HARPS Na I index confirms the anticorrelation between \ha and Na I for GJ 581 observed in the HARPS survey \citep{gds11}.

\subsubsection{GJ 708}

With a mass $M_* = 0.77 M_{\odot}$, GJ 708 is at the upper limit of where the \citet{delfosse00} mass calibration is valid.  While the Orion Spiral Arm Catalog lists it as spectral type M1, its mass indicates GJ 708 is most likely a late K dwarf.  Our RV series shows a monotonic decreasing trend indicative of a binary companion, for which we are currently unable to estimate a period.  In many ways, then, GJ 708 is quite similar to GJ 270.

GJ 708 is unique among the objects presented here because it exhibits a long-term trend in \ha along with a periodic signal.  We note that we have fitted and removed a linear slope from the \iha series prior to creating the phase plot shown in Figure \ref{gj708_cycle}.  Once removing the trend, the periodogram on the residual \iha values (Figure \ref{gj708_ps}) shows significant power at 296 days, although we note that the peak appears regardless of whether the trend has been removed.  In $10^4$ bootstrap resamplings of the residual \iha values for GJ 708, we observed 282 false alarms, for a FAP of 0.03 on the 296-day period.  It is also important to point out that, while the period observed for GJ 708 is close to the 1 year alias, neither the periodogram of our time sampling (the window function) nor the Ca I periodogram showed periodicity near 296 days, eliminating the possibility that the observed signal is caused by our sampling.  

The 296 day period of GJ 708's activity cycle is too long to be the rotation period of a typical old dwarf star (normally weeks or months), but is also somewhat short to be reminiscent of a long, solar-type cycle.  On the other hand, it is quite similar to the $\sim 442$ day period observed for the M dwarf Proxima Centauri \citep{cincunegui07a}.  It is possible, then, that these two stars represent the first examples of a new class of intermediate-duration activity cycle that occurs in low-mass stars.  Alternatively, it may be the case that the linear trend we observe is the star's true ``activity cycle," and that the 296-day signal is an intermediate-period cycle that exists alongside the longer signal.  Such sub-cycles have been observed for the Sun \citep[e.g.][]{tan12}, suggesting similar physics may be at work here.

\subsubsection{GJ 730}

The periodogram for the time series \iha of GJ 730 (Figure \ref{gj730_ps}) shows interesting peaks at 3.06 days and 993 days.  While the longer periods of the other \iha signals in this paper make the longer period peak seem intuitively more likely to be the true signal, the 3-day signal continues to increase in power as we acquire additional data, while the 993-day peak remains constant.  While many of our stars exhibit short-term variability, only GJ 730 displays a coherent signal of such statistical significance; our bootstrap FAP estimate gives an upper limit of $10^{-4}$ for the 3-day peak, making it much more significant than the long-period peak.  We note that the 993-day peak is no longer present in the residuals around the 3-day fit, and also that the 3-day peak disappears from the residuals to a 993-day fit, giving further evidence to the two signals being aliases.  Because it is possible for an alias to display stronger periodogram power than a true signal, and because the longer period is much more typical of stellar activity cycles, we will not make a definitive argument as to which is the true period.  We include fits to both periods in Table \ref{fits}, and show phase plots of both signals in Figure \ref{gj730_cycle}.


If the true period for GJ 730 is 3 days, it is by far the shortest periodicity presented herein.  While 3 days could concievably be a rotation period for a very young, active star, it is almost certainly too short for a main-sequence M star, although we are unaware of any $v \sin i$ measurements which might confirm or reject the rotation hypothesis.  Hipparcos photometry \citep{esa97} is not sampled densely enough to be sensitive to a 3-day period.  At spectral type M2, with a mass of $0.62 M_{\odot}$ and [M/H] = 0.014, GJ 730 appears more or less typical of our sample, leaving no indication as to why such an unusually short periodicity should be present.  Its average \iha value of 0.064 is also close to the overall average for our targets, so the star is not especially active.

We note that 3 days is common for the periods of hot Jupiter planets.  Previous results \citep{shkolnik08,pillitteri11} have shown evidence that hot Jupiters may induce star-planet interaction (SPI), with measurable periodicity in stellar activity tracers matching the period of the planet's orbit.  Our RVs do not currently show evidence for planetary companions to GJ 730, and the RMS scatter of 7.4 m/s shows definitively that any hot Jupiter planet would need to be in an orbit which is highly inclined relative to the line of sight.  While the inclination requirement in combination with the inherent paucity of hot Jupiters around M stars \citep{endl06,johnson12} makes SPI an unlikely explanation for the 3-day activity signal observed for GJ 730, we present it as an interesting possibility.

On the other hand, if the $\sim 1000$ day signal is the true period, GJ 730 fits nicely with GJ 270, GJ 476, and GJ 581 as a star displaying solar-type activity cycles.  Further study will be required to determine the nature of this star's periodic activity.

\subsubsection{GJ 552}

GJ 552 is an extreme example of non-sinusoidal behavior in \ha activity cycles, as evidenced by the sharp upturn in \iha observed between July 2006 and April 2008.  As mentioned previously, while we recognize the inadequacy of a sine wave fit, we do not have a sufficient number of observations to fit a long Fourier series.  Furthermore, we see in Figure \ref{gj552_fourier} that our different fits offer quite dissimilar estimates for the period of the cycle.  While it is possible we have recorded close to a full period, some models result in a period far exceeding our observational time baseline.  For this reason, while we believe the behavior seen in our \ha index is likely periodic, we do not claim a specific period, and do not include a model fit in Figure \ref{gj552_trend}.

\subsection{Long-Term Activity Trends}

For a second subset of our target stars, we see long-term slopes or curvature in \ihap, but are unable to confirm any periodicity for these objects because our observational time baseline is shorter than the duration of any periodic signals.  We show the time-series \iha data for these stars--GJ 16, GJ 521, GJ 96, GJ 3023, GJ 3801, GJ 611.3, and GJ 630--in Figure \ref{trends}, along with the corresponding Ca I index.  The details and statistics of our fitting are included in Table \ref{fits} as well.  We separate the stars in Figure \ref{trends} into two groups: stars showing monotonic slopes, and objects for which we observe some curvature or turnaround.

\subsubsection{GJ 16, GJ 521, GJ 96, and GJ 3023}

GJ 16, GJ 521, GJ 96, and GJ 3023 all display a linear trend in \iha over our entire observational time baseline.  To ensure the statistical significance of these slopes, we perform a linear least-squares fit to each time series, and calculate the Pearson correlation coefficient.  We include these fits in Figures \ref{gj16_trend}-\ref{gj3023_trend}.  In all four cases, the correlation coefficient indicates the relations are highly significant (with the probability of a null result $p << 0.01$).

\subsubsection{GJ 3801, GJ 611.3, and GJ 630}

GJ 3801, GJ 611.3, and GJ 630 all show significant variability, but are poorly fit by linear regression.  For GJ 3801 and GJ 611.3, we have compared the quality of linear and quadratic fits with an F-test using 

$F_{\textrm{poly}} = (N-3)(\chi^2_{\textrm{slope}} - \chi^2_{\textrm{poly}})/\chi^2_{\textrm{poly}}$ 

\noindent as described in \citet{gds11}.  Here, $\chi^2_{\textrm{slope}}$ is the chi-squared value of a straight-line fit, while $\chi^2_{\textrm{poly}}$ is the chi-squared of a quadratic fit.  We find $P(F_{\textrm{poly}}) = 0.04$ for GJ 3801 and $P(F_{\textrm{poly}}) = 0.01$ for GJ 611.3, indicating the quadratic fits are a statistically significant improvement over straight-line models.  Our fits are included in Figures \ref{gj3801_trend}-\ref{gj611_trend}. In GJ 630, the recent dramatic upturn in \iha looks very similar to that of GJ 552.  If the \ha activity seen for GJ 630 does prove to be another highly non-sinusoidal periodic signal, its period will easily exceed 10 years, making it a direct analog to the 11-year solar cycle.  Since the behavior of \iha is approximately linear prior to this rapid upturn, we have not included a fit for GJ 630, pending future observations to more accurately characterize the shape of the curve.

\subsection{RV correlation}

Among our motivations for examining the stellar activity of our targets was to evaluate to what level, if any, chromospheric variability influenced our RV measurements.  To this end, we have searched for correlations between RV and \iha for our entire sample.  In cases where we see evidence for substellar companions (planets or brown dwarfs) in the RVs, we have subtracted those signals and examined the residual velocities.

Overall, we see little correlation between \iha and RV.  We find this result unsurprising for our data; \citet{gds12} report correlations between stellar activity indicators and RV typically occur at or below the 5 m/s level.  Such an effect will doubtless be significant for our sample following the HET/HRS upgrade, which will yield a throughput gain of approximately 2 magnitudes  at our R = 60,000 setting, dramatically improving the RV precision for our faint M dwarf targets.  However, with our current RV precision limited to $\sim 6$ m/s for the majority of our sample, it is understandable that we see relatively few correlations between \iha and RV.  Furthermore, \citet{lovis11} observe a decrease in correlation between RV and $R'_{\textrm{HK}}$ with decreasing $T_{eff}$, so it is likely we will see relatively fewer M stars with RV-\ha correlation regardless of precision.

On the other hand, we do see two examples--GJ 1170 and GJ 3801--for which RV is anticorrelated with \iha at a level higher than 5 m/s.  We show our measured RVs and the corresponding \ha indices for these stars in Figure \ref{harv}.  For GJ 1170, we derive a Pearson correlation coefficient of -0.48 for the relation between RV and \ihap, yielding a probability $p = 0.005$ of no correlation.  Similarly, for GJ 3801 the Pearson correlation coefficient is -0.51, indicating the probability of no correlation is just 0.03.  

GJ 1170 is a particularly interesting case, as the \ha emission is reflected in the RVs, resulting in a signal which initially appeared to be a Keplerian planet orbit with RV amplitude $K \sim 20$ m/s.  Assuming a stellar mass of $M_* = 0.52 M_{\odot}$, this signal corresponds to a planetary minimum mass of approximately 0.9 Jupiter masses.  We therefore conclude that, while \iha does not frequently correlate with RV for M dwarfs at a precision level higher than 5 m/s, there does exist a small subset of stars where \ha activity corresponds to large RV shifts.  With such a small number of objects, though, we must concede the possibility that these stars actually have giant planets causing the observed RVs, and coincidentally exhibit activity cycles with periods similar to those of the planets (as with Jupiter and the solar cycle).

\subsection{Stellar Activity Levels}

In addition to examining time-series activity of individual stars, we have also considered the mean activity levels of our targets relative to the overall sample.  To do so, we average our \ha index over all observations of a star.

Based on their observed positive correlation between time-averaged \ha index and $V - I$ color, \citet{gds11} report that mean activity is a function of temperature.  We attempted to confirm this correlation with our sample, but find considerably larger scatter than previously observed, as shown in Figure \ref{VmI_Ha}. While the relation holds relatively well within the $1.6 \le V - I \le 2.8$ color range covered by the \citet{gds11} sample, it is much less suitable outside that regime.  While we looked at mean \iha as a function of several color indices, we found that no color was a satisfactory tracer of \hap.

We considered the possibility that stellar mass might be a better predictor of mean \ha activity than temperature or color.  In Figure \ref{sha_mass}, we show our time-averaged \ha index as a function of stellar mass.  The resulting anticorrelation is obvious, and the scatter is greatly reduced from that of the color relations.  It is apparent that \ha emission is much more tightly connected to mass than temperature.  Performing a linear least squares fit to the data, we find the relation

\begin{equation}
\label{sha_mass_eq} 
\langle I_{\textrm{H}\alpha} \rangle = 0.094 - 0.050\frac{\textrm{M}_*}{\textrm{M}_\odot}
\end{equation}

\noindent Our fit gives $1 \sigma$ uncertainties of 0.003 on the slope and 0.001 on the intercept.

We explored the residuals to the fit above to determine whether additional dependencies exist.  As before, we see no convincing correlation to any color index.  On the other hand, stellar metallicity does appear to play a role.  Figure \ref{sha_z} shows the residual \ha index after subtracting the \ha-M$_*$ relation.  The resulting correlation is a clear indicator that stellar [M/H] is the primary contributor to the scatter around the mass-activity fit.  Again performing a linear least-squares fit to our data, we find

\begin{equation}
\label{sha_z_eq}
\langle I_{\textrm{H}\alpha} \rangle_{\textrm{res}} = 7.6 \times 10^{-4} + 9.7 \times 10^{-3}\textrm{[M/H]}
\end{equation}

\noindent with $1 \sigma$ uncertainties of $10^{-3}$ on the slope and $3 \times 10^{-4}$ on the intercept.

Before drawing conclusions as to the physical interpretation behind Figure \ref{sha}, it is important to understand how the stellar continuum changes as a function of mass and metallicity.  Since \iha is normalized by the continuum adjacent to the \ha line, it is possible that the observed trends are due to changes in continuum flux rather than \ha emission.  We considered this possibility by converting our \ha index to an equivalent width (EW) using a transformation basically identical to the one outlined in Appendix A of \citet{zechmeister09}, then to the quantity \lhap, the logarithm of the luminosity in \hap, normalized by the star's bolometric luminosity.  For this transformation, we used the method of \citet{walkowicz04}, which calibrates the ratio of the continuum near \ha to the bolometric luminosity as a function of color for M dwarf stars.  Briefly, we use the transformations

\begin{equation}
\label{ewha}
\textrm{EW}_{\textrm{H}\alpha} = \Delta \lambda_{\textrm{H}\alpha}(1 - \frac{2\Delta \lambda_C}{\lambda_{\textrm{H}\alpha}}I_{\textrm{H}\alpha})
\end{equation}

\begin{equation}
\label{lha_trans}
[L_{\textrm{H}\alpha}/L_{\textrm{bol}}] = \chi \textrm{EW}_{\textrm{H}\alpha}
\end{equation}

\noindent Where $\lambda_C = 8.75$ \AA~is the width of each reference band used to compute \ihap, $\lambda_{\textrm{H}\alpha} = 6562.808$ \AA, $\Delta \lambda_{\textrm{H}\alpha} = 1.6$ \AA~is the width of the spectral window in which we sum the \ha flux, and $\chi$ is the luminosity scaling factor, calculated according to \citet{walkowicz04}. In Figure \ref{lha}, we again show \ha emission as a function of stellar mass and metallicity, but measure \ha emission with \lha instead of \ihap.

Perhaps surprisingly, both of the trends seen in Figure \ref{sha} are reversed in Figure \ref{lha}.  Our fit to \lha as a function of stellar mass gives

\begin{equation}
\label{lha_mass_eq}
\langle [L_{\textrm{H}\alpha}/L_{\textrm{bol}}] \rangle = -4.17 + 0.87\frac{\textrm{M}_*}{\textrm{M}_\odot}
\end{equation}

with residuals anticorrelated with stellar metallicity.  A linear least squares fit to Figure \ref{lha_z} yields

\begin{equation}
\label{lha_z_eq}
\langle [L_{\textrm{H}\alpha}/L_{\textrm{bol}}] \rangle_{\textrm{res}} = -0.019 - 0.27\textrm{[M/H]}.
\end{equation}

\noindent Our $1\sigma$ uncertainties on the slopes and intercepts, respectively, are 0.08 and 0.04 for Equation \ref{lha_mass_eq}, and 0.01 and 0.04 for Equation \ref{lha_z_eq}.

At first, the results of Equations \ref{sha_mass_eq}-\ref{sha_z_eq} and \ref{lha_mass_eq}-\ref{lha_z_eq} appear to present conflicting information regarding the dependence of stellar activity on mass and metallicity.  However, we suggest a conceptually simple hypothesis consistent with all four relations.  It has been firmly established that the stellar continuum near \ha varies along with bolometric luminosity \citep{hall96,walkowicz04,cincunegui07b}.  According to the \citet{cincunegui07b} $F_{\lambda 6605}$ versus $B - V$ relation, we expect the continuum around \ha to differ by up to a factor of 5 across the range of $B - V$ typical of our sample.  Thus, since \lha spans a factor of $\sim 4$ over our sampled range of stellar mass, we would expect a $\sim 80$ percent decrease in \iha over the same range (assuming $\Delta I_{\textrm{H}\alpha} \sim \frac{\Delta L_{\textrm{H}\alpha}}{\Delta F_{\lambda 6605}}$).  The observed \iha decrease of $\sim 60$ percent is fully consistent with this expectation, considering the significant uncertainty in the Cincunegui relation at high $B - V$.  We therefore conclude that chromospheric \ha emission is increasing with stellar mass, but the effect does not appear in our original \ha index because of the increasing continuum flux.

Understanding the metallicity dependence for \ha emission requires a similar deconstruction of stellar luminosity/continuum effects.  When we subtract a linear fit to the \iha - M$_*$ relation, the effects of the varying stellar continuum should be removed, suggesting higher-metallicity stars are more active.  However, verifying this demands explaining the decrease in \lha with [M/H].  Previous observations of M dwarf activity have shown that more active stars are more luminous, possibly due to an increase in stellar radius \citep{bochanski11}.  Furthermore, stars with high [M/H] will be more luminous at a given $B - V$ or $V - K$ color than metal-poor stars due to line blanketing of blue light in the more metal-rich objects.  Figures \ref{sha_z} and \ref{lha_z} are therefore consistent with the argument that metal-rich M dwarfs display higher \ha emission, but their increased activity and opacity in the blue cause $L_{\textrm{bol}}$ to increase faster than $L_{\textrm{H}\alpha}$, causing the negative slope seen in Figure \ref{lha_z}.

Earlier studies of M dwarf activity \citep[e.g.][]{west08,browning10} emphasize the importance of considering the various subtypes of M stars, particularly around M4, where stars typically become fully convective.  Figure \ref{type_LHa} shows the time-averaged \lha values for objects of known subtype in our sample.  In the Figure, the error bars represent the RMS scatter within a given bin, rather than measurement error.  The decrease in \lha towards later types essentially reflects the observed mass dependence.  However, we see more clearly in Figure \ref{type_LHa} the tendency for more active bins to have more scatter as well.

\section{\bf Discussion}

\subsection{Stellar Cycles}

Our observed stellar activity cycles indicate that even for an old, quiet stellar population, magnetic activity may appear at essentially any period.  The long-period subset of stars shown in Figure \ref{cycles}, particularly GJ 270 ($P \sim 7.5$ years), demonstrate stellar periodicity on the order of the 11-year solar cycle is far from unique; stars with cyclic activity of period $\gtrsim 1$ year appear in at least $5\%$ of our sample.  In future work we will include additional tracers \citep[e.g. the Na I D feature, see][]{diaz07a}, both to increase sensitivity to additional activity cycles, and to fully evaluate the impact of stellar variability on our RV survey.  The addition of other tracer indices will almost surely increase the number of candidate signals among our targets. 

While the periods of the stellar cycles observed in Figure \ref{cycles} fall towards the low end of the period distribution found by \citet{baliunas95} for solar-type stars, our results are entirely consistent with the Mount Wilson survey.  While the Baliunas distribution peaks between $\sim 5-12$ years, there are a significant number of detections between 1 and 5 years.  Furthermore, our observational time baseline, while extensive, is still too short to properly characterize stellar cycles with $P \ge 10$ years.  It is quite likely that the six stars shown in Figure \ref{trends} are examples of partial detections of such periods.  Clearly, it is essential that both the durations and number of targets be increased for M dwarf activity surveys for a proper comparison to FGK stars.

\citet{lovis11} find activity cycles with amplitude 0.04 or higher in $R'_{\textrm{HK}}$ for $61 \pm 8 \%$ of quiet FGK stars.  At present, the percentage of M dwarfs with observed magnetic cycles is significantly lower.  Using the Na I D feature, \citet{gds12} find periodic signals for 5 of 27 M stars surveyed, for a rate of $19 \%$.  With 6 periodic signals found in 93 stars, our resulting ``cyclic fraction" is even lower, at $6.5 \%$.  If we assume the 7 stars for which we observe long-term trends are exhibiting activity cycles, our rate increases to $14 \%$, which is more in agreement with the \citet{gds12} value.  It is possible that the difference in the periodic fractions is due to the relative sensitivites of the activity indices.  Because the sodium feature is emitted deeper in the chromosphere than \hap, where the local density is higher, a change in the magnetic field should produce relatively more emission in sodium than hydrogen, making the amplitude of a given stellar cycle higher in the Na I index than in \ihap.  Our overall sample size, and the number of observed stellar cycles and trends in \ha will allow us to make a thorough comparison between the two indices in future work, as we will have a robust set of comparison cases to establish whether the Na I D feature is simply more sensitive to weaker signals, or in fact probes different stellar physics.

In any case, current data suggests M dwarfs are less likely to display measurable magnetic periodicity than earlier spectral types.  A possible explanation for the discrepancy lies in the lack of a tachocline for many M stars.  At a spectral subtype of M4 and below, stars become fully convective \citep{chabrier97} and therefore do not possess a solar-type dynamo, which is driven by differential rotation between the radiative and convective layers.  If the solar cycle and its analogs are driven by this dynamo,  then we should expect to see only low-amplitude cycles or quasi-periodic oscillations for fully convective stars.  Indeed, of the stars we identify as having cycles or trends, only GJ 476 and GJ 581 have spectral subtypes later than M3 according to \citet{bobylev06}.  Of those two, GJ 476 has an estimated mass of $0.47 M_{\odot}$, which is massive enough to have a tachocline despite its listed type of M4, and the activity cycle of GJ 581 may be connected to its planetary system (see below).  Furthermore, Figure \ref{type_LHa} shows a distinct dropoff in \lha around spectral subtype M3-M4, suggesting the fully convective stars in our sample are characteristically less active.  If continued study shows stellar activity cycles in M stars to be confined mainly to subtypes M0-M3, it will be an intriguing piece of evidence in favor of the tachocline dynamo as the origin of stellar magnetic cycles.

The period of the \iha cycle seen for GJ 270 is nearly equal to that of GJ 752 A \citep{buccino11}, making it one of the two longest M dwarf activity cycles observed to date.  It is worth noting that both of these stars are members of binary systems \citep{endl06,buccino11}.  Among the stars in Figures \ref{cycles} and \ref{trends}, we see evidence in our RVs for a brown-dwarf binary companion to GJ 708, and GJ 581 hosts a known multi-planet system.  \citet{morgan12} have shown that M dwarfs with close binary companions display increased magnetic activity, and the SB2 GJ 375 has been shown to exhibit a long-period activity cycle in the Ca K index \citep{diaz07}.  Likewise, \citet{krejcova12} see tentative evidence of increasing Ca II emission for stars in known exoplanet systems.  Interestingly, of the binary/planetary systems for which stellar cycles have been observed, none of the stellar activity periods match the companion orbital periods or their harmonics.  Further study will be required to definitively establish or reject a physical connection between more distant/less massive substellar companions and stellar activity cycles, but it is worth noting that a number of our candidates exhibit both, as do those of previous studies.

\subsection{Magnetic Activity and RV Planet Surveys}

Aside from the obvious value to the study of low-mass stellar physics, the understanding of M dwarf activity cycles has important implications for RV planet surveys.  It has been well established that periodic RV signals can be caused by stellar chromospheric variability, as traced by photometry or activity indicators such as Ca H and K \citep{queloz01,forveille09,robertson12a}.  While activity-induced RV signals are typically expected to have amplitudes of just a few m/s, Figure \ref{harv} shows that exceptionally large-amplitude effects will occasionally appear.  In general, though, these results will be increasingly important as M dwarf RV surveys approach precisions of 1 m/s or less.  GJ 581 stands as an excellent example.  We notice a striking similarity between our periodogram of the GJ 581 \ha index (Figure \ref{gj581_ps}) and the periodogram of the residual RVs to a 4-planet fit shown in \citet[][Figure 3, panel 5]{vogt10}, particularly at the $\sim 450$- and $\sim 1600$-day peaks.  The 433-day period of the unconfirmed planet f is very close to our observed alias at 448 days, as is the candidate 400-day planet identified in \citet{gregory11}.  While fitting a Keplerian orbit to an activity-induced RV signal would essentially serve to correct the RVs for stellar activity, if the true period of GJ 581's \iha cycle is 1300-1600 days, then removing a 433-day orbit would lead to incorrect residuals, complicating the analysis of any additional planets.  Clearly, any RV detection of low-mass planets around M dwarf stars should be accompanied by a careful analysis of the corresponding stellar activity indices.

\subsection{Mean Activity Levels}

Our results for the comparative activity levels between the stars in our sample are interesting in the context of previous surveys of more active M stars.  Figure \ref{type_LHa} shows essentially identical behavior to the findings of \citet[][Figure 5]{west04} and \citet[][Figure 7]{kruse10}.  We note a slight discrepancy from \citet[][Figure 12]{morgan12}, though, who do not see a decrease in \lha until spectral type M4.  Whether this behavior extends to solar-type stars is somewhat unclear based on current results.  \citet{lovis11} report that most G stars have $\log R'_{\textrm{HK}}$ around $-5$, while stars of K type cluster around $\log R'_{\textrm{HK}} \sim -4.7$, suggesting an inversion of the relation observed for M stars.  These results could still be reconciled, since stellar bolometric luminosity increases at earlier spectral types, so some normalization may be required to properly understand how mean activity levels vary across the entire main sequence.  Of course, it is known that \ha behaves differently from Ca H and K, especially for cool dwarfs such as those in our sample \citep[e.g.][]{cincunegui07b}.  A more harmonious result can be seen in \citet{baliunas95}, who plot mean $\log S_{\textrm{HK}}$ against $(B - V)$ color for FGK stars, finding an increasing trend (and scatter) similar to Figure \ref{VmI_Ha}.

The data presented here are perhaps more compelling when shown as a function of stellar mass, though.  Because our targets are pre-selected to be ``quiet," and do not show \ha in emission, they should all be older than their ``activity lifetime" \citep{west08}.  Figure \ref{lha_mass} seems to suggest that for these quiet stars, the dynamo responsible for driving the residual magnetic actvity following the spin-down phase is more effective in more massive stars, which presumably are not fully convective.  Again, it is possible that the presence of a more significant radiative zone in the interiors of the more massive stars leads to the generation of a solar-type magnetic dynamo at the interface of the convective and radiative layers, rather than the rotation-dominated magnetic activity of younger, highly active M dwarfs.

As explained above, our data indicate that metal-rich stars tend to be more active than average at a given stellar mass.  One possible cause for this relation is that our measured [M/H] is--in a statistical sense--a tracer of age, and we are simply observing the previously-confirmed connection between stellar age and chromospheric variability \citep[e.g.][]{west08,bochanski11}.  To test this hypothesis, we have obtained space velocities for our sample from \citet{bobylev06}.  In Figure \ref{sigma}, we plot the mean galactic velocity dispersions of various metallicity bins in the $UVW$ coordinates, excluding very high and low metallicities due to small numbers of stars.  In general, we see that low-metallicity stars have higher velocity dispersions in all three coordinates, lending support to the proposal that the metal-rich stars in our sample are younger.  Further evidence of the age-metallicity relationship is discussed in \citet{west08}.  The fact that we see higher average \iha values towards increasing [M/H] is therefore likely a reflection of the younger members of our sample being more active.  Again, all of the stars discussed herein are old, and should have evolved past their early active stages.  It is remarkable, then, that the effects of stellar age are still visible in the activity levels of these objects.  It is worth pointing out that, while the $UVW$ velocities suggest a spread in age among our objects, the probability distributions of \citet{reddy06} indicate all of our targets are members of the Galactic thin disk with $\ge 99 \%$ probability, so our observed mass and metallicity/age dependencies may or may not apply as cleanly to stars in the thick disk or halo.

The implications of this study are potentially encouraging for the prospects of current and future near-infrared (NIR) RV surveys for planets around late M stars.  The old, low-mass stars which will make up the majority of the targets should have relatively low chromospheric activity levels, and will be less likely to exhibit activity cycles.  Furthermore, we have shown that the stellar cycles which may be confused with planetary signals appear in \hap, which is much more likely to be accessible to a NIR spectrograph than Ca H and K.  The CARMENES spectrograph \citep{quirrenbach11}, in particular, will be able to acquire \hap.  Therefore, problematic signals may be properly identified and corrected.  While the Habitable Zone Planet Finder on HET \citep{mahadevan10} will not have wavelength coverage at \hap, our results illustrate the necessity of identifying a suitable activity tracer in the NIR for RV surveys.

\section{\bf Conclusion}

We have measured and analyzed time-series \ha emission for $\sim 11$ years of spectra for 93 low-mass stars.  Our primary results may be summarized as follows:  

\emph{1}.  We find strong periodic \ha variability for 6 stars in our sample, and long-term trends or curvature for 7 others.  Among these, we confirm the discovery of the activity cycle for GJ 581 found by \citet{gds12}.  Our observed stellar cycles are analogous to those previously observed for FGK stars.

\emph{2}.  While activity-induced RV signals are relatively rare in our sample at its current precision level, we identify two stars--GJ 1170 and GJ 3801--for which our \ha index is anticorrelated with RV, indicating activity cycles causing RV variations of nearly 20 m/s.  Additionally, we see evidence that the signal of the purported planet f in the GJ 581 system may in fact be caused by an alias of the $\sim 1600$-day stellar activity cycle, further emphasizing the need to consider stellar RV modulation at the $\le 3$ m/s level of precision.

\emph{3}.  The mean \ha activity levels of our targets correlate with stellar mass.  This result is qualitatively consistent with previous observations of both solar-type and low-mass stars, although the correlation has more often been expressed in terms of stellar temperature or color, rather than mass.  We confirm that \lha decreases towards later stellar subtypes, a phenomenon that appears to hold true for M stars regardless of whether or not they are in the ``active" phase.  Lower \lha values and an apparent lack of observable stellar cycles for fully convective stars may be related to the absence of a tachocline for those stars.

\emph{4}.  In addition to the mass-activity relation, we see a metallicity dependence on average \ha emission.  Lower average galacitc velocity dispersions indicate the more metal-rich stars in our survey are in fact younger.  Even though all the stars in our sample have aged beyond their ``active" phase, it appears some residual age effect still contributes to their mean activity levels, and that the younger of our targets are still more active at a given stellar mass.

\begin{acknowledgements}
P.~R. thanks Andrew West, Jason Wright, and Edward Robinson for valuable discussions.  P.~R. is supported by a University of Texas Continuing Fellowship.  This material is based on work supported by the National Aeronautics and Space Administration under Grant NNX09AB30G through the Origins of Solar Systems program, and Grant NNG04G141G.  The Hobby-Eberly Telescope (HET) is a joint project of the University of Texas at Austin, the Pennsylvania State University, Stanford University, Ludwig-Maximilians-Universit\"{a}t M\"{u}nchen, and Georg-August-Universit\"{a}t G\"{o}ttingen.  The HET is named in honor of its principal benefactors, William P. Hobby and Robert E. Eberly.  We would like to thank the McDonald Observatory TAC for generous allocation of observing time.  We are grateful to the HET Resident Astronomers and Telescope Operators for their valuable assistance in gathering our HET/HRS data.
\end{acknowledgements}

\clearpage

\begin{sidewaysfigure}
\subfigure[\label{gj270_cycle}]{\includegraphics[scale=0.3]{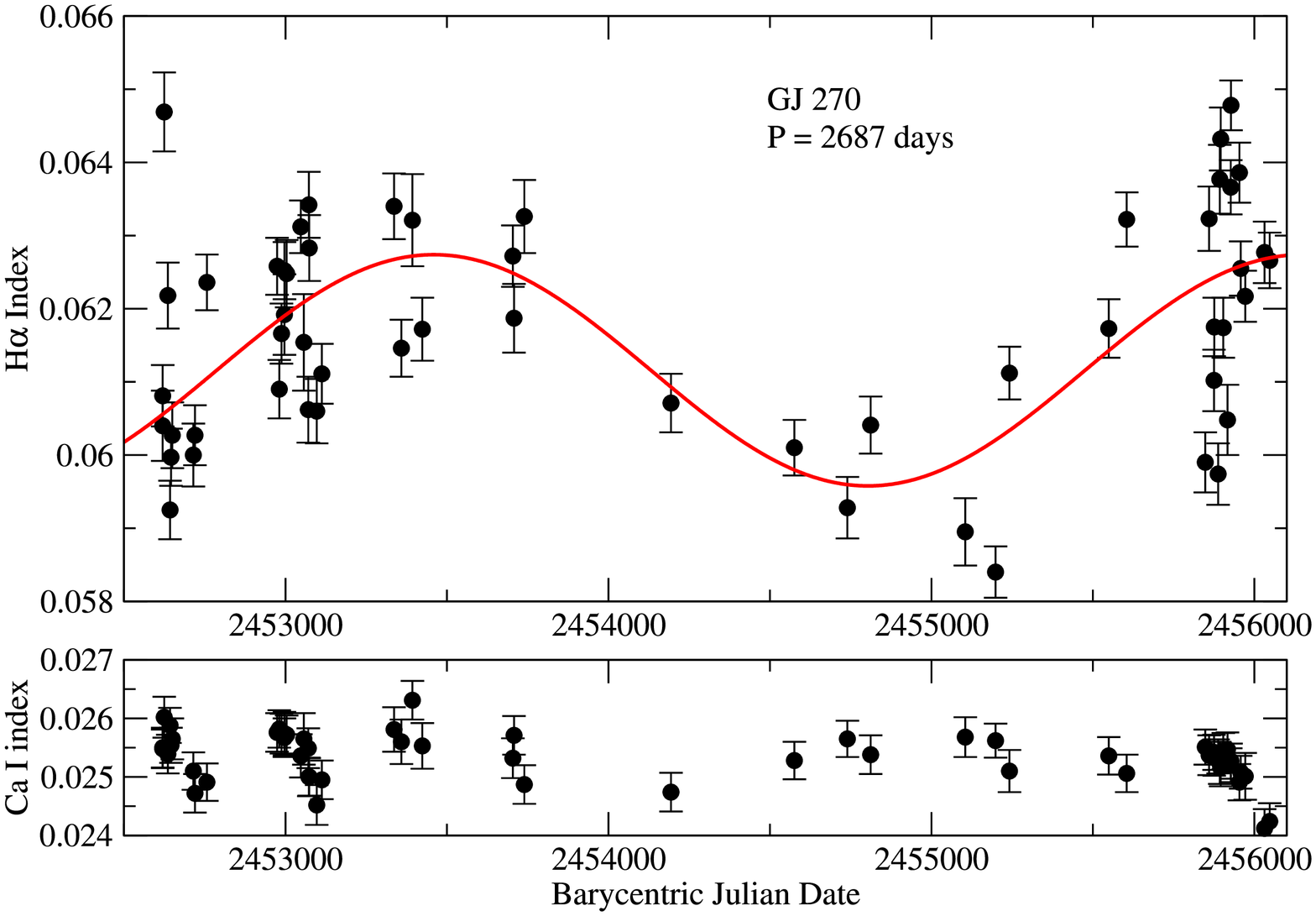}}
\subfigure[\label{gj476_cycle}]{\includegraphics[scale=0.3]{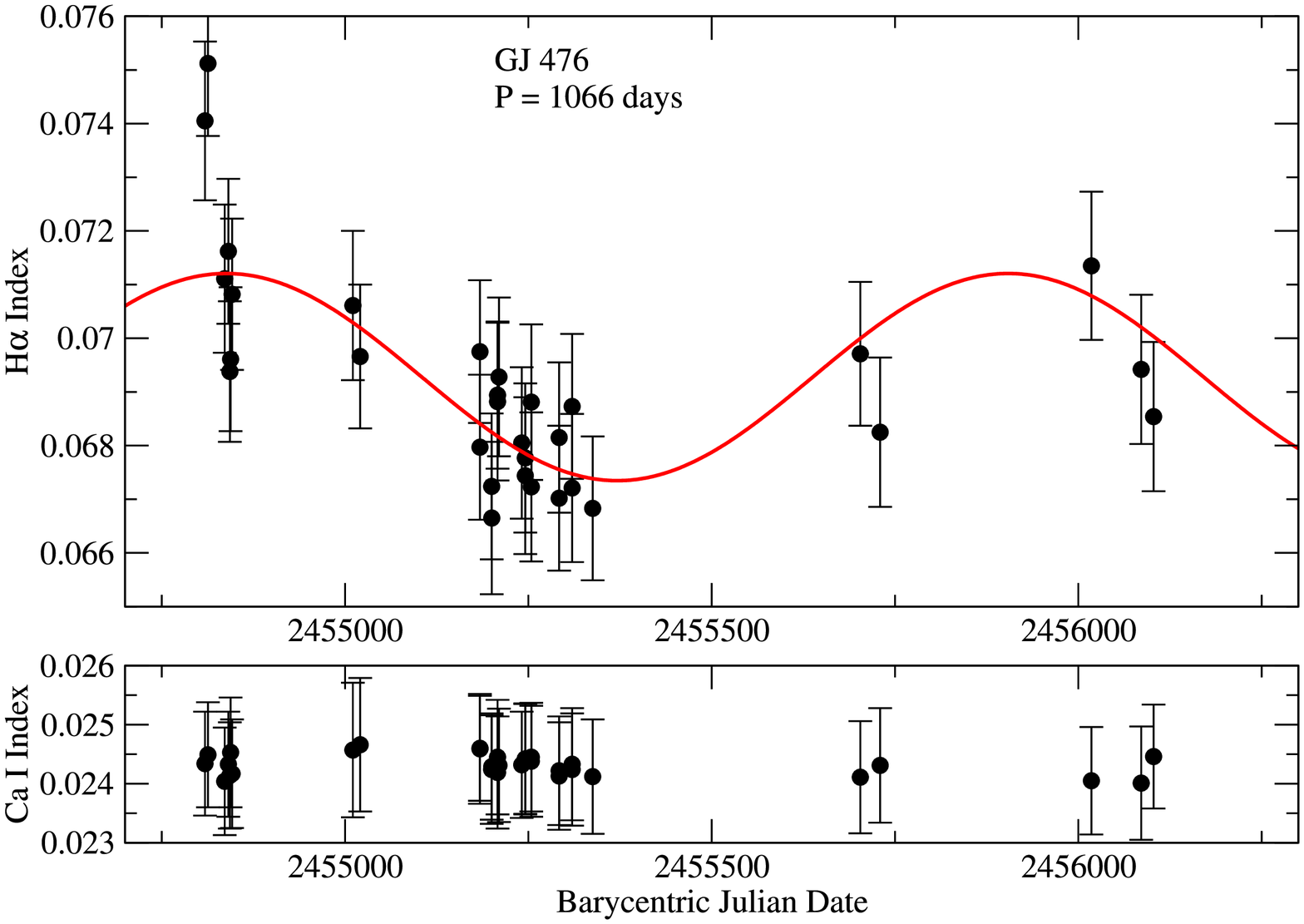}}
\subfigure[\label{gj581_cycle}]{\includegraphics[scale=0.3]{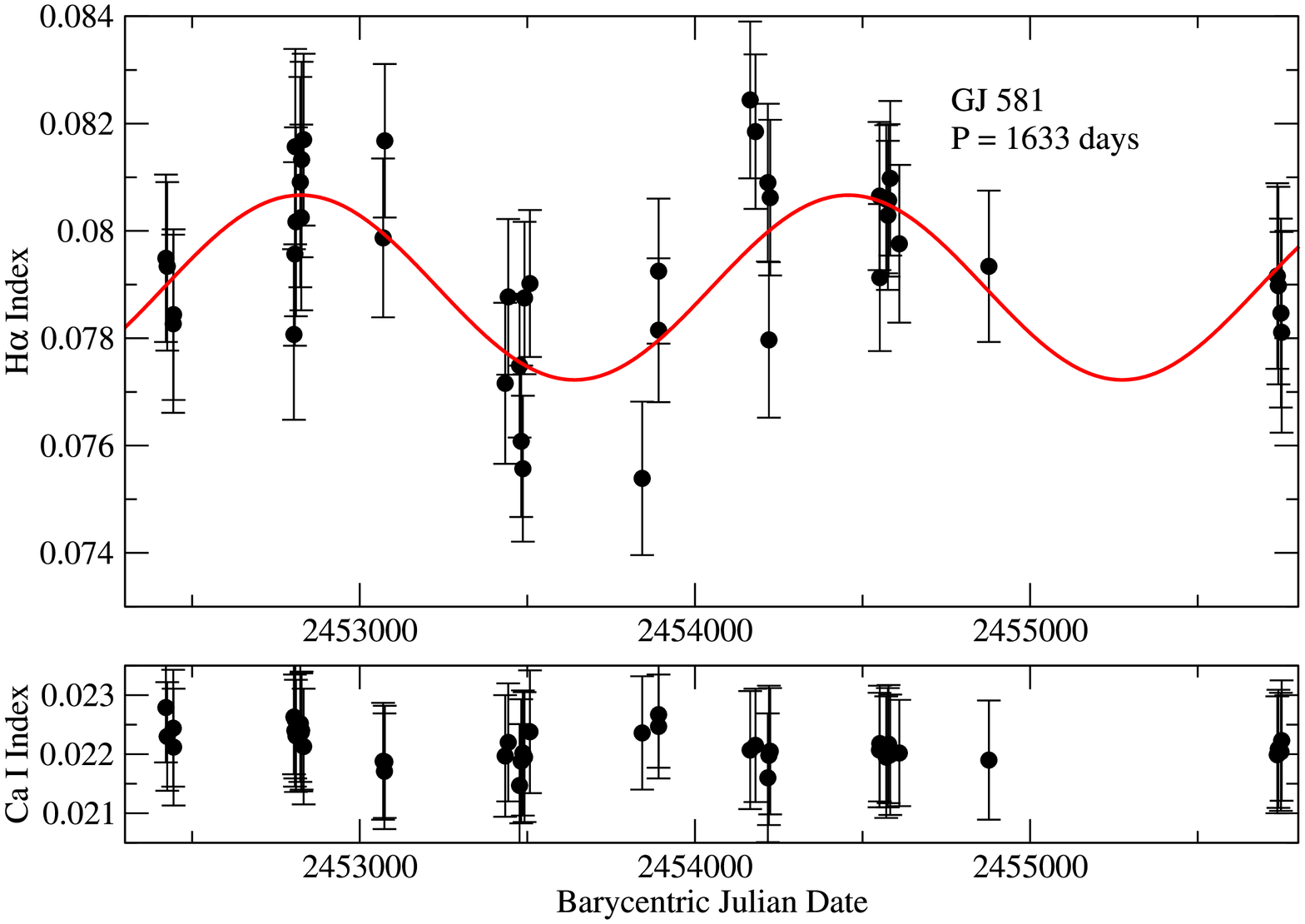}}
\subfigure[\label{gj708_cycle}]{\includegraphics[scale=0.3]{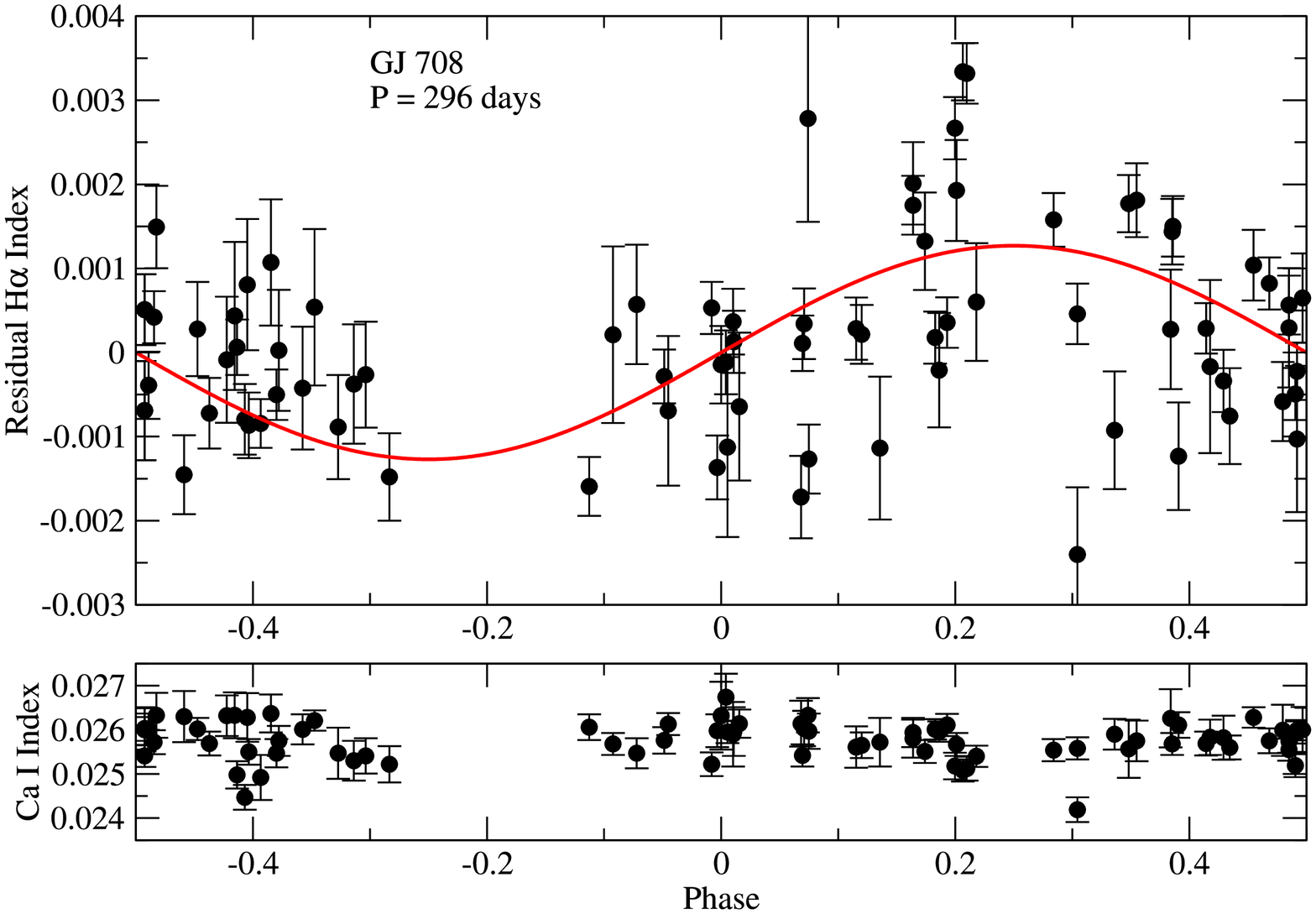}}
\subfigure[\label{gj730_cycle}]{\includegraphics[scale=0.3]{gj730_ha_both.eps}}
\subfigure[\label{gj552_trend}]{\includegraphics[scale=0.3]{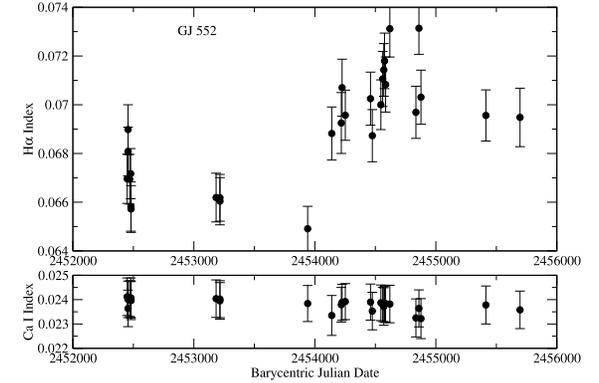}}
\caption{We have searched the time-series \ha variability of our entire M dwarf survey for periodic activity, particularly analogs to the solar cycle.  The stars in this Figure show definitive evidence of periodic activity in the \ha line.  The data have been phase-folded for shorter periods for clarity.  The red curves show our best fit to the data.  For GJ 708, we have also removed a linear trend.  In the case of GJ 730, we include two possible fits (see Section 3.1.4).  We do not include a fit for GJ 552 because we are unable to satisfactorily constrain the period of the observed cycle.  Below each \ha series is the corresponding Ca I index, which we use as a control.}
\label{cycles}
\end{sidewaysfigure}

\begin{figure}
\begin{center}
\subfigure[\label{gj270_fourier}]{\includegraphics[scale=0.4]{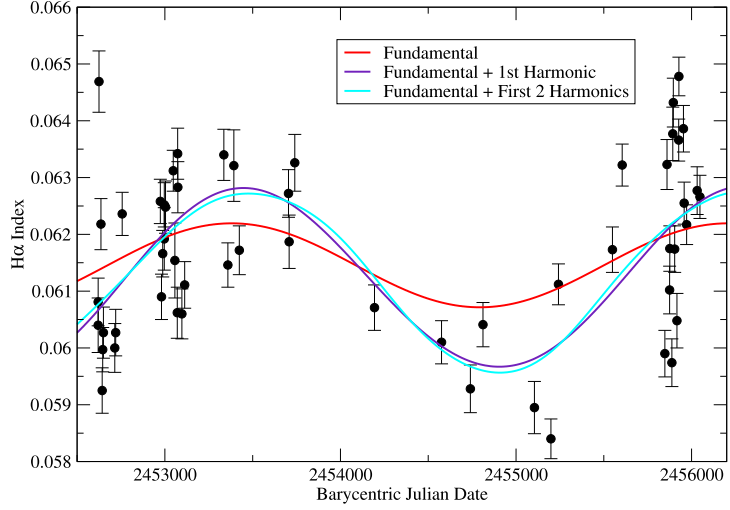}}
\subfigure[\label{gj552_fourier}]{\includegraphics[scale=0.4]{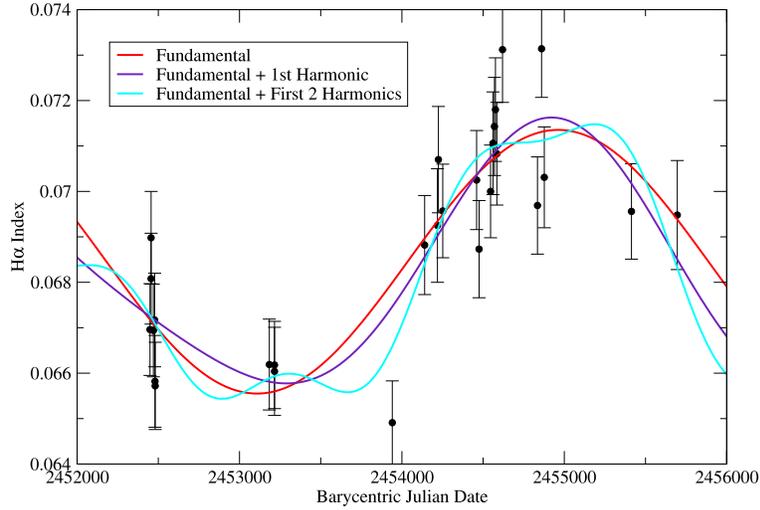}}
\caption{Several of our observed activity cycles exhibit non-sinusoidal \iha variability.  Here, we show the results of fitting \ha activity cycles with Fourier series instead of sine curves for \emph{a} GJ 270 and \emph{b} GJ 552.  We see that our fits improve with the inclusion of higher-frequency harmonics, but that for few data points (as seen in \emph{b}), we cannot constrain enough free parameters to justify adding a large number of terms.}
\label{fourier}
\end{center}
\end{figure}

\begin{sidewaysfigure}
\begin{center}
\subfigure[\label{gj16_trend}]{\includegraphics[scale=0.25]{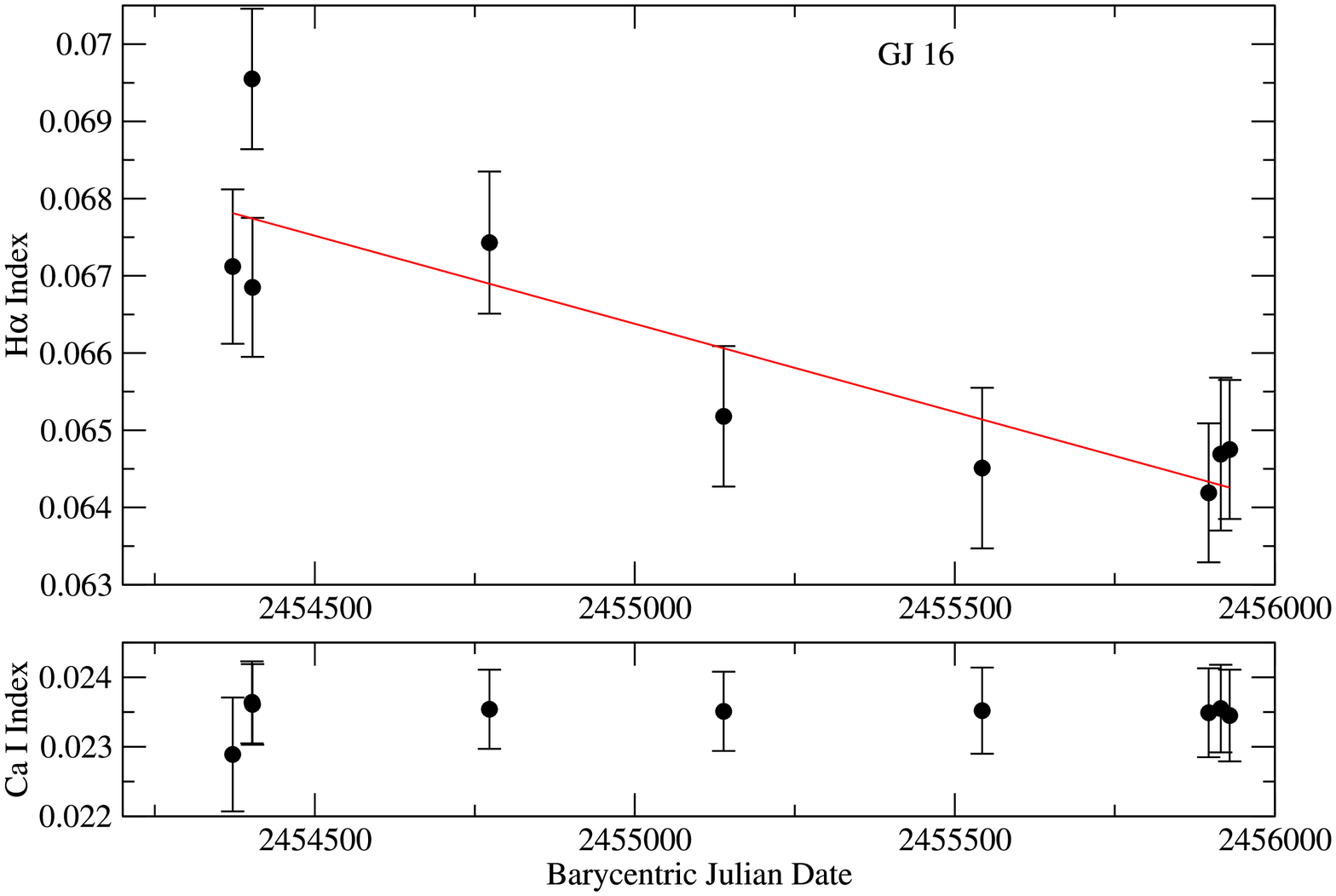}}
\subfigure[\label{gj521_trend}]{\includegraphics[scale=0.25]{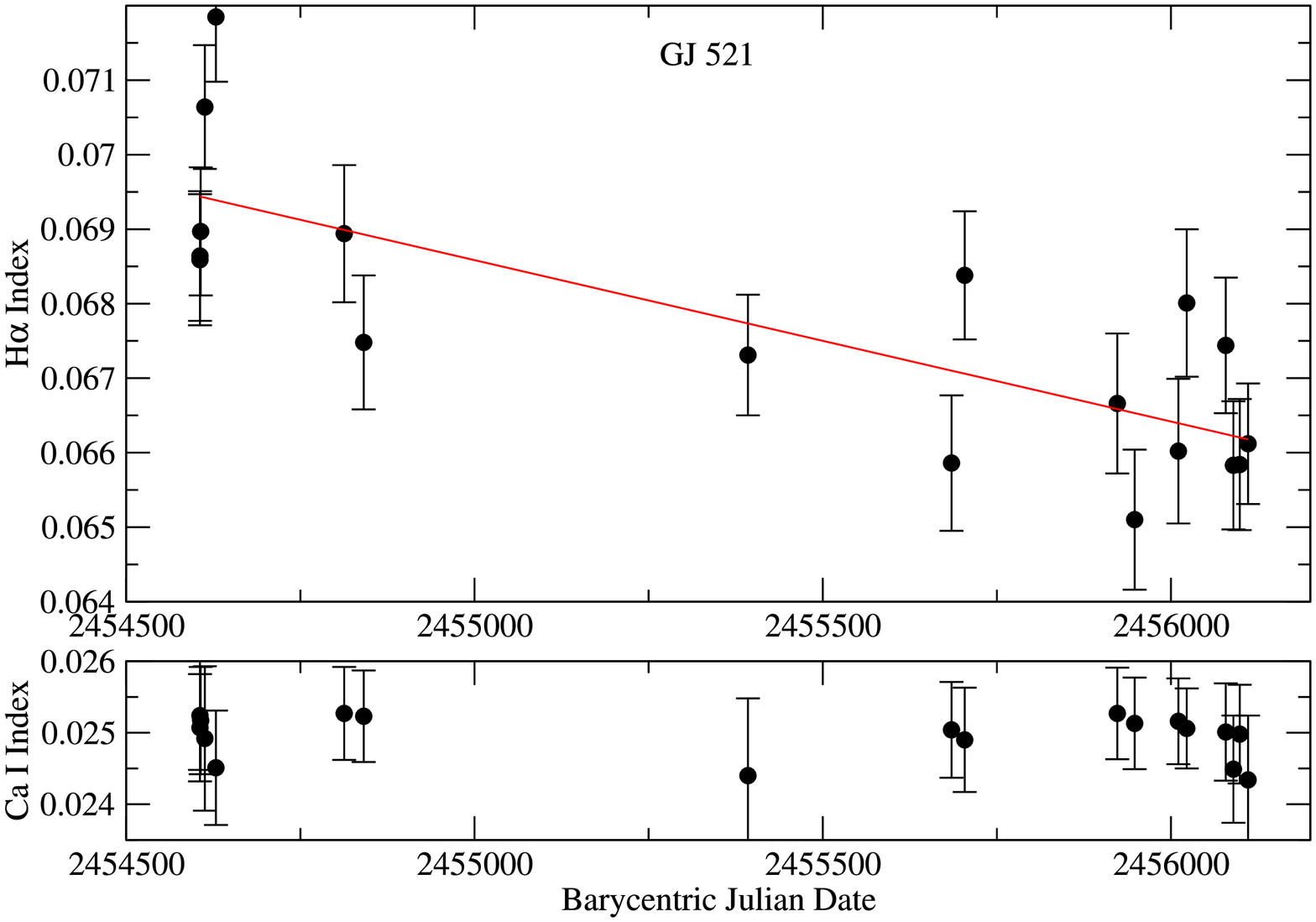}}
\subfigure[\label{gj96_trend}]{\includegraphics[scale=0.25]{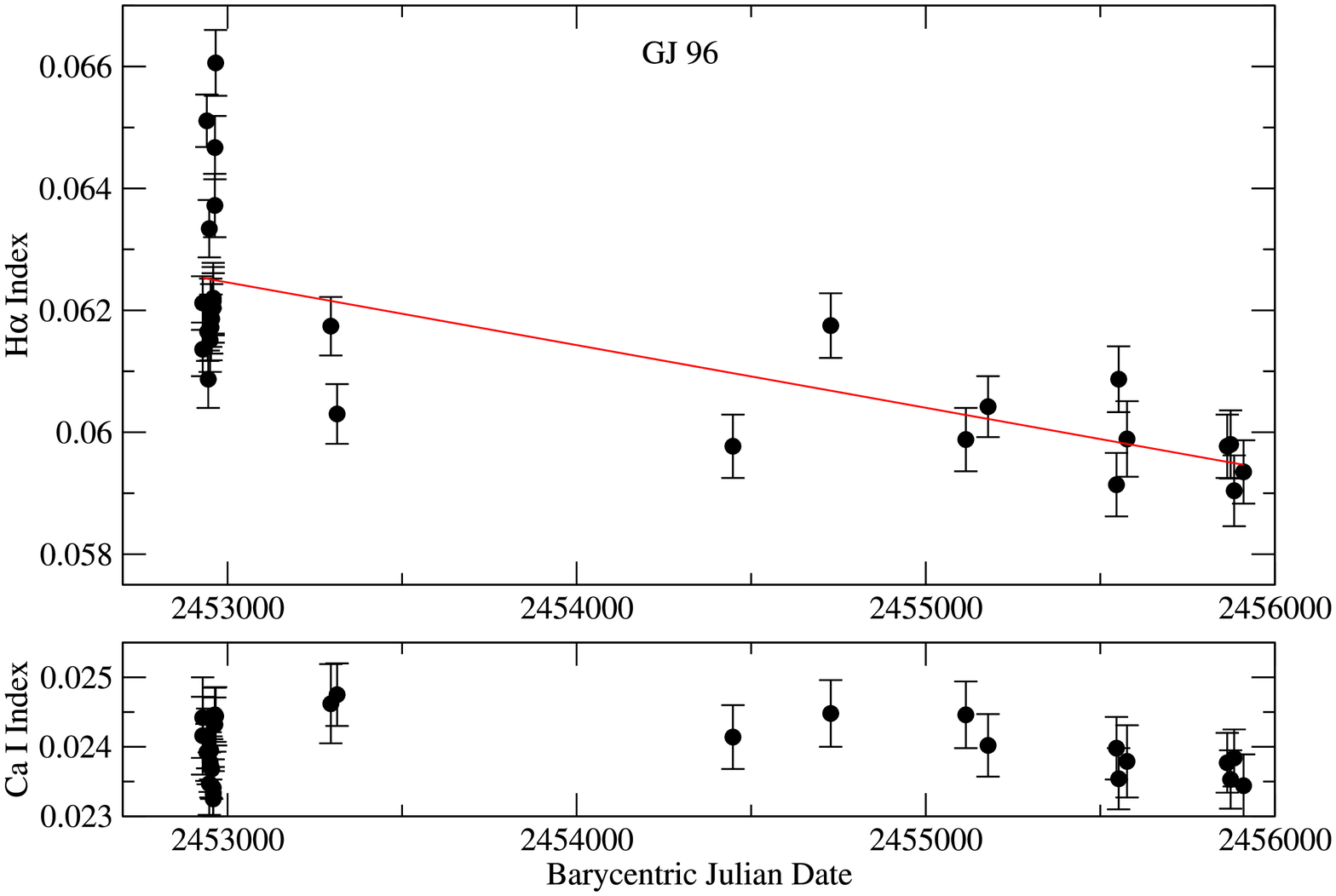}}
\subfigure[\label{gj3023_trend}]{\includegraphics[scale=0.25]{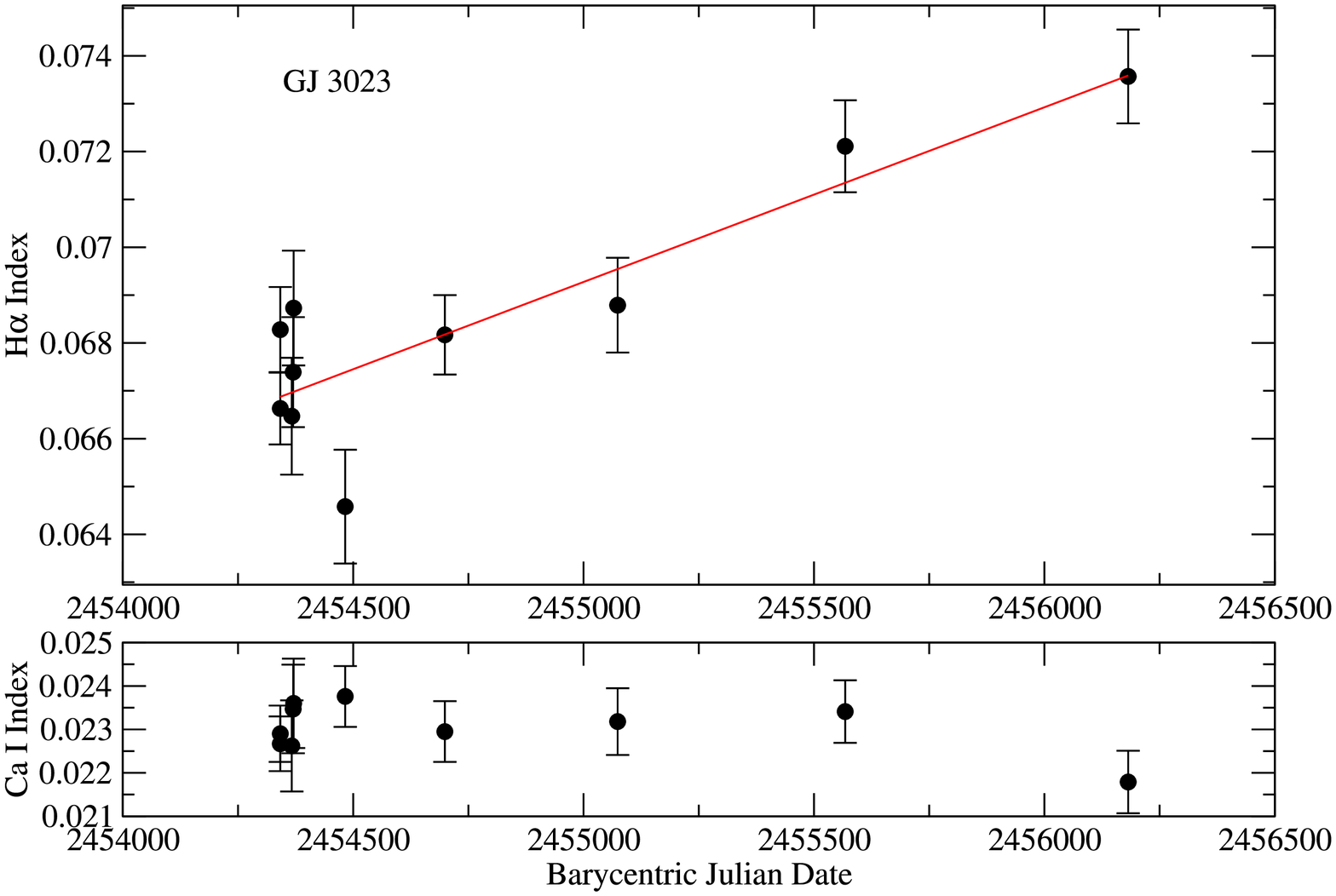}}
\subfigure[\label{gj3801_trend}]{\includegraphics[scale=0.25]{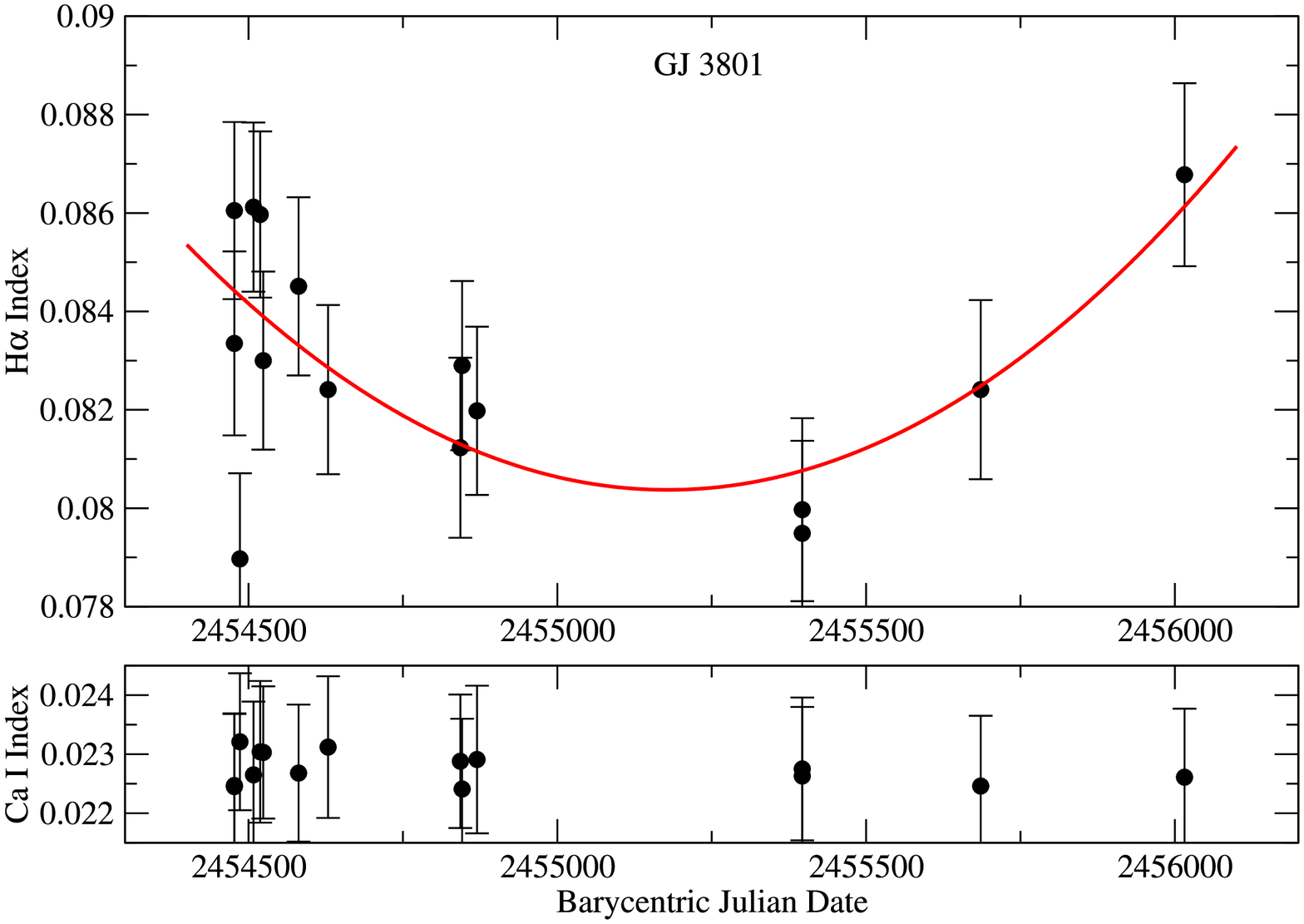}}
\subfigure[\label{gj611_trend}]{\includegraphics[scale=0.25]{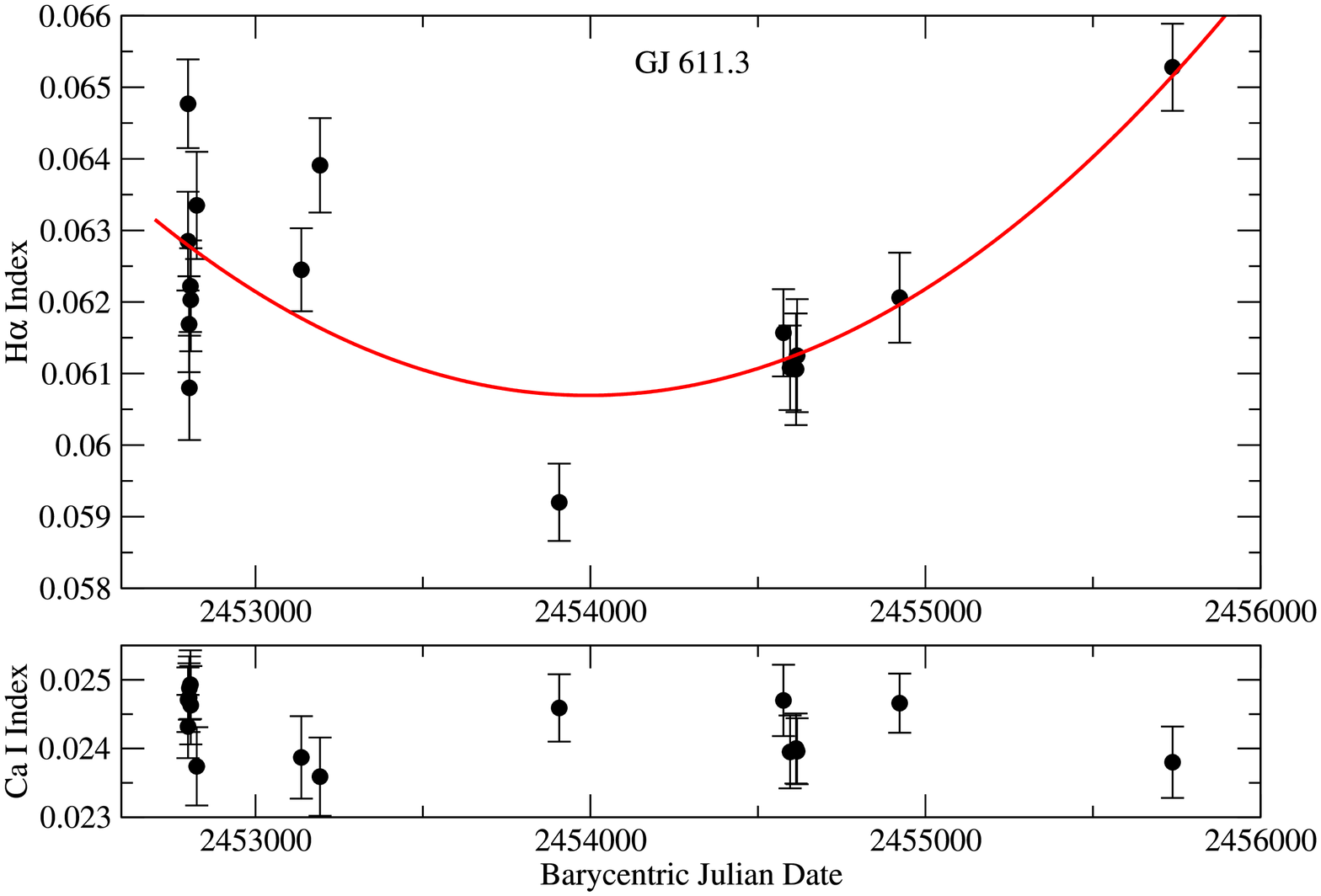}}
\subfigure[\label{gj630_trend}]{\includegraphics[scale=0.25]{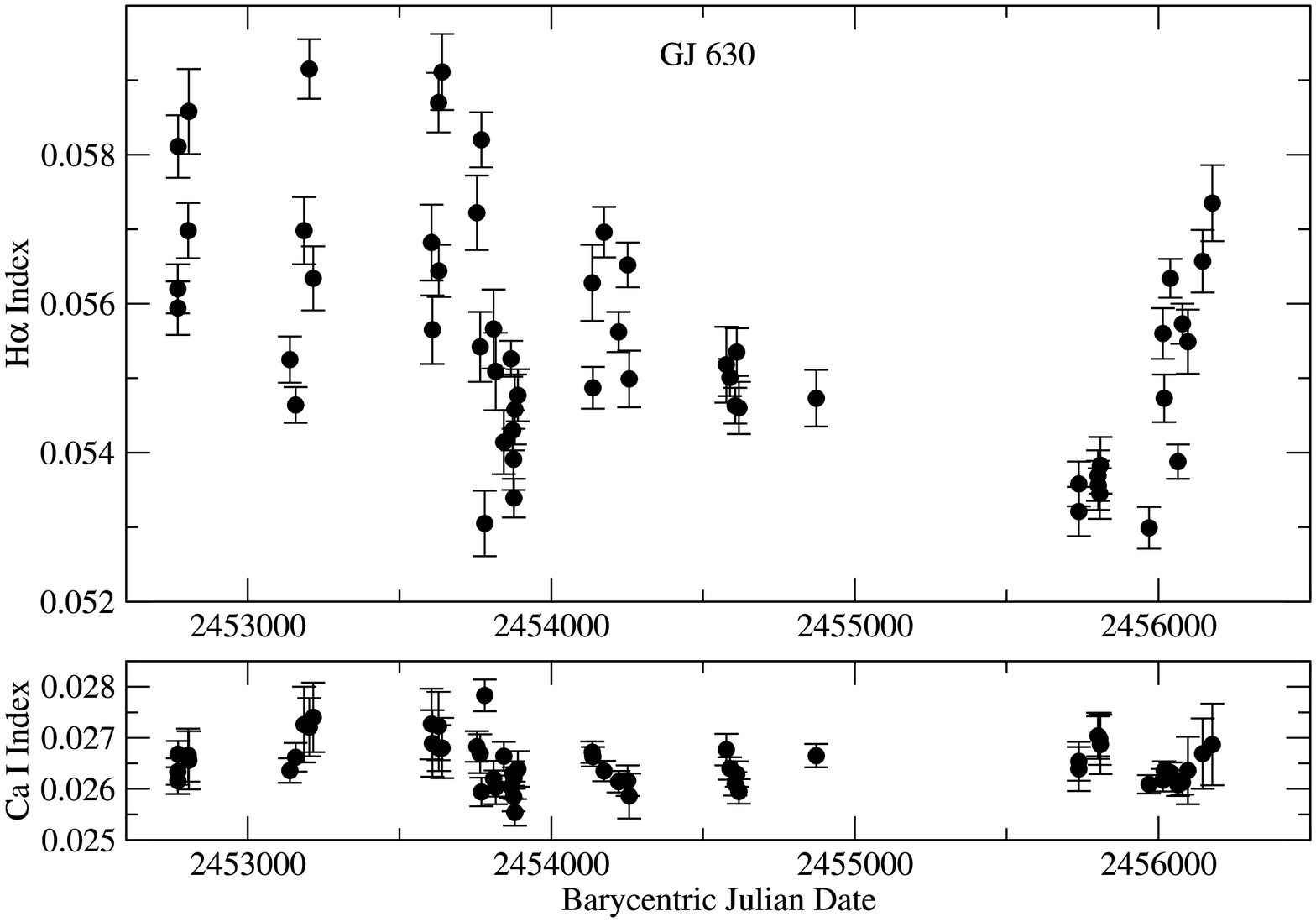}}

\caption{In addition to clearly periodic signals, we observe seven stars with long-term linear trends or curvature in \iha.  In this Figure, we show the time-series \ha indices for those stars.  Below each \ha series is the corresponding Ca I index. which we use as a control.  In \emph{a-d}, we show the linear least squares fit to the data in red.  For \emph{e} and \emph{f}, the red curve indicates our best-fit quadratic function.}
\label{trends}
\end{center}
\end{sidewaysfigure}

\begin{figure}
\begin{center}
\subfigure[\label{gj1170_harv}]{\includegraphics[scale=0.4]{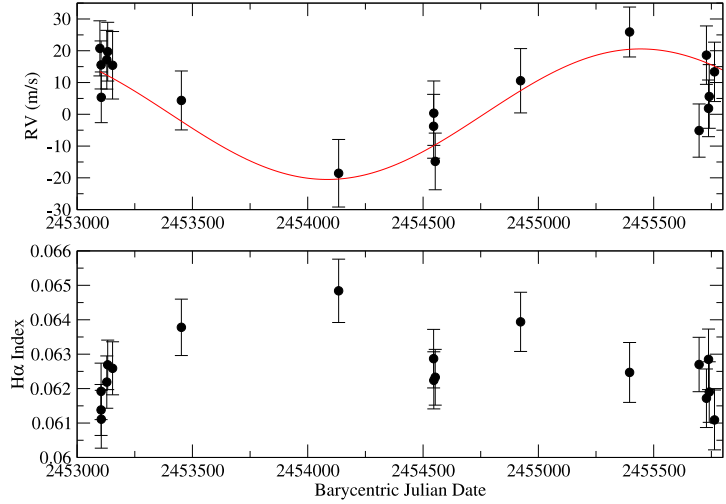}}
\subfigure[\label{gj3801_harv}]{\includegraphics[scale=0.4]{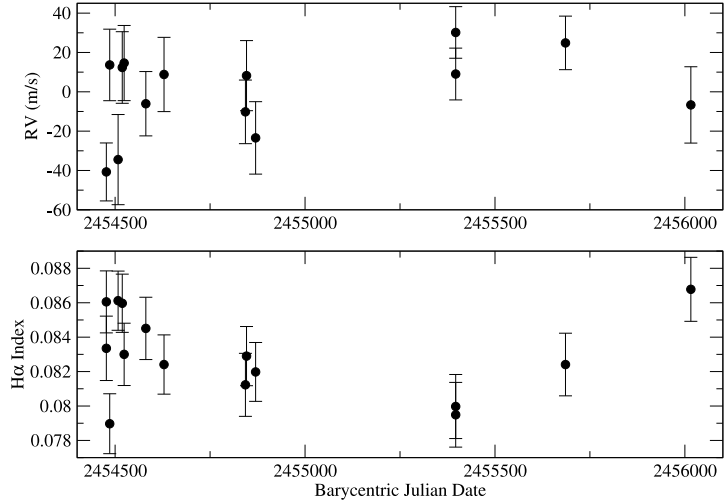}}
\caption{Previous results suggest the effects of stellar activity on RV measurements should be at or below the 5 m/s level, but we find two examples of activity-RV correlation at larger amplitudes.  Here, we show radial velocity measurements (\emph{Top}) and their corresponding \ha indices (\emph{Bottom}) for GJ 1170 (a) and GJ 3801 (b).  The red curve over the GJ 1170 RVs shows a Keplerian fit to the data.  The RV variation appears to be caused by a stellar activity cycle, as seen by the anticorrelation between RV and \ihap.}
\label{harv}
\end{center}
\end{figure}

\begin{figure}
\begin{center}
\subfigure[\label{VmI_Ha_full}]{\includegraphics[scale=0.4]{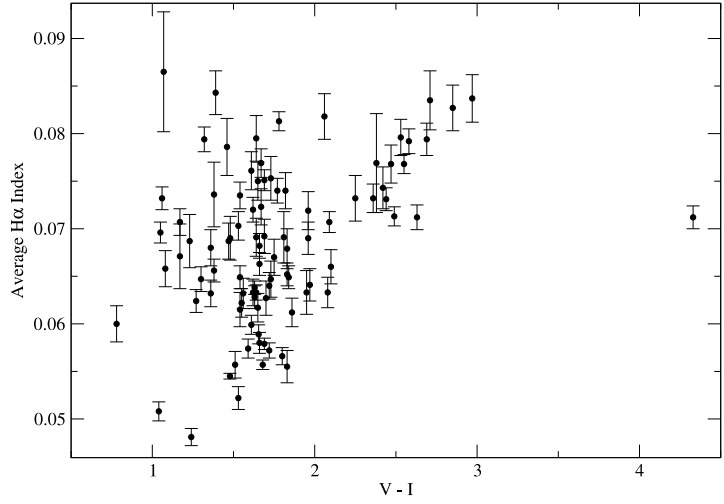}}
\subfigure[\label{VmI_Ha_cut}]{\includegraphics[scale=0.4]{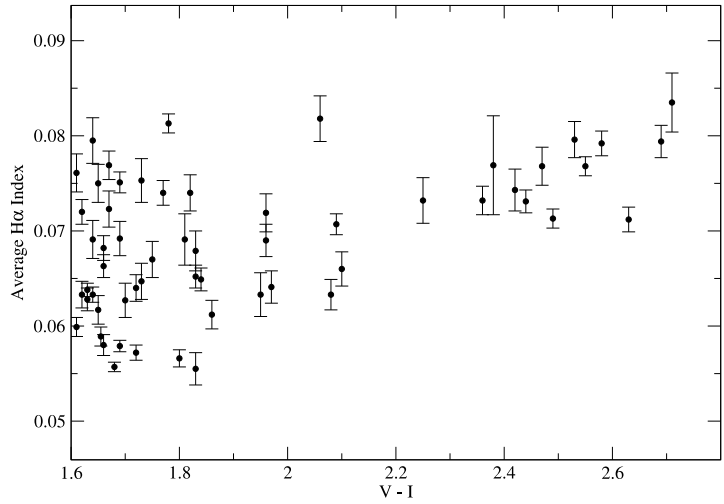}}
\end{center}
\caption{If stellar activity varies as a function of $T_{eff}$, the average \ha index should correlate with $V - I$ color.  In \emph{a}, we show our time-averaged \iha as a function of $V - I$ for our entire sample.  \emph{b} shows the same data, but restricted to the same range in $V - I$ as covered by the \citet{gds11} sample.}
\label{VmI_Ha}
\end{figure}

\begin{figure}
\begin{center}
\subfigure[\label{sha_mass}]{\includegraphics[scale=0.4]{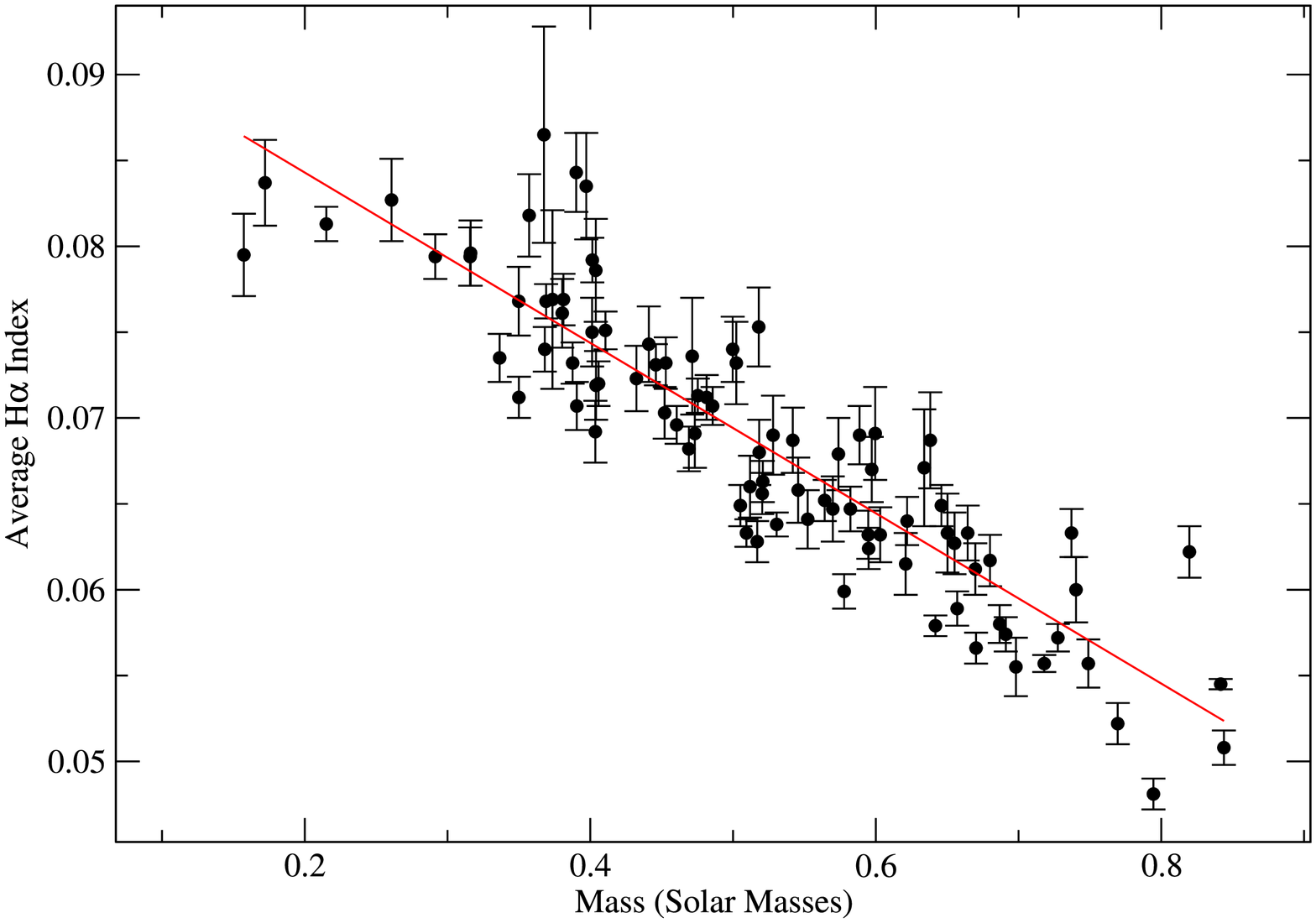}}
\subfigure[\label{sha_z}]{\includegraphics[scale=0.4]{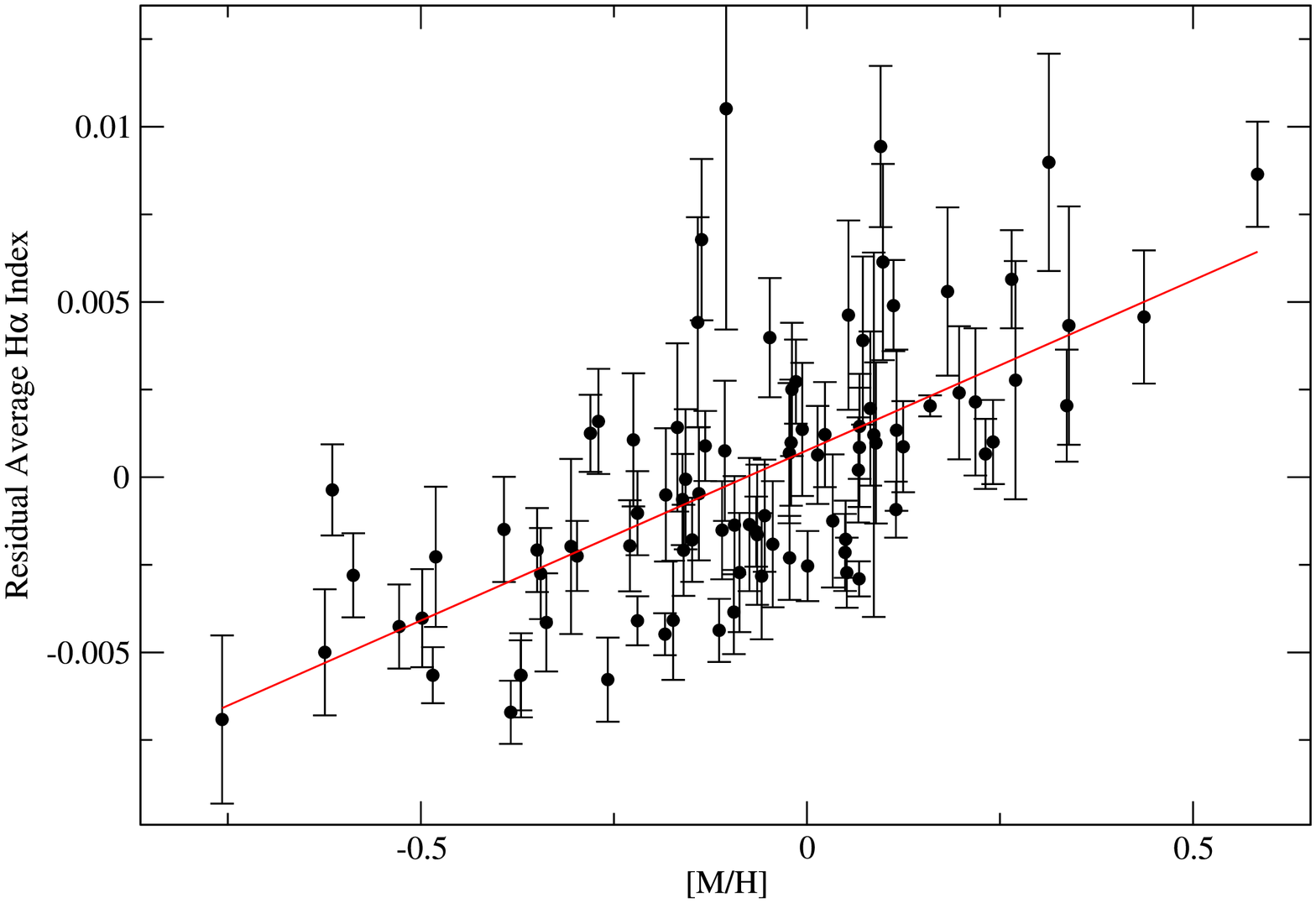}}
\end{center}
\caption{Since mean \ha activity does not correlate well with color (and therefore presumably $T_{eff}$, we investigate whether there are other stellar parameters which better predict \ha behavior.  \emph{a}. shows our time-averaged \ha index as a function of stellar mass for our sample.  The red line shows our linear least squares fit to the data.  In \emph{b}, we give the residuals to the fit shown in \emph{a}, plotted as a function of stellar metallicity.  The red line shows our linear least-squares fit to the data.}
\label{sha}
\end{figure}

\begin{figure}
\begin{center}
\subfigure[\label{lha_mass}]{\includegraphics[scale=0.4]{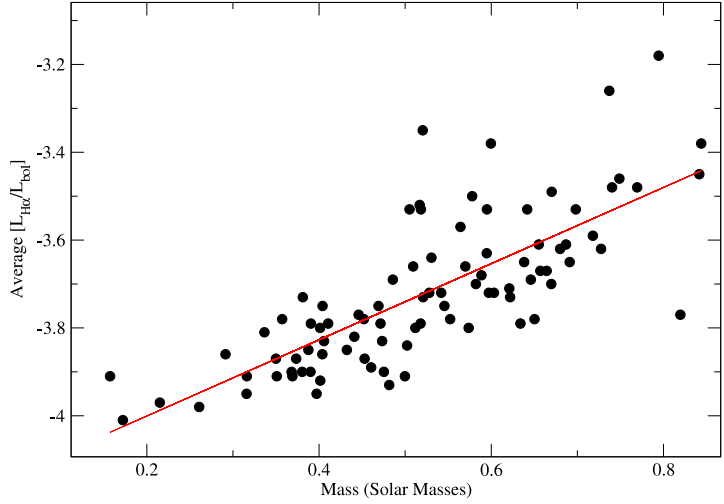}}
\subfigure[\label{lha_z}]{\includegraphics[scale=0.4]{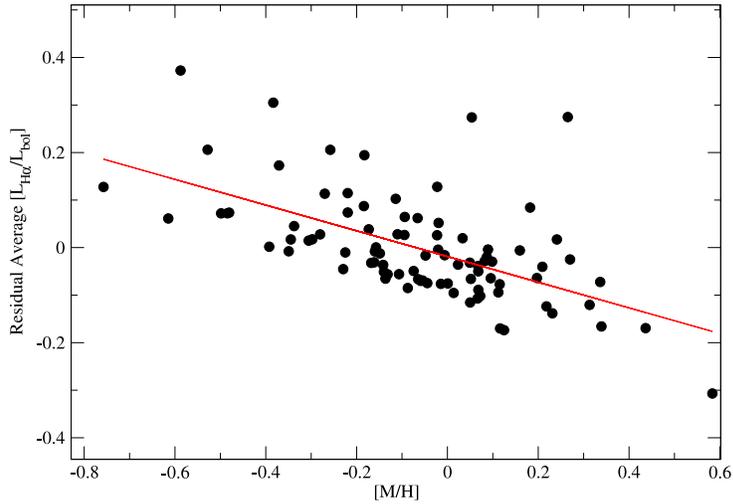}}
\end{center}
\caption{Because our \ha index is normalized to the nearby stellar continuum, variability in continuum flux as a function of stellar mass or $T_{eff}$ may affect the interpretation of Figure \ref{sha}.  Here, we transform our \iha values to \lha to remove the influence of the continuum.  \emph{a}. shows the time-averaged \lha as a function of stellar mass for our sample.  The red line shows our linear least squares fit to the data.  \emph{b}. again gives the residuals to the fit shown in \emph{a}, plotted as a function of stellar metallicity.  The red line shows our linear least-squares fit to the data.}
\label{lha}
\end{figure}

\begin{figure}
\begin{center}
\includegraphics[scale=0.5]{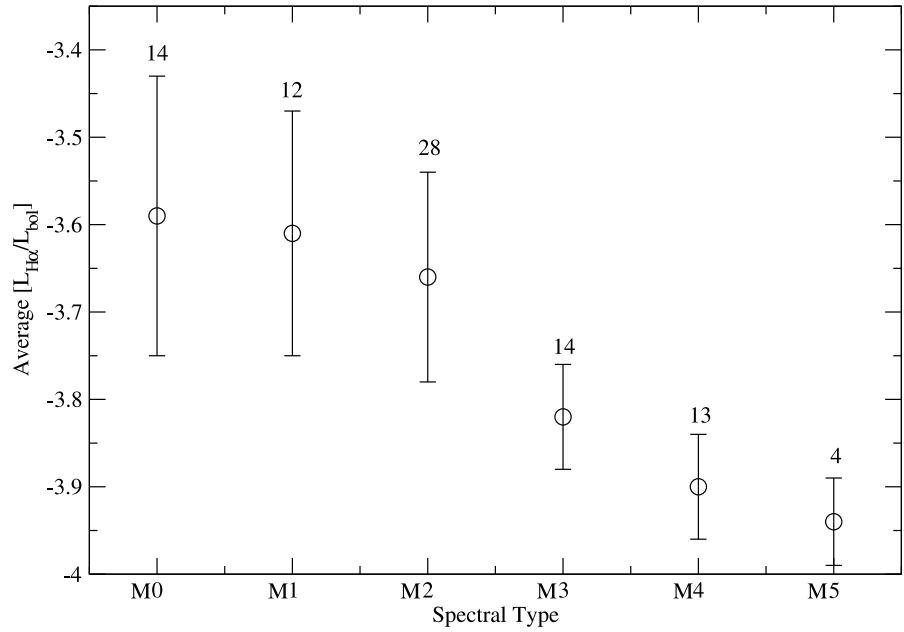}
\caption{Activity levels of M stars has previously been shown to vary with spectral subtype.  In this Figure, we have grouped our time-averaged \lha values as a function of spectral subtype for our sample.  The error bars indicate the RMS scatter in each bin, and the numbers above each point give the number of stars in the bin.}
\label{type_LHa}
\end{center}
\end{figure}

\begin{sidewaysfigure}
\begin{center}
\subfigure[\label{sigU}]{\includegraphics[scale=0.4]{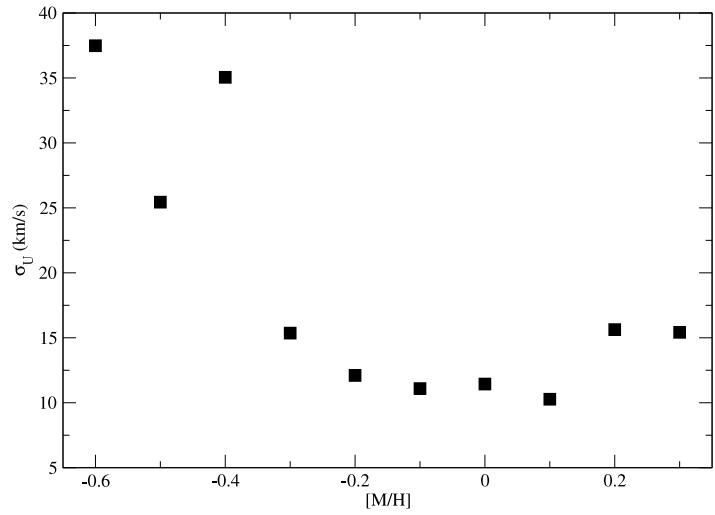}}
\subfigure[\label{sigV}]{\includegraphics[scale=0.4]{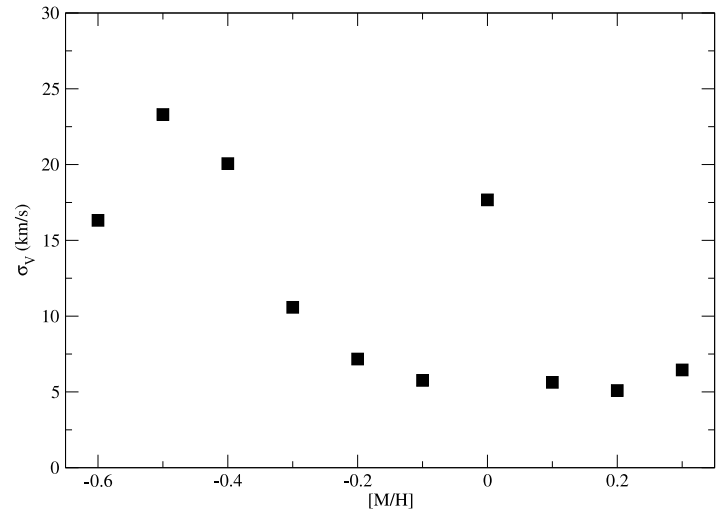}}
\subfigure[\label{sigW}]{\includegraphics[scale=0.4]{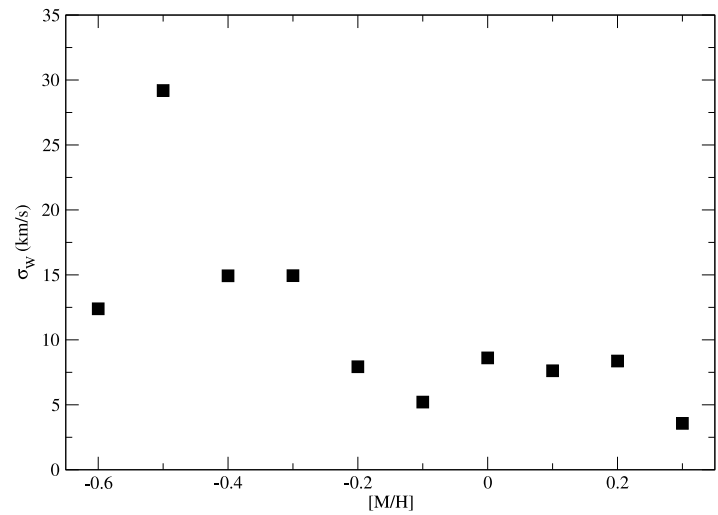}}
\caption{We investigate whether there is any indication that the more metal-poor stars in our sample are older than the metal-rich stars.  Here, we plot galactocentric velocity dispersion in $UVW$ coordinates as a function of stellar metallicity.  Metal-poor stars tend to have higher velocity dispersions in all three coordinates, suggesting they are older, potentially explaining why our metal-poor targets exhibit lower mean \iha values.}
\label{sigma}
\end{center}
\end{sidewaysfigure}

\begin{table}
\begin{center}
\begin{tabular}{| l | l l l |}

\hline & & &\\
 & \multicolumn{3}{| c |}{\emph{Periodic Signals}} \\
Star & Period (days) & Amplitude (\ihap) & FAP \\
\hline & & &\\
GJ 270 & $2687 \pm 16$ & $0.00158 \pm 0.0003$ & $< 10^{-4}$ \\
GJ 476 & $1066 \pm 200$ & $0.00193 \pm 0.0005$ & $<10^{-4}$ \\
GJ 581 & $1633 \pm 93$ & $0.00172 \pm 0.0003$ & $0.12$ \\
GJ 708 & $296 \pm 1$ & $0.00127 \pm 0.0002$ & 0.03 \\
GJ 730 (3-day fit) & $3.06 \pm 0.01$ & $0.00128 \pm 0.0002$ & $<10^{-4}$ \\
GJ 730 (994-day fit) & $994 \pm 40$ & $0.00134 \pm 0.0002$ & $0.10$ \\
GJ 552 & $>2000$ & $\cdots$ & $<10^{-4}$ \\
\hline & & & \\
 & \multicolumn{3}{| c |}{\emph{Linear Trends}} \\
 & $I_{\textrm{H}\alpha}(t)$ & $r$ & $P(r)$ \\
\hline & & & \\
GJ 16 & $5.7_{\pm 1} - 2.3_{\pm 0.5} \times 10^{-6}t$ & $-0.868$ & $0.0012$ \\
GJ 521 & $5.4_{\pm 1} - 2.2_{\pm 0.4} \times 10^{-6}t$ & $-0.783$ & $6.1 \times 10^{-5}$ \\
GJ 96 & $2.6_{\pm 0.5} - 1.0_{\pm 0.2} \times 10^{-6}t$ & $-0.720$ & $5.3 \times 10^{-6}$ \\
GJ 3023 & $-8.9_{\pm 2} - 3.6_{\pm 0.7} \times 10^{-6}t$ & $+0.877$ & $4.3 \times 10^{-4}$ \\
\hline & & & \\
 & \multicolumn{3}{| c |}{\emph{Quadratic Trends}} \\
 & $I_{\textrm{H}\alpha}(t)$ & $F_{\textrm{poly}}$ & $P(F_{\textrm{poly}})$ \\
\hline & & & \\
GJ 3801 & $50000_{\pm 16000} - 0.04_{\pm 0.01}t + 8_{\pm 3} \times 10^{-9}t^2$ & 9.31 & 0.04 \\
GJ 611.3 & $9000_{\pm 2000} - 0.007_{\pm 0.002}t + 1.5_{\pm 0.4} \times 10^{-9}t^2$ & 12.4 & 0.01 \\
GJ 630 & \multicolumn{3}{| c |}{Not Fit} \\
\hline

\end{tabular}
\caption{Parameters of our fits to the \iha time series discussed in Section 3.  For the periodic fits, the FAP listed is taken from the results of our bootstrap resampling analyses.  With the linear fits, we include the Pearson correlation coefficient $r$ and the resulting probability of no correlation $P(r)$.  Similarly, the quadratic trends include the results of our F-test, as detailed in Section 3.2.2.  Our fits to $I_{\textrm{H}\alpha}(t)$ are given as a function of barycentric Julian date.}
\label{fits}
\end{center}
\end{table}


\clearpage

\begin{center}
\footnotesize

\tablecaption{\ha and Ca I indices shown in Figures \ref{cycles} and \ref{trends}}
\label{hatab}	
\tablefirsthead{\hline
BJD - 2450000 & \ha index & Ca I index \\
\hline}
		
\tablehead{\hline
\emph{Table \ref{hatab} cont'd.} & & \\ \hline
BJD - 2450000 & \ha index  & Ca I index \\
\hline}
		
\tabletail{\hline}

\begin{supertabular}{| l l l |}

\multicolumn{3}{| c |}{GJ 270} \\
\hline & & \\

2619.74495602 	 & $ 0.06040 \pm 0.00048 $ & 	 $ 0.02550 \pm 0.00034 $ \\
2619.76246122 	 & $ 0.06081 \pm 0.00042 $ & 	 $ 0.02548 \pm 0.00033 $ \\
2624.73217882 	 & $ 0.06469 \pm 0.00054 $ & 	 $ 0.02602 \pm 0.00035 $ \\
2635.93473395 	 & $ 0.06218 \pm 0.00045 $ & 	 $ 0.02539 \pm 0.00033 $ \\
2642.69473208 	 & $ 0.05925 \pm 0.00040 $ & 	 $ 0.02588 \pm 0.00030 $ \\
2645.67800006 	 & $ 0.05997 \pm 0.00039 $ & 	 $ 0.02554 \pm 0.00030 $ \\
2649.66792308 	 & $ 0.06027 \pm 0.00045 $ & 	 $ 0.02565 \pm 0.00035 $ \\
2714.71477040 	 & $ 0.06000 \pm 0.00043 $ & 	 $ 0.02510 \pm 0.00032 $ \\
2719.71740232 	 & $ 0.06027 \pm 0.00041 $ & 	 $ 0.02472 \pm 0.00033 $ \\
2756.61978882 	 & $ 0.06236 \pm 0.00038 $ & 	 $ 0.02491 \pm 0.00032 $ \\
2974.01609612 	 & $ 0.06258 \pm 0.00039 $ & 	 $ 0.02576 \pm 0.00033 $ \\
2980.76850420 	 & $ 0.06090 \pm 0.00040 $ & 	 $ 0.02582 \pm 0.00032 $ \\
2987.75676078 	 & $ 0.06166 \pm 0.00041 $ & 	 $ 0.02574 \pm 0.00034 $ \\
2996.71864146 	 & $ 0.06192 \pm 0.00055 $ & 	 $ 0.02574 \pm 0.00037 $ \\
2996.73387092 	 & $ 0.06252 \pm 0.00039 $ & 	 $ 0.02567 \pm 0.00033 $ \\
3003.71574529 	 & $ 0.06248 \pm 0.00046 $ & 	 $ 0.02572 \pm 0.00033 $ \\
3047.59314927 	 & $ 0.06312 \pm 0.00036 $ & 	 $ 0.02536 \pm 0.00037 $ \\
3056.79170774 	 & $ 0.06154 \pm 0.00066 $ & 	 $ 0.02565 \pm 0.00044 $ \\
3070.74665118 	 & $ 0.06062 \pm 0.00045 $ & 	 $ 0.02549 \pm 0.00034 $ \\
3072.73438156 	 & $ 0.06342 \pm 0.00045 $ & 	 $ 0.02500 \pm 0.00033 $ \\
3073.73829100 	 & $ 0.06283 \pm 0.00045 $ & 	 $ 0.02500 \pm 0.00032 $ \\
3096.66830179 	 & $ 0.06060 \pm 0.00044 $ & 	 $ 0.02452 \pm 0.00034 $ \\
3112.62306948 	 & $ 0.06111 \pm 0.00041 $ & 	 $ 0.02495 \pm 0.00033 $ \\
3336.02956980 	 & $ 0.06340 \pm 0.00045 $ & 	 $ 0.02581 \pm 0.00038 $ \\
3358.75899075 	 & $ 0.06146 \pm 0.00039 $ & 	 $ 0.02560 \pm 0.00038 $ \\
3392.89441676 	 & $ 0.06321 \pm 0.00063 $ & 	 $ 0.02631 \pm 0.00033 $ \\
3423.80129696 	 & $ 0.06172 \pm 0.00043 $ & 	 $ 0.02553 \pm 0.00039 $ \\
3704.00760396 	 & $ 0.06272 \pm 0.00042 $ & 	 $ 0.02532 \pm 0.00034 $ \\
3707.76840342 	 & $ 0.06187 \pm 0.00047 $ & 	 $ 0.02571 \pm 0.00033 $ \\
3739.70652391 	 & $ 0.06326 \pm 0.00050 $ & 	 $ 0.02487 \pm 0.00033 $ \\
4193.67268488 	 & $ 0.06071 \pm 0.00040 $ & 	 $ 0.02474 \pm 0.00033 $ \\
4575.63333239 	 & $ 0.06010 \pm 0.00038 $ & 	 $ 0.02528 \pm 0.00032 $ \\
4739.94479093 	 & $ 0.05928 \pm 0.00042 $ & 	 $ 0.02565 \pm 0.00031 $ \\
4811.76596343 	 & $ 0.06041 \pm 0.00039 $ & 	 $ 0.02538 \pm 0.00033 $ \\
5104.94743407 	 & $ 0.05895 \pm 0.00046 $ & 	 $ 0.02568 \pm 0.00034 $ \\
5198.70136009 	 & $ 0.05840 \pm 0.00035 $ & 	 $ 0.02562 \pm 0.00029 $ \\
5241.58241178 	 & $ 0.06112 \pm 0.00036 $ & 	 $ 0.02510 \pm 0.00036 $ \\
5548.96506060 	 & $ 0.06173 \pm 0.00040 $ & 	 $ 0.02536 \pm 0.00032 $ \\
5604.60284049 	 & $ 0.06322 \pm 0.00037 $ & 	 $ 0.02506 \pm 0.00032 $ \\
5847.91548311 	 & $ 0.05990 \pm 0.00041 $ & 	 $ 0.02551 \pm 0.00030 $ \\
5859.87415031 	 & $ 0.06323 \pm 0.00044 $ & 	 $ 0.02537 \pm 0.00034 $ \\
5874.85664609 	 & $ 0.06102 \pm 0.00042 $ & 	 $ 0.02542 \pm 0.00030 $ \\
5875.84165490 	 & $ 0.06175 \pm 0.00040 $ & 	 $ 0.02533 \pm 0.00031 $ \\
5887.82421737 	 & $ 0.05974 \pm 0.00042 $ & 	 $ 0.02535 \pm 0.00029 $ \\
5892.80041024 	 & $ 0.06377 \pm 0.00047 $ & 	 $ 0.02524 \pm 0.00036 $ \\
5895.78595998 	 & $ 0.06432 \pm 0.00043 $ & 	 $ 0.02516 \pm 0.00032 $ \\
5903.76162361 	 & $ 0.06174 \pm 0.00041 $ & 	 $ 0.02544 \pm 0.00031 $ \\
5916.73727825 	 & $ 0.06048 \pm 0.00048 $ & 	 $ 0.02547 \pm 0.00029 $ \\
5926.71009501 	 & $ 0.06366 \pm 0.00037 $ & 	 $ 0.02520 \pm 0.00029 $ \\
5927.70885023 	 & $ 0.06478 \pm 0.00034 $ & 	 $ 0.02527 \pm 0.00030 $ \\
5953.62135339 	 & $ 0.06386 \pm 0.00041 $ & 	 $ 0.02491 \pm 0.00031 $ \\
5957.62588797 	 & $ 0.06255 \pm 0.00037 $ & 	 $ 0.02508 \pm 0.00028 $ \\
5971.59748963 	 & $ 0.06217 \pm 0.00035 $ & 	 $ 0.02501 \pm 0.00040 $ \\
6031.64221681 	 & $ 0.06277 \pm 0.00042 $ & 	 $ 0.02412 \pm 0.00033 $ \\
6047.60139792 	 & $ 0.06266 \pm 0.00038 $ & 	 $ 0.02424 \pm 0.00031 $ \\

\hline & & \\
\multicolumn{3}{| c |}{GJ 476} \\
\hline & & \\

4809.00935640 	 & $ 0.07405 \pm 0.00148 $ & 	 $ 0.02434 \pm 0.00088 $ \\
4813.00078938 	 & $ 0.07512 \pm 0.00135 $ & 	 $ 0.02449 \pm 0.00089 $ \\
4835.93825372 	 & $ 0.07111 \pm 0.00138 $ & 	 $ 0.02404 \pm 0.00091 $ \\
4840.93266452 	 & $ 0.07162 \pm 0.00135 $ & 	 $ 0.02433 \pm 0.00089 $ \\
4842.93199363 	 & $ 0.06938 \pm 0.00131 $ & 	 $ 0.02414 \pm 0.00090 $ \\
4843.91050638 	 & $ 0.06961 \pm 0.00134 $ & 	 $ 0.02453 \pm 0.00093 $ \\
4845.91645282 	 & $ 0.07082 \pm 0.00141 $ & 	 $ 0.02417 \pm 0.00092 $ \\
5010.63249904 	 & $ 0.07061 \pm 0.00139 $ & 	 $ 0.02457 \pm 0.00114 $ \\
5020.62065338 	 & $ 0.06966 \pm 0.00134 $ & 	 $ 0.02466 \pm 0.00113 $ \\
5183.98313987 	 & $ 0.06975 \pm 0.00133 $ & 	 $ 0.02459 \pm 0.00093 $ \\
5183.99443947 	 & $ 0.06797 \pm 0.00135 $ & 	 $ 0.02460 \pm 0.00089 $ \\
5199.94723445 	 & $ 0.06724 \pm 0.00136 $ & 	 $ 0.02424 \pm 0.00092 $ \\
5199.95853867 	 & $ 0.06665 \pm 0.00142 $ & 	 $ 0.02429 \pm 0.00090 $ \\
5207.92782339 	 & $ 0.06894 \pm 0.00137 $ & 	 $ 0.02445 \pm 0.00097 $ \\
5207.93911867 	 & $ 0.06882 \pm 0.00147 $ & 	 $ 0.02419 \pm 0.00095 $ \\
5209.91363282 	 & $ 0.06928 \pm 0.00148 $ & 	 $ 0.02431 \pm 0.00096 $ \\
5240.99126123 	 & $ 0.06805 \pm 0.00141 $ & 	 $ 0.02432 \pm 0.00090 $ \\
5245.82151704 	 & $ 0.06744 \pm 0.00146 $ & 	 $ 0.02442 \pm 0.00094 $ \\
5245.83282300 	 & $ 0.06777 \pm 0.00139 $ & 	 $ 0.02442 \pm 0.00093 $ \\
5253.95868114 	 & $ 0.06881 \pm 0.00145 $ & 	 $ 0.02438 \pm 0.00094 $ \\
5253.97154526 	 & $ 0.06723 \pm 0.00139 $ & 	 $ 0.02445 \pm 0.00092 $ \\
5291.85653799 	 & $ 0.06815 \pm 0.00140 $ & 	 $ 0.02413 \pm 0.00091 $ \\
5291.86784578 	 & $ 0.06702 \pm 0.00135 $ & 	 $ 0.02422 \pm 0.00092 $ \\
5309.65430856 	 & $ 0.06721 \pm 0.00138 $ & 	 $ 0.02424 \pm 0.00095 $ \\
5309.66572957 	 & $ 0.06873 \pm 0.00135 $ & 	 $ 0.02433 \pm 0.00095 $ \\
5337.73268927 	 & $ 0.06683 \pm 0.00134 $ & 	 $ 0.02412 \pm 0.00097 $ \\
5702.73796815 	 & $ 0.06971 \pm 0.00134 $ & 	 $ 0.02411 \pm 0.00095 $ \\
5729.65679328 	 & $ 0.06825 \pm 0.00139 $ & 	 $ 0.02431 \pm 0.00097 $ \\
6017.87319403 	 & $ 0.07135 \pm 0.00138 $ & 	 $ 0.02405 \pm 0.00091 $ \\
6085.70114197 	 & $ 0.06942 \pm 0.00139 $ & 	 $ 0.02401 \pm 0.00096 $ \\
6102.64742851 	 & $ 0.06854 \pm 0.00139 $ & 	 $ 0.02446 \pm 0.00088 $ \\

\hline & & \\
\multicolumn{3}{| c |}{GJ 581} \\
\hline & & \\

2421.74447247 	 & $ 0.07949 \pm 0.00156 $ & 	 $ 0.02279 \pm 0.00093 $ \\
2425.73640218 	 & $ 0.07934 \pm 0.00157 $ & 	 $ 0.02230 \pm 0.00092 $ \\
2443.70418288 	 & $ 0.07827 \pm 0.00166 $ & 	 $ 0.02244 \pm 0.00099 $ \\
2444.68379899 	 & $ 0.07844 \pm 0.00159 $ & 	 $ 0.02212 \pm 0.00099 $ \\
2803.70286496 	 & $ 0.07807 \pm 0.00159 $ & 	 $ 0.02263 \pm 0.00097 $ \\
2806.69497578 	 & $ 0.07957 \pm 0.00171 $ & 	 $ 0.02240 \pm 0.00095 $ \\
2807.70023704 	 & $ 0.08157 \pm 0.00182 $ & 	 $ 0.02258 \pm 0.00099 $ \\
2809.69168253 	 & $ 0.08017 \pm 0.00176 $ & 	 $ 0.02231 \pm 0.00095 $ \\
2822.66406479 	 & $ 0.08091 \pm 0.00196 $ & 	 $ 0.02252 \pm 0.00099 $ \\
2825.64546199 	 & $ 0.08025 \pm 0.00173 $ & 	 $ 0.02237 \pm 0.00100 $ \\
2826.65113681 	 & $ 0.08133 \pm 0.00182 $ & 	 $ 0.02240 \pm 0.00100 $ \\
2832.64663205 	 & $ 0.08170 \pm 0.00160 $ & 	 $ 0.02213 \pm 0.00098 $ \\
3069.97261868 	 & $ 0.07987 \pm 0.00148 $ & 	 $ 0.02188 \pm 0.00099 $ \\
3074.95259041 	 & $ 0.08168 \pm 0.00143 $ & 	 $ 0.02187 \pm 0.00095 $ \\
3433.98226317 	 & $ 0.07716 \pm 0.00150 $ & 	 $ 0.02197 \pm 0.00103 $ \\
3442.95918646 	 & $ 0.07877 \pm 0.00145 $ & 	 $ 0.02220 \pm 0.00100 $ \\
3476.85336355 	 & $ 0.07749 \pm 0.00134 $ & 	 $ 0.02147 \pm 0.00104 $ \\
3481.84858218 	 & $ 0.07608 \pm 0.00141 $ & 	 $ 0.02188 \pm 0.00105 $ \\
3486.84197848 	 & $ 0.07557 \pm 0.00136 $ & 	 $ 0.02202 \pm 0.00106 $ \\
3491.81260996 	 & $ 0.07875 \pm 0.00142 $ & 	 $ 0.02195 \pm 0.00110 $ \\
3507.79346518 	 & $ 0.07902 \pm 0.00137 $ & 	 $ 0.02238 \pm 0.00104 $ \\
3842.86894559 	 & $ 0.07539 \pm 0.00143 $ & 	 $ 0.02236 \pm 0.00096 $ \\
3891.72779286 	 & $ 0.07815 \pm 0.00134 $ & 	 $ 0.02247 \pm 0.00088 $ \\
3891.74278810 	 & $ 0.07925 \pm 0.00135 $ & 	 $ 0.02267 \pm 0.00090 $ \\
4164.97842690 	 & $ 0.08244 \pm 0.00146 $ & 	 $ 0.02207 \pm 0.00100 $ \\
4180.93453894 	 & $ 0.08185 \pm 0.00144 $ & 	 $ 0.02215 \pm 0.00096 $ \\
4217.83656080 	 & $ 0.08090 \pm 0.00147 $ & 	 $ 0.02160 \pm 0.00109 $ \\
4220.83911388 	 & $ 0.07797 \pm 0.00145 $ & 	 $ 0.02198 \pm 0.00118 $ \\
4224.82363852 	 & $ 0.08062 \pm 0.00145 $ & 	 $ 0.02205 \pm 0.00107 $ \\
4550.92235669 	 & $ 0.08065 \pm 0.00138 $ & 	 $ 0.02207 \pm 0.00097 $ \\
4551.91403140 	 & $ 0.07913 \pm 0.00137 $ & 	 $ 0.02218 \pm 0.00098 $ \\
4570.87339939 	 & $ 0.08059 \pm 0.00138 $ & 	 $ 0.02195 \pm 0.00103 $ \\
4575.84797092 	 & $ 0.08029 \pm 0.00139 $ & 	 $ 0.02211 \pm 0.00101 $ \\
4577.84773598 	 & $ 0.08057 \pm 0.00142 $ & 	 $ 0.02217 \pm 0.00100 $ \\
4582.82691079 	 & $ 0.08098 \pm 0.00144 $ & 	 $ 0.02199 \pm 0.00102 $ \\
4609.77239519 	 & $ 0.07976 \pm 0.00147 $ & 	 $ 0.02202 \pm 0.00090 $ \\
4877.02218442 	 & $ 0.07934 \pm 0.00141 $ & 	 $ 0.02190 \pm 0.00101 $ \\
5737.66382144 	 & $ 0.07916 \pm 0.00173 $ & 	 $ 0.02199 \pm 0.00099 $ \\
5740.66331690 	 & $ 0.07898 \pm 0.00184 $ & 	 $ 0.02209 \pm 0.00100 $ \\
5748.64872936 	 & $ 0.07847 \pm 0.00176 $ & 	 $ 0.02204 \pm 0.00100 $ \\
5750.64396108 	 & $ 0.07811 \pm 0.00187 $ & 	 $ 0.02223 \pm 0.00102 $ \\

\hline & & \\
\multicolumn{3}{| c |}{GJ 708} \\
\hline & & \\

2421.76346943 	 & $ 0.05248 \pm 0.00052 $ & 	 $ 0.02632 \pm 0.00077 $ \\
2421.77979618 	 & $ 0.05280 \pm 0.00046 $ & 	 $ 0.02590 \pm 0.00073 $ \\
2424.78077980 	 & $ 0.05331 \pm 0.00039 $ & 	 $ 0.02596 \pm 0.00055 $ \\
2424.95998937 	 & $ 0.05307 \pm 0.00037 $ & 	 $ 0.02614 \pm 0.00052 $ \\
2441.92344871 	 & $ 0.05121 \pm 0.00049 $ & 	 $ 0.02603 \pm 0.00040 $ \\
2442.71502056 	 & $ 0.05327 \pm 0.00042 $ & 	 $ 0.02597 \pm 0.00040 $ \\
2443.90548522 	 & $ 0.05166 \pm 0.00041 $ & 	 $ 0.02590 \pm 0.00035 $ \\
2816.69540677 	 & $ 0.05171 \pm 0.00070 $ & 	 $ 0.02626 \pm 0.00066 $ \\
2830.85081407 	 & $ 0.05290 \pm 0.00071 $ & 	 $ 0.02611 \pm 0.00029 $ \\
2832.86129953 	 & $ 0.05139 \pm 0.00064 $ & 	 $ 0.02611 \pm 0.00025 $ \\
3069.98786632 	 & $ 0.05281 \pm 0.00030 $ & 	 $ 0.02518 \pm 0.00030 $ \\
3071.98819887 	 & $ 0.05512 \pm 0.00037 $ & 	 $ 0.02508 \pm 0.00025 $ \\
3073.96759840 	 & $ 0.05579 \pm 0.00034 $ & 	 $ 0.02512 \pm 0.00027 $ \\
3074.96878585 	 & $ 0.05577 \pm 0.00036 $ & 	 $ 0.02419 \pm 0.00028 $ \\
3102.91202937 	 & $ 0.05289 \pm 0.00036 $ & 	 $ 0.02519 \pm 0.00026 $ \\
3157.95295985 	 & $ 0.05190 \pm 0.00031 $ & 	 $ 0.02601 \pm 0.00036 $ \\
3163.94220809 	 & $ 0.05200 \pm 0.00040 $ & 	 $ 0.02633 \pm 0.00051 $ \\
3165.92941835 	 & $ 0.05388 \pm 0.00049 $ & 	 $ 0.02630 \pm 0.00058 $ \\
3172.91264854 	 & $ 0.05093 \pm 0.00047 $ & 	 $ 0.02632 \pm 0.00046 $ \\
3183.68902746 	 & $ 0.05229 \pm 0.00075 $ & 	 $ 0.02633 \pm 0.00052 $ \\
3185.68858568 	 & $ 0.05281 \pm 0.00088 $ & 	 $ 0.02628 \pm 0.00055 $ \\
3188.87176498 	 & $ 0.05318 \pm 0.00078 $ & 	 $ 0.02637 \pm 0.00043 $ \\
3194.85931903 	 & $ 0.05344 \pm 0.00075 $ & 	 $ 0.02601 \pm 0.00034 $ \\
3202.85173544 	 & $ 0.05194 \pm 0.00073 $ & 	 $ 0.02621 \pm 0.00023 $ \\
3205.84298424 	 & $ 0.05290 \pm 0.00093 $ & 	 $ 0.02628 \pm 0.00023 $ \\
3442.97424666 	 & $ 0.05325 \pm 0.00042 $ & 	 $ 0.02575 \pm 0.00028 $ \\
3446.96351257 	 & $ 0.05303 \pm 0.00031 $ & 	 $ 0.02574 \pm 0.00026 $ \\
3451.94395604 	 & $ 0.05277 \pm 0.00035 $ & 	 $ 0.02594 \pm 0.00025 $ \\
3453.94293104 	 & $ 0.05198 \pm 0.00034 $ & 	 $ 0.02572 \pm 0.00027 $ \\
3460.92021514 	 & $ 0.05262 \pm 0.00031 $ & 	 $ 0.02569 \pm 0.00027 $ \\
3474.87600025 	 & $ 0.05147 \pm 0.00042 $ & 	 $ 0.02498 \pm 0.00031 $ \\
3481.86415042 	 & $ 0.05225 \pm 0.00033 $ & 	 $ 0.02550 \pm 0.00029 $ \\
3484.86326323 	 & $ 0.05132 \pm 0.00039 $ & 	 $ 0.02492 \pm 0.00051 $ \\
3487.86418360 	 & $ 0.05134 \pm 0.00029 $ & 	 $ 0.02547 \pm 0.00032 $ \\
3491.84926346 	 & $ 0.05168 \pm 0.00030 $ & 	 $ 0.02547 \pm 0.00034 $ \\
3582.80419835 	 & $ 0.05270 \pm 0.00071 $ & 	 $ 0.02613 \pm 0.00025 $ \\
3590.76075956 	 & $ 0.05143 \pm 0.00089 $ & 	 $ 0.02595 \pm 0.00025 $ \\
3605.72212062 	 & $ 0.05099 \pm 0.00107 $ & 	 $ 0.02614 \pm 0.00032 $ \\
3608.73157162 	 & $ 0.05147 \pm 0.00088 $ & 	 $ 0.02594 \pm 0.00033 $ \\
3652.58567419 	 & $ 0.05410 \pm 0.00049 $ & 	 $ 0.02551 \pm 0.00026 $ \\
3655.59108942 	 & $ 0.05341 \pm 0.00058 $ & 	 $ 0.02567 \pm 0.00026 $ \\
3663.56322439 	 & $ 0.05401 \pm 0.00060 $ & 	 $ 0.02540 \pm 0.00024 $ \\
3668.57311897 	 & $ 0.05268 \pm 0.00070 $ & 	 $ 0.02575 \pm 0.00034 $ \\
3788.01559027 	 & $ 0.05204 \pm 0.00072 $ & 	 $ 0.02547 \pm 0.00058 $ \\
3802.99870980 	 & $ 0.05112 \pm 0.00062 $ & 	 $ 0.02530 \pm 0.00045 $ \\
3806.97523726 	 & $ 0.05163 \pm 0.00071 $ & 	 $ 0.02541 \pm 0.00040 $ \\
3809.95817637 	 & $ 0.05174 \pm 0.00063 $ & 	 $ 0.02522 \pm 0.00041 $ \\
3815.94969305 	 & $ 0.05052 \pm 0.00052 $ & 	 $ 0.02598 \pm 0.00037 $ \\
3898.72199851 	 & $ 0.05059 \pm 0.00038 $ & 	 $ 0.02674 \pm 0.00053 $ \\
3900.91993971 	 & $ 0.05184 \pm 0.00038 $ & 	 $ 0.02633 \pm 0.00039 $ \\
3921.66287372 	 & $ 0.05473 \pm 0.00123 $ & 	 $ 0.02572 \pm 0.00055 $ \\
3939.82730454 	 & $ 0.05080 \pm 0.00085 $ & 	 $ 0.02596 \pm 0.00023 $ \\
3954.77105354 	 & $ 0.05172 \pm 0.00068 $ & 	 $ 0.02558 \pm 0.00025 $ \\
3989.68370535 	 & $ 0.04951 \pm 0.00080 $ & 	 $ 0.02606 \pm 0.00029 $ \\
4161.99612571 	 & $ 0.05024 \pm 0.00035 $ & 	 $ 0.02576 \pm 0.00030 $ \\
4180.94946715 	 & $ 0.05154 \pm 0.00032 $ & 	 $ 0.02522 \pm 0.00027 $ \\
4192.91495195 	 & $ 0.05235 \pm 0.00031 $ & 	 $ 0.02542 \pm 0.00025 $ \\
4215.84574815 	 & $ 0.05192 \pm 0.00033 $ & 	 $ 0.02565 \pm 0.00029 $ \\
4230.81320097 	 & $ 0.05202 \pm 0.00035 $ & 	 $ 0.02580 \pm 0.00043 $ \\
4243.77943634 	 & $ 0.05355 \pm 0.00035 $ & 	 $ 0.02583 \pm 0.00040 $ \\
4318.77353674 	 & $ 0.05160 \pm 0.00103 $ & 	 $ 0.02599 \pm 0.00023 $ \\
4340.72126095 	 & $ 0.05073 \pm 0.00087 $ & 	 $ 0.02602 \pm 0.00025 $ \\
4358.67459053 	 & $ 0.05203 \pm 0.00056 $ & 	 $ 0.02601 \pm 0.00023 $ \\
4544.94380190 	 & $ 0.05186 \pm 0.00031 $ & 	 $ 0.02554 \pm 0.00025 $ \\
4574.87470590 	 & $ 0.05325 \pm 0.00032 $ & 	 $ 0.02568 \pm 0.00025 $ \\
4604.79589362 	 & $ 0.05310 \pm 0.00039 $ & 	 $ 0.02582 \pm 0.00050 $ \\
4617.75601950 	 & $ 0.05132 \pm 0.00037 $ & 	 $ 0.02599 \pm 0.00058 $ \\
4632.71943325 	 & $ 0.05107 \pm 0.00047 $ & 	 $ 0.02600 \pm 0.00052 $ \\
4637.71503991 	 & $ 0.05230 \pm 0.00053 $ & 	 $ 0.02603 \pm 0.00048 $ \\
4640.90649759 	 & $ 0.05096 \pm 0.00059 $ & 	 $ 0.02600 \pm 0.00049 $ \\
4908.95539173 	 & $ 0.05186 \pm 0.00030 $ & 	 $ 0.02541 \pm 0.00025 $ \\
5054.78046302 	 & $ 0.05175 \pm 0.00105 $ & 	 $ 0.02602 \pm 0.00027 $ \\
4640.90649759 	 & $ 0.05216 \pm 0.00042 $ & 	 $ 0.02569 \pm 0.00027 $ \\
5801.72358549 	 & $ 0.05069 \pm 0.00057 $ & 	 $ 0.02568 \pm 0.00025 $ \\
5816.66339932 	 & $ 0.05174 \pm 0.00071 $ & 	 $ 0.02560 \pm 0.00027 $ \\
5848.58886667 	 & $ 0.05065 \pm 0.00042 $ & 	 $ 0.02555 \pm 0.00025 $ \\
6002.95662503 	 & $ 0.05173 \pm 0.00037 $ & 	 $ 0.02447 \pm 0.00028 $ \\
6071.78666575 	 & $ 0.05322 \pm 0.00034 $ & 	 $ 0.02561 \pm 0.00047 $ \\
6073.77503727 	 & $ 0.05326 \pm 0.00044 $ & 	 $ 0.02557 \pm 0.00066 $ \\
6082.95147847 	 & $ 0.05295 \pm 0.00036 $ & 	 $ 0.02575 \pm 0.00046 $ \\

\hline & & \\
\multicolumn{3}{| c |}{GJ 730} \\
\hline & & \\

2806.77210976 	 & $ 0.06383 \pm 0.00065 $ & 	 $ 0.02493 \pm 0.00077 $ \\
2807.78411331 	 & $ 0.06422 \pm 0.00059 $ & 	 $ 0.02478 \pm 0.00076 $ \\
2823.73656281 	 & $ 0.06327 \pm 0.00066 $ & 	 $ 0.02452 \pm 0.00051 $ \\
2824.72252734 	 & $ 0.06199 \pm 0.00077 $ & 	 $ 0.02449 \pm 0.00049 $ \\
2824.73816894 	 & $ 0.06226 \pm 0.00061 $ & 	 $ 0.02416 \pm 0.00046 $ \\
2824.85921550 	 & $ 0.06360 \pm 0.00070 $ & 	 $ 0.02458 \pm 0.00047 $ \\
3157.83409461 	 & $ 0.06689 \pm 0.00062 $ & 	 $ 0.02434 \pm 0.00056 $ \\
3182.87988530 	 & $ 0.06254 \pm 0.00061 $ & 	 $ 0.02471 \pm 0.00059 $ \\
3195.84876306 	 & $ 0.06318 \pm 0.00077 $ & 	 $ 0.02473 \pm 0.00056 $ \\
3205.81091498 	 & $ 0.06252 \pm 0.00098 $ & 	 $ 0.02486 \pm 0.00051 $ \\
3215.78964708 	 & $ 0.06323 \pm 0.00116 $ & 	 $ 0.02481 \pm 0.00041 $ \\
3601.71619604 	 & $ 0.06781 \pm 0.00115 $ & 	 $ 0.02440 \pm 0.00048 $ \\
3605.70795136 	 & $ 0.06377 \pm 0.00126 $ & 	 $ 0.02440 \pm 0.00045 $ \\
3609.71594164 	 & $ 0.06570 \pm 0.00122 $ & 	 $ 0.02481 \pm 0.00048 $ \\
3609.72618630 	 & $ 0.06452 \pm 0.00121 $ & 	 $ 0.02454 \pm 0.00049 $ \\
3612.69830942 	 & $ 0.06382 \pm 0.00096 $ & 	 $ 0.02431 \pm 0.00048 $ \\
3615.70319690 	 & $ 0.06379 \pm 0.00101 $ & 	 $ 0.02443 \pm 0.00049 $ \\
3625.65935946 	 & $ 0.06616 \pm 0.00106 $ & 	 $ 0.02430 \pm 0.00046 $ \\
3628.65472372 	 & $ 0.06643 \pm 0.00089 $ & 	 $ 0.02444 \pm 0.00045 $ \\
3899.77908026 	 & $ 0.06165 \pm 0.00058 $ & 	 $ 0.02511 \pm 0.00069 $ \\
3905.77441403 	 & $ 0.06106 \pm 0.00057 $ & 	 $ 0.02482 \pm 0.00066 $ \\
3956.74984765 	 & $ 0.06233 \pm 0.00121 $ & 	 $ 0.02443 \pm 0.00044 $ \\
3966.72803865 	 & $ 0.06378 \pm 0.00103 $ & 	 $ 0.02420 \pm 0.00045 $ \\
3997.64703601 	 & $ 0.06266 \pm 0.00049 $ & 	 $ 0.02399 \pm 0.00045 $ \\
4205.94360136 	 & $ 0.06234 \pm 0.00055 $ & 	 $ 0.02444 \pm 0.00047 $ \\
4207.93884490 	 & $ 0.06354 \pm 0.00055 $ & 	 $ 0.02425 \pm 0.00046 $ \\
4249.95357766 	 & $ 0.06339 \pm 0.00068 $ & 	 $ 0.02386 \pm 0.00063 $ \\
4251.81349095 	 & $ 0.06321 \pm 0.00061 $ & 	 $ 0.02403 \pm 0.00062 $ \\
4255.93527667 	 & $ 0.06705 \pm 0.00067 $ & 	 $ 0.02421 \pm 0.00071 $ \\
4329.61300676 	 & $ 0.06384 \pm 0.00126 $ & 	 $ 0.02457 \pm 0.00045 $ \\
4565.95433274 	 & $ 0.06565 \pm 0.00056 $ & 	 $ 0.02380 \pm 0.00047 $ \\
4574.93691937 	 & $ 0.06498 \pm 0.00061 $ & 	 $ 0.02433 \pm 0.00045 $ \\
4586.90741429 	 & $ 0.06459 \pm 0.00056 $ & 	 $ 0.02385 \pm 0.00048 $ \\
4601.86950800 	 & $ 0.06633 \pm 0.00064 $ & 	 $ 0.02372 \pm 0.00047 $ \\
4616.83025685 	 & $ 0.06367 \pm 0.00075 $ & 	 $ 0.02358 \pm 0.00075 $ \\
4663.70331949 	 & $ 0.06496 \pm 0.00108 $ & 	 $ 0.02466 \pm 0.00073 $ \\
4684.63406145 	 & $ 0.06336 \pm 0.00121 $ & 	 $ 0.02439 \pm 0.00043 $ \\
4688.63578915 	 & $ 0.06286 \pm 0.00136 $ & 	 $ 0.02439 \pm 0.00044 $ \\
4732.62582364 	 & $ 0.06284 \pm 0.00062 $ & 	 $ 0.02419 \pm 0.00045 $ \\
4735.62142047 	 & $ 0.06266 \pm 0.00067 $ & 	 $ 0.02417 \pm 0.00047 $ \\
5290.97110702 	 & $ 0.06358 \pm 0.00054 $ & 	 $ 0.02374 \pm 0.00047 $ \\
5290.97894777 	 & $ 0.06366 \pm 0.00057 $ & 	 $ 0.02380 \pm 0.00047 $ \\
5293.95863282 	 & $ 0.06402 \pm 0.00059 $ & 	 $ 0.02396 \pm 0.00047 $ \\
5293.96646211 	 & $ 0.06459 \pm 0.00058 $ & 	 $ 0.02388 \pm 0.00047 $ \\
5294.96195596 	 & $ 0.06410 \pm 0.00057 $ & 	 $ 0.02396 \pm 0.00047 $ \\
5294.96979451 	 & $ 0.06258 \pm 0.00057 $ & 	 $ 0.02398 \pm 0.00047 $ \\
5308.92314266 	 & $ 0.06553 \pm 0.00056 $ & 	 $ 0.02361 \pm 0.00050 $ \\
5308.93096705 	 & $ 0.06555 \pm 0.00060 $ & 	 $ 0.02360 \pm 0.00047 $ \\
5324.88564373 	 & $ 0.06497 \pm 0.00063 $ & 	 $ 0.02331 \pm 0.00049 $ \\
5324.89347973 	 & $ 0.06348 \pm 0.00066 $ & 	 $ 0.02321 \pm 0.00047 $ \\
5351.81733642 	 & $ 0.06626 \pm 0.00069 $ & 	 $ 0.02400 \pm 0.00066 $ \\
5351.82515888 	 & $ 0.06463 \pm 0.00069 $ & 	 $ 0.02412 \pm 0.00065 $ \\
5369.77055715 	 & $ 0.06651 \pm 0.00066 $ & 	 $ 0.02445 \pm 0.00083 $ \\
5369.77838613 	 & $ 0.06435 \pm 0.00067 $ & 	 $ 0.02451 \pm 0.00081 $ \\
5372.75425063 	 & $ 0.06479 \pm 0.00064 $ & 	 $ 0.02448 \pm 0.00064 $ \\
5372.76207992 	 & $ 0.06579 \pm 0.00060 $ & 	 $ 0.02446 \pm 0.00062 $ \\
5800.71138893 	 & $ 0.06316 \pm 0.00083 $ & 	 $ 0.02424 \pm 0.00048 $ \\
5802.70649889 	 & $ 0.06358 \pm 0.00097 $ & 	 $ 0.02437 \pm 0.00049 $ \\
5805.68692039 	 & $ 0.06347 \pm 0.00103 $ & 	 $ 0.02423 \pm 0.00049 $ \\
5808.68091699 	 & $ 0.06326 \pm 0.00091 $ & 	 $ 0.02420 \pm 0.00047 $ \\
5812.68401024 	 & $ 0.06275 \pm 0.00079 $ & 	 $ 0.02421 \pm 0.00047 $ \\
5814.66188152 	 & $ 0.06398 \pm 0.00084 $ & 	 $ 0.02426 \pm 0.00048 $ \\
6064.86563127 	 & $ 0.06316 \pm 0.00059 $ & 	 $ 0.02371 \pm 0.00045 $ \\
6072.95041837 	 & $ 0.06291 \pm 0.00065 $ & 	 $ 0.02346 \pm 0.00060 $ \\
6081.94417888 	 & $ 0.06305 \pm 0.00071 $ & 	 $ 0.02375 \pm 0.00073 $ \\
6082.93894380 	 & $ 0.06432 \pm 0.00072 $ & 	 $ 0.02403 \pm 0.00063 $ \\
6088.92823555 	 & $ 0.06389 \pm 0.00066 $ & 	 $ 0.02404 \pm 0.00062 $ \\
6089.92354455 	 & $ 0.06566 \pm 0.00067 $ & 	 $ 0.02432 \pm 0.00066 $ \\
6170.69649448    & $ 0.06346 \pm 0.00079 $ &     $ 0.02394 \pm 0.00046 $ \\
6170.71493655    & $ 0.06120 \pm 0.00077 $ &     $ 0.02415 \pm 0.00047 $ \\

\hline & & \\
\multicolumn{3}{| c |}{GJ 552} \\
\hline & & \\

2448.75113066 	 & $ 0.06696 \pm 0.00101 $ & 	 $ 0.02412 \pm 0.00077 $ \\
2455.71782301 	 & $ 0.06808 \pm 0.00100 $ & 	 $ 0.02401 \pm 0.00075 $ \\
2455.73274519 	 & $ 0.06898 \pm 0.00102 $ & 	 $ 0.02364 \pm 0.00075 $ \\
2469.68925359 	 & $ 0.06694 \pm 0.00102 $ & 	 $ 0.02404 \pm 0.00073 $ \\
2477.66327025 	 & $ 0.06717 \pm 0.00103 $ & 	 $ 0.02402 \pm 0.00075 $ \\
2479.66601240 	 & $ 0.06582 \pm 0.00101 $ & 	 $ 0.02407 \pm 0.00082 $ \\
2480.66247793 	 & $ 0.06572 \pm 0.00096 $ & 	 $ 0.02396 \pm 0.00078 $ \\
3183.72216292 	 & $ 0.06619 \pm 0.00100 $ & 	 $ 0.02404 \pm 0.00077 $ \\
3215.63097455 	 & $ 0.06604 \pm 0.00097 $ & 	 $ 0.02402 \pm 0.00076 $ \\
3215.64649163 	 & $ 0.06618 \pm 0.00096 $ & 	 $ 0.02395 \pm 0.00075 $ \\
3940.65683466 	 & $ 0.06491 \pm 0.00092 $ & 	 $ 0.02384 \pm 0.00074 $ \\
4139.90394966 	 & $ 0.06882 \pm 0.00109 $ & 	 $ 0.02335 \pm 0.00082 $ \\
4218.69833653 	 & $ 0.06925 \pm 0.00125 $ & 	 $ 0.02379 \pm 0.00071 $ \\
4224.87675309 	 & $ 0.07070 \pm 0.00117 $ & 	 $ 0.02389 \pm 0.00074 $ \\
4251.79630645 	 & $ 0.06957 \pm 0.00103 $ & 	 $ 0.02392 \pm 0.00074 $ \\
4460.02422784 	 & $ 0.07025 \pm 0.00109 $ & 	 $ 0.02390 \pm 0.00074 $ \\
4474.99206740 	 & $ 0.06873 \pm 0.00107 $ & 	 $ 0.02353 \pm 0.00077 $ \\
4545.81077781 	 & $ 0.07000 \pm 0.00102 $ & 	 $ 0.02387 \pm 0.00073 $ \\
4560.76617405 	 & $ 0.07106 \pm 0.00113 $ & 	 $ 0.02380 \pm 0.00070 $ \\
4569.75287007 	 & $ 0.07143 \pm 0.00108 $ & 	 $ 0.02372 \pm 0.00077 $ \\
4575.92073547 	 & $ 0.07180 \pm 0.00114 $ & 	 $ 0.02380 \pm 0.00074 $ \\
4584.69254115 	 & $ 0.07083 \pm 0.00113 $ & 	 $ 0.02384 \pm 0.00075 $ \\
4619.80412414 	 & $ 0.07312 \pm 0.00116 $ & 	 $ 0.02382 \pm 0.00077 $ \\
4835.00770689 	 & $ 0.06969 \pm 0.00107 $ & 	 $ 0.02325 \pm 0.00078 $ \\
4859.95361018 	 & $ 0.07314 \pm 0.00107 $ & 	 $ 0.02364 \pm 0.00076 $ \\
4876.89689302 	 & $ 0.07031 \pm 0.00111 $ & 	 $ 0.02322 \pm 0.00082 $ \\
5413.62589217 	 & $ 0.06956 \pm 0.00105 $ & 	 $ 0.02378 \pm 0.00078 $ \\
5695.65820816 	 & $ 0.06948 \pm 0.00120 $ & 	 $ 0.02358 \pm 0.00077 $ \\

\hline & & \\
\multicolumn{3}{| c |}{GJ 16} \\
\hline & & \\

4371.71690550 	 & $ 0.06712 \pm 0.00100 $ & 	 $ 0.02289 \pm 0.00082 $ \\
4401.79042154 	 & $ 0.06955 \pm 0.00091 $ & 	 $ 0.02364 \pm 0.00059 $ \\
4402.62106200 	 & $ 0.06685 \pm 0.00090 $ & 	 $ 0.02361 \pm 0.00058 $ \\
4772.77189820 	 & $ 0.06743 \pm 0.00092 $ & 	 $ 0.02354 \pm 0.00057 $ \\
5138.77685076 	 & $ 0.06518 \pm 0.00091 $ & 	 $ 0.02351 \pm 0.00057 $ \\
5542.66322352 	 & $ 0.06451 \pm 0.00104 $ & 	 $ 0.02352 \pm 0.00062 $ \\
5896.70222250 	 & $ 0.06419 \pm 0.00090 $ & 	 $ 0.02349 \pm 0.00064 $ \\
5915.64475395 	 & $ 0.06469 \pm 0.00099 $ & 	 $ 0.02355 \pm 0.00063 $ \\
5929.59721433 	 & $ 0.06475 \pm 0.00090 $ & 	 $ 0.02345 \pm 0.00066 $ \\

\hline & & \\
\multicolumn{3}{| c |}{GJ 521} \\
\hline & & \\

4605.81420282 	 & $ 0.06859 \pm 0.00088 $ & 	 $ 0.02507 \pm 0.00075 $ \\
4605.82945467 	 & $ 0.06864 \pm 0.00087 $ & 	 $ 0.02524 \pm 0.00076 $ \\
4606.81447579 	 & $ 0.06897 \pm 0.00086 $ & 	 $ 0.02517 \pm 0.00075 $ \\
4612.80109233 	 & $ 0.07064 \pm 0.00083 $ & 	 $ 0.02492 \pm 0.00101 $ \\
4628.73797731 	 & $ 0.07185 \pm 0.00087 $ & 	 $ 0.02451 \pm 0.00080 $ \\
4813.02061638 	 & $ 0.06894 \pm 0.00092 $ & 	 $ 0.02527 \pm 0.00065 $ \\
4840.95145462 	 & $ 0.06748 \pm 0.00090 $ & 	 $ 0.02523 \pm 0.00064 $ \\
5392.66208754 	 & $ 0.06731 \pm 0.00081 $ & 	 $ 0.02440 \pm 0.00108 $ \\
5684.88482109 	 & $ 0.06586 \pm 0.00091 $ & 	 $ 0.02504 \pm 0.00067 $ \\
5703.80173796 	 & $ 0.06838 \pm 0.00086 $ & 	 $ 0.02490 \pm 0.00073 $ \\
5922.97859508 	 & $ 0.06666 \pm 0.00094 $ & 	 $ 0.02527 \pm 0.00064 $ \\
5947.91120096 	 & $ 0.06510 \pm 0.00094 $ & 	 $ 0.02513 \pm 0.00064 $ \\
6010.73812761 	 & $ 0.06602 \pm 0.00097 $ & 	 $ 0.02516 \pm 0.00060 $ \\
6022.71712893 	 & $ 0.06801 \pm 0.00099 $ & 	 $ 0.02506 \pm 0.00056 $ \\
6078.77366322 	 & $ 0.06744 \pm 0.00091 $ & 	 $ 0.02501 \pm 0.00068 $ \\
6089.72880678 	 & $ 0.06583 \pm 0.00086 $ & 	 $ 0.02449 \pm 0.00075 $ \\
6098.71812098 	 & $ 0.06584 \pm 0.00088 $ & 	 $ 0.02498 \pm 0.00069 $ \\
6110.68392028 	 & $ 0.06612 \pm 0.00081 $ & 	 $ 0.02434 \pm 0.00090 $ \\

\hline & & \\
\multicolumn{3}{| c |}{GJ 96} \\
\hline & & \\

2928.71737766 	 & $ 0.06136 \pm 0.00044 $ & 	 $ 0.02442 \pm 0.00058 $ \\
2928.72907718 	 & $ 0.06212 \pm 0.00044 $ & 	 $ 0.02416 \pm 0.00056 $ \\
2940.68691338 	 & $ 0.06511 \pm 0.00043 $ & 	 $ 0.02392 \pm 0.00041 $ \\
2942.90722459 	 & $ 0.06165 \pm 0.00048 $ & 	 $ 0.02412 \pm 0.00043 $ \\
2944.67763094 	 & $ 0.06087 \pm 0.00047 $ & 	 $ 0.02391 \pm 0.00045 $ \\
2947.67704335 	 & $ 0.06334 \pm 0.00047 $ & 	 $ 0.02347 \pm 0.00045 $ \\
2949.89161468 	 & $ 0.06151 \pm 0.00052 $ & 	 $ 0.02378 \pm 0.00043 $ \\
2951.63202344 	 & $ 0.06196 \pm 0.00056 $ & 	 $ 0.02395 \pm 0.00042 $ \\
2952.63215148 	 & $ 0.06172 \pm 0.00054 $ & 	 $ 0.02371 \pm 0.00044 $ \\
2954.64677573 	 & $ 0.06186 \pm 0.00057 $ & 	 $ 0.02368 \pm 0.00043 $ \\
2958.84818141 	 & $ 0.06215 \pm 0.00056 $ & 	 $ 0.02341 \pm 0.00041 $ \\
2958.86048483 	 & $ 0.06204 \pm 0.00057 $ & 	 $ 0.02325 \pm 0.00040 $ \\
2958.87197529 	 & $ 0.06220 \pm 0.00058 $ & 	 $ 0.02332 \pm 0.00039 $ \\
2963.64218924 	 & $ 0.06372 \pm 0.00052 $ & 	 $ 0.02432 \pm 0.00039 $ \\
2963.82361268 	 & $ 0.06467 \pm 0.00052 $ & 	 $ 0.02446 \pm 0.00039 $ \\
2965.84855790 	 & $ 0.06606 \pm 0.00054 $ & 	 $ 0.02444 \pm 0.00042 $ \\
3295.68562983 	 & $ 0.06174 \pm 0.00048 $ & 	 $ 0.02462 \pm 0.00057 $ \\
3313.86451512 	 & $ 0.06030 \pm 0.00049 $ & 	 $ 0.02475 \pm 0.00045 $ \\
4447.77277951 	 & $ 0.05977 \pm 0.00052 $ & 	 $ 0.02414 \pm 0.00046 $ \\
4727.99644880 	 & $ 0.06175 \pm 0.00053 $ & 	 $ 0.02448 \pm 0.00048 $ \\
5114.97547929 	 & $ 0.05988 \pm 0.00052 $ & 	 $ 0.02446 \pm 0.00048 $ \\
5178.55049329 	 & $ 0.06042 \pm 0.00050 $ & 	 $ 0.02402 \pm 0.00045 $ \\
5546.56412660 	 & $ 0.05914 \pm 0.00052 $ & 	 $ 0.02398 \pm 0.00045 $ \\
5552.76540903 	 & $ 0.06087 \pm 0.00054 $ & 	 $ 0.02354 \pm 0.00044 $ \\
5576.69626674 	 & $ 0.05989 \pm 0.00062 $ & 	 $ 0.02379 \pm 0.00052 $ \\
5863.90279976 	 & $ 0.05977 \pm 0.00052 $ & 	 $ 0.02377 \pm 0.00043 $ \\
5873.64370409 	 & $ 0.05980 \pm 0.00056 $ & 	 $ 0.02353 \pm 0.00042 $ \\
5883.83733067 	 & $ 0.05904 \pm 0.00058 $ & 	 $ 0.02384 \pm 0.00041 $ \\
5910.53810524 	 & $ 0.05935 \pm 0.00052 $ & 	 $ 0.02344 \pm 0.00045 $ \\

\hline & & \\
\multicolumn{3}{| c |}{GJ 3023} \\
\hline & & \\

4341.75211215 	 & $ 0.06828 \pm  0.00089 $ 	 & $ 0.02267 \pm  0.00063 $ \\
4341.76757184 	 & $ 0.06663 \pm  0.00075 $ 	 & $ 0.02290 \pm  0.00065 $ \\
4366.67935366 	 & $ 0.06647 \pm  0.00122 $ 	 & $ 0.02262 \pm  0.00105 $ \\
4369.68014645 	 & $ 0.06739 \pm  0.00115 $ 	 & $ 0.02347 \pm  0.00102 $ \\
4370.68601039 	 & $ 0.06873 \pm  0.00120 $ 	 & $ 0.02360 \pm  0.00103 $ \\
4482.58914037 	 & $ 0.06458 \pm  0.00119 $ 	 & $ 0.02376 \pm  0.00070 $ \\
4698.78616303 	 & $ 0.06817 \pm  0.00083 $ 	 & $ 0.02295 \pm  0.00070 $ \\
5073.75275416 	 & $ 0.06879 \pm  0.00099 $ 	 & $ 0.02318 \pm  0.00077 $ \\
5567.63575564 	 & $ 0.07211 \pm  0.00096 $ 	 & $ 0.02341 \pm  0.00072 $ \\
6181.94542826 	 & $ 0.07357 \pm  0.00098 $ 	 & $ 0.02179 \pm  0.00072 $ \\

\hline & & \\
\multicolumn{3}{| c |}{GJ 3801} \\
\hline & & \\

4476.93940118 	 & $ 0.08335 \pm 0.00187 $ & 	 $ 0.02245 \pm 0.00123 $ \\
4476.95469620 	 & $ 0.08605 \pm 0.00180 $ & 	 $ 0.02247 \pm 0.00122 $ \\
4485.90972080 	 & $ 0.07897 \pm 0.00174 $ & 	 $ 0.02321 \pm 0.00116 $ \\
4507.85448032 	 & $ 0.08612 \pm 0.00172 $ & 	 $ 0.02265 \pm 0.00124 $ \\
4518.82118420 	 & $ 0.08597 \pm 0.00169 $ & 	 $ 0.02304 \pm 0.00120 $ \\
4523.81256612 	 & $ 0.08300 \pm 0.00181 $ & 	 $ 0.02303 \pm 0.00112 $ \\
4580.87665907 	 & $ 0.08451 \pm 0.00181 $ & 	 $ 0.02268 \pm 0.00116 $ \\
4628.76518832 	 & $ 0.08241 \pm 0.00172 $ & 	 $ 0.02312 \pm 0.00120 $ \\
4842.95049497 	 & $ 0.08123 \pm 0.00183 $ & 	 $ 0.02288 \pm 0.00113 $ \\
4845.93432657 	 & $ 0.08290 \pm 0.00172 $ & 	 $ 0.02241 \pm 0.00119 $ \\
4869.86597289 	 & $ 0.08198 \pm 0.00171 $ & 	 $ 0.02291 \pm 0.00125 $ \\
5396.64145117 	 & $ 0.07997 \pm 0.00186 $ & 	 $ 0.02263 \pm 0.00117 $ \\
5396.65275859 	 & $ 0.07949 \pm 0.00188 $ & 	 $ 0.02275 \pm 0.00121 $ \\
5685.64790599 	 & $ 0.08241 \pm 0.00182 $ & 	 $ 0.02246 \pm 0.00119 $ \\
6015.73038865 	 & $ 0.08678 \pm 0.00186 $ & 	 $ 0.02261 \pm 0.00116 $ \\

\hline & & \\
\multicolumn{3}{| c |}{GJ 611.3} \\
\hline & & \\

2798.66234075 	 & $ 0.06477 \pm 0.00062 $ & 	 $ 0.02432 \pm 0.00046 $ \\
2798.68495852 	 & $ 0.06285 \pm 0.00069 $ & 	 $ 0.02471 \pm 0.00047 $ \\
2801.82913296 	 & $ 0.06169 \pm 0.00067 $ & 	 $ 0.02474 \pm 0.00050 $ \\
2802.82265109 	 & $ 0.06080 \pm 0.00073 $ & 	 $ 0.02488 \pm 0.00046 $ \\
2805.81202651 	 & $ 0.06222 \pm 0.00064 $ & 	 $ 0.02493 \pm 0.00050 $ \\
2806.80481737 	 & $ 0.06203 \pm 0.00072 $ & 	 $ 0.02463 \pm 0.00057 $ \\
2824.75534886 	 & $ 0.06335 \pm 0.00075 $ & 	 $ 0.02374 \pm 0.00057 $ \\
3136.92020795 	 & $ 0.06245 \pm 0.00058 $ & 	 $ 0.02387 \pm 0.00060 $ \\
3192.74203561 	 & $ 0.06391 \pm 0.00066 $ & 	 $ 0.02359 \pm 0.00057 $ \\
3906.79408624 	 & $ 0.05920 \pm 0.00054 $ & 	 $ 0.02459 \pm 0.00049 $ \\
4575.95962878 	 & $ 0.06157 \pm 0.00061 $ & 	 $ 0.02470 \pm 0.00052 $ \\
4595.91849395 	 & $ 0.06108 \pm 0.00059 $ & 	 $ 0.02395 \pm 0.00053 $ \\
4613.70348818 	 & $ 0.06106 \pm 0.00078 $ & 	 $ 0.02400 \pm 0.00051 $ \\
4616.85256418 	 & $ 0.06125 \pm 0.00079 $ & 	 $ 0.02396 \pm 0.00048 $ \\
4922.84993631 	 & $ 0.06206 \pm 0.00063 $ & 	 $ 0.02466 \pm 0.00043 $ \\
5737.63188445 	 & $ 0.06528 \pm 0.00061 $ & 	 $ 0.02380 \pm 0.00052 $ \\

\hline & & \\
\multicolumn{3}{| c |}{GJ 630} \\
\hline & & \\

2769.72843715 	 & $ 0.05594 \pm 0.00036 $ & 	 $ 0.02634 \pm 0.00026 $ \\
2769.74429402 	 & $ 0.05620 \pm 0.00033 $ & 	 $ 0.02616 \pm 0.00026 $ \\
2770.74390336 	 & $ 0.05811 \pm 0.00042 $ & 	 $ 0.02668 \pm 0.00026 $ \\
2803.87105450 	 & $ 0.05698 \pm 0.00037 $ & 	 $ 0.02666 \pm 0.00052 $ \\
2805.85276790 	 & $ 0.05858 \pm 0.00057 $ & 	 $ 0.02656 \pm 0.00057 $ \\
3138.95335192 	 & $ 0.05525 \pm 0.00031 $ & 	 $ 0.02636 \pm 0.00024 $ \\
3157.89220154 	 & $ 0.05464 \pm 0.00024 $ & 	 $ 0.02662 \pm 0.00028 $ \\
3185.82205461 	 & $ 0.05698 \pm 0.00045 $ & 	 $ 0.02726 \pm 0.00074 $ \\
3202.78860526 	 & $ 0.05915 \pm 0.00040 $ & 	 $ 0.02721 \pm 0.00057 $ \\
3215.74414667 	 & $ 0.05634 \pm 0.00043 $ & 	 $ 0.02740 \pm 0.00068 $ \\
3605.66947853 	 & $ 0.05682 \pm 0.00051 $ & 	 $ 0.02727 \pm 0.00069 $ \\
3608.65651004 	 & $ 0.05565 \pm 0.00046 $ & 	 $ 0.02689 \pm 0.00065 $ \\
3628.61207839 	 & $ 0.05870 \pm 0.00040 $ & 	 $ 0.02723 \pm 0.00067 $ \\
3629.60629054 	 & $ 0.05644 \pm 0.00035 $ & 	 $ 0.02680 \pm 0.00045 $ \\
3640.58140678 	 & $ 0.05911 \pm 0.00051 $ & 	 $ 0.02680 \pm 0.00059 $ \\
3755.02925396 	 & $ 0.05722 \pm 0.00050 $ & 	 $ 0.02683 \pm 0.00030 $ \\
3766.01195160 	 & $ 0.05542 \pm 0.00047 $ & 	 $ 0.02669 \pm 0.00038 $ \\
3770.00652326 	 & $ 0.05820 \pm 0.00037 $ & 	 $ 0.02594 \pm 0.00028 $ \\
3780.96926864 	 & $ 0.05305 \pm 0.00044 $ & 	 $ 0.02783 \pm 0.00031 $ \\
3809.89071786 	 & $ 0.05566 \pm 0.00053 $ & 	 $ 0.02620 \pm 0.00035 $ \\
3816.88025027 	 & $ 0.05509 \pm 0.00052 $ & 	 $ 0.02603 \pm 0.00033 $ \\
3843.79011280 	 & $ 0.05414 \pm 0.00043 $ & 	 $ 0.02664 \pm 0.00028 $ \\
3867.71826947 	 & $ 0.05526 \pm 0.00024 $ & 	 $ 0.02604 \pm 0.00022 $ \\
3872.93219238 	 & $ 0.05430 \pm 0.00027 $ & 	 $ 0.02630 \pm 0.00024 $ \\
3875.92137530 	 & $ 0.05391 \pm 0.00041 $ & 	 $ 0.02585 \pm 0.00029 $ \\
3876.93499510 	 & $ 0.05339 \pm 0.00026 $ & 	 $ 0.02621 \pm 0.00021 $ \\
3879.93254381 	 & $ 0.05458 \pm 0.00047 $ & 	 $ 0.02554 \pm 0.00026 $ \\
3889.89922516 	 & $ 0.05477 \pm 0.00035 $ & 	 $ 0.02639 \pm 0.00035 $ \\
4134.99091051 	 & $ 0.05628 \pm 0.00051 $ & 	 $ 0.02672 \pm 0.00021 $ \\
4136.98099813 	 & $ 0.05487 \pm 0.00028 $ & 	 $ 0.02663 \pm 0.00019 $ \\
4173.87775570 	 & $ 0.05696 \pm 0.00034 $ & 	 $ 0.02635 \pm 0.00020 $ \\
4221.75954594 	 & $ 0.05562 \pm 0.00027 $ & 	 $ 0.02614 \pm 0.00021 $ \\
4251.67206540 	 & $ 0.05652 \pm 0.00030 $ & 	 $ 0.02616 \pm 0.00030 $ \\
4256.89485013 	 & $ 0.05499 \pm 0.00038 $ & 	 $ 0.02586 \pm 0.00044 $ \\
4576.79143792 	 & $ 0.05518 \pm 0.00051 $ & 	 $ 0.02677 \pm 0.00031 $ \\
4588.75535565 	 & $ 0.05501 \pm 0.00025 $ & 	 $ 0.02640 \pm 0.00022 $ \\
4605.92940501 	 & $ 0.05463 \pm 0.00024 $ & 	 $ 0.02610 \pm 0.00023 $ \\
4610.91897503 	 & $ 0.05535 \pm 0.00032 $ & 	 $ 0.02629 \pm 0.00025 $ \\
4617.89417931 	 & $ 0.05460 \pm 0.00035 $ & 	 $ 0.02595 \pm 0.00024 $ \\
4872.97577035 	 & $ 0.05473 \pm 0.00038 $ & 	 $ 0.02665 \pm 0.00023 $ \\
5737.83092890 	 & $ 0.05321 \pm 0.00033 $ & 	 $ 0.02654 \pm 0.00038 $ \\
5737.84606950 	 & $ 0.05358 \pm 0.00030 $ & 	 $ 0.02639 \pm 0.00043 $ \\
5800.67495127 	 & $ 0.05369 \pm 0.00034 $ & 	 $ 0.02704 \pm 0.00045 $ \\
5801.66454511 	 & $ 0.05356 \pm 0.00033 $ & 	 $ 0.02703 \pm 0.00039 $ \\
5806.64375181 	 & $ 0.05345 \pm 0.00034 $ & 	 $ 0.02698 \pm 0.00051 $ \\
5808.64447093 	 & $ 0.05383 \pm 0.00038 $ & 	 $ 0.02687 \pm 0.00058 $ \\
5968.98110307 	 & $ 0.05299 \pm 0.00028 $ & 	 $ 0.02609 \pm 0.00018 $ \\
6014.83479136 	 & $ 0.05560 \pm 0.00034 $ & 	 $ 0.02617 \pm 0.00022 $ \\
6018.83054636 	 & $ 0.05473 \pm 0.00032 $ & 	 $ 0.02636 \pm 0.00018 $ \\
6038.78501409 	 & $ 0.05634 \pm 0.00026 $ & 	 $ 0.02632 \pm 0.00020 $ \\
6063.93996078 	 & $ 0.05388 \pm 0.00023 $ & 	 $ 0.02608 \pm 0.00022 $ \\
6078.90154931 	 & $ 0.05573 \pm 0.00027 $ & 	 $ 0.02613 \pm 0.00024 $ \\
6096.85995871 	 & $ 0.05549 \pm 0.00043 $ & 	 $ 0.02636 \pm 0.00066 $ \\
6145.71936074	 & $ 0.05657 \pm 0.00042 $ &	 $ 0.02669 \pm 0.00069 $ \\
6177.63694756	 & $ 0.05735 \pm 0.00051 $ &	 $ 0.02687 \pm 0.00080 $ \\

\end{supertabular}

\end{center}


\clearpage

\begin{thebibliography}{}
%
\bibitem[Baliunas et al.(1995)]{baliunas95} Baliunas, S.~L., Donahue, R.~A., Soon, W.~H. et al. \ 1995, \apj, 438, 269
%
%
\bibitem[Bobylev et al.(2006)]{bobylev06} Bobylev, V.~V.,Bajkova, A.~T., \& Gontcharov, G.~A. \ 2006, A\&AT, 25, 143
%
\bibitem[Bochanski et al.(2011)]{bochanski11} Bochanski, J.~J., Hawley, S.~L., \& West, A.~A. \ 2011, \aj, 141, 98
%
\bibitem[Bonfils et al.(2005a)]{bonfils05} Bonfils, X., Delfosse, X., Udry, S. et al. \ 2005, \aap, 442, 635
%
\bibitem[Bonfils et al.(2005b)]{bonfils05b} Bonfils, X., Forveille, T., Delfosse, X. et al. \ 2005, \aap, 443, L15
%
\bibitem[Brown et al.(2011)]{brown11} Brown, B.~P., Miesch, M.~S., Browning, M.~K. et al. \ 2011, \apj, 731, 69
%
\bibitem[Browning et al.(2010)]{browning10} Browning, M.~K., Basri, G., Marcy, G.~W., West, A.~A. \& Zhang, J. \ 2010, \aj, 139, 504
%
\bibitem[Buccino et al.(2011)]{buccino11} Buccino, A.~P., D\'{i}az, R.~F., Luoni, M.~L., Abrevaya, X.~C., \& Mauas, P.~J.~D. \ 2011, \aj, 141, 34
%
\bibitem[Cameron \& Sch\"{u}ssler(2008)]{cameron08} Cameron, R. \& Sch\"{u}ssler, M. \ 2008, \apj, 685, 1291
%
\bibitem[Chabrier \& Baraffe(1997)]{chabrier97} Chabrier, G. \& Baraffe, I. \ 1997, \aap, 327, 1039
%
\bibitem[Cincunegui et al.(2007a)]{cincunegui07a} Cincunegui, C., D\'{i}az, R.~F., \& Mauas, P.~J.~D. \ 2007, \aap, 461, 1107
%
\bibitem[Cincunegui et al.(2007b)]{cincunegui07b} Cincunegui, C., D\'{i}az, R.~F., \& Mauas, P.~J.~D. \ 2007, \aap, 469, 309
%
\bibitem[Cochran \& Hatzes(1993)]{cochran93} Cochran, W.~D. \& Hatzes, A.~P. \ 1993, ASP Conference Series, 36, 267
%
\bibitem[Delfosse et al.(2000)]{delfosse00} Delfosse, X., Forveille, T., S\'{e}gransan, D. et al. \ 2000, \aap, 364, 217
%
\bibitem[D\'{i}az et al.(2007a)]{diaz07a} D\'{i}az, R.~F., Cincunegui, C., \& Mauas, P.~J.~D. \ 2007, \mnras, 378, 1007
%
\bibitem[D\'{i}az et al.(2007b)]{diaz07} D\'{i}az, R.~F., Gonz\'{a}lez, J.~F., Cincunegui, C., \& Mauas, P.~J.~D. \ 2007, \aap, 474, 345
%
\bibitem[Endl et al.(2003)]{endl03} Endl, M., Cochran, W.~D., Tull, R.~G., \& MacQueen, P. \ 2003, \aj, 126, 3099
%
\bibitem[Endl et al.(2006)]{endl06} Endl, M., Cochran, W.~D., K\"{u}rster, M. et al. \ 2006, \apj, 649, 436
%
\bibitem[ESA(1997)]{esa97} ESA \ 1997, The Hipparcos and Tycho Catalogues (ESA SP-1200) (Noordwijk: ESA)
%
\bibitem[Forveille et al.(2009)]{forveille09} Forveille, T., Bonfils, X., Delfosse, X. et al. \ 2009, \aap, 493, 645
%
\bibitem[Gomes da Silva et al.(2011)]{gds11} Gomes da Silva, J., Santos, N.~C., Bonfils, X. et al. \ 2011, \aap, 534A, 30
%
\bibitem[Gomes da Silva et al.(2012)]{gds12} Gomes da Silva, J., Santos, N.~C., Bonfils, X. et al. \ 2012, \aap, 541A, 9
%
\bibitem[Gregory(2011)]{gregory11} Gregory, P.~C. \ 2011, \mnras, 415, 2523
%
\bibitem[Hall(1996)]{hall96} Hall, J.~C. \ 1996, \pasp, 108, 313
%
\bibitem[H\"{u}nsch et al.(1999)]{hunsch99} H\"{u}nsch, M., Schmitt, J.~H.~M.~M., Sterzik, M.~F., \& Voges, W. \ 1999, A\&AS, 135, 319
%
\bibitem[Isaacson \& Fischer(2010)]{if10} Isaacson, H. \& Fischer, D. \ 2010, \apj, 725, 875
%
\bibitem[Johnson \& Apps(2009)]{johnsonapps09} Johnson, J.~A. \& Apps, K. \ 2009, \apj, 699, 933
%
\bibitem[Johnson et al.(2012)]{johnson12} Johnson, J.~A., Gazak, J.~Z., Apps, K. et al. \ 2012, \aj, 143, 111
%
\bibitem[Krej\v{c}ov\'{a} \& Budaj(2012)]{krejcova12} Krej\v{c}ov\'{a}, T. \& Budaj, J. \ 2012, \aap, 540, A82
%
\bibitem[Kruse et al.(2010)]{kruse10} Kruse, E.~A., Berger, E., Knapp, G.~R. et al. \ 2010, \apj, 722, 1352
%
\bibitem[K\"{u}rster et al.(1997)]{kurster97} K\"{u}rster, M., Schmitt, J.~H.~M.~M., Cutispoto, G. \& Dennerl, K. \ 1997, \aap, 320, 831
%
\bibitem[K\"{u}rster et al.(2003)]{kurster03} K\"{u}rster, M., Endl, M., Rouesnel, F. et al. \ 2003, \aap, 403, 1077
%
\bibitem[L\'{o}pez-Santiago et al.(2010)]{ls10} L\'{o}pez-Santiago, J., Montes, D., G\'{a}lvez-Ortiz, M.~C. et al. \ 2010, \aap, 514, A97
%
\bibitem[Lovis et al.(2011)]{lovis11} Lovis, C., Dumusque, X., Santos, N.~C. et al. \ 2011, arXiv:1107.5325v1
%
\bibitem[Mahadevan et al.(2010)]{mahadevan10} Mahadevan, S., Ramsey, L, Wright, J. et al. \ 2010, Proc. SPIE, 7735, 227
%
\bibitem[Mayor et al.(2009)]{mayor09} Mayor, M., Bonfils, X., Forveille, T. et al. \ 2009, \aap, 507, 487
%
\bibitem[Morgan et al.(2012)]{morgan12} Morgan, D.~P., West, A.~A., Garc\'{e}s, A. et al. \ 2012, arXiv:1205.6806v2
%
\bibitem[Ossendrijver(2003)]{ossendrijver03} Ossendrijver, M. \ 2003, \aap Rev., 11, 287
%
\bibitem[Paulson et al.(2004)]{paulson04} Paulson, D.~B., Cochran, W.~D., \& Hatzes, A.~P. \ 2004, \aj, 127, 3579
%
\bibitem[Pillitteri et al.(2011)]{pillitteri11} Pillitteri, I., G\"{u}nther, H.~M., Wolk, S.~J., Kashyap, V.~L., \& Cohen, O. \ 2011, \apj, 741, 18
%
\bibitem[Pipin \& Kosovichev(2011)]{pipin11} Pipin, V.~V. \& Kosovichev, A.~G. \ 2011, \apj, 741, 1
%
\bibitem[Queloz et al.(2001)]{queloz01} Queloz, D., Henry, G.~W., Sivan, J.~P. et al. \ 2001, \aap, 379, 279
%
\bibitem[Quirrenbach et al.(2011)]{quirrenbach11} Quirrenbach, A., Amado, P.~J., Caballero, J.~A. et al. \ 2011, Proc. IAU, 276, 545
%
\bibitem[Ramsey et al.(1998)]{ramsey98} Ramsey, L.~W. et al. \ 1998, Proc. SPIE, 3352, 34
%
\bibitem[Reddy et al.(2006)]{reddy06} Reddy, B.~E., Lambert, D.~L., \& Prieto, C.~A. \ 2006, \mnras, 367, 1329
%
\bibitem[Reiners et al.(2012)]{reiners12} Reiners, A., Joshi, N., \& Goldman, B. \ 2012, \aj, 143, 93
%
\bibitem[Robertson et al.(2012a)]{robertson12a} Robertson, P., Endl, M., Cochran, W.~D. et al. \ 2012, \apj, 749, 39
%
\bibitem[Robertson et al.(2012b)]{robertson12b} Robertson, P., Horner, J., Wittenmyer, R.~A. et al. \ 2012, \apj, 754, 50
%
\bibitem[Santos et al.(2010)]{santos10} Santos, N.~C., Gomes da Silva, J., Lovis, C., \& Melo, C.\ 2010, \aap, 511, A54 
%
\bibitem[Schlaufman \& Laughlin(2010)]{schlaufman10} Schlaufman, K.~C. \& Laughlin, G. \ 2010, \aap, 519A, 105
%
\bibitem[Shkolnik et al.(2008)]{shkolnik08} Shkolnik, E., Bohlender, D.~A., Walker, G.~A., \& Collier Cameron, A. \ 2008, \apj, 676, 628
%
\bibitem[Sturrock \& Scargle(2010)]{sturrock10} Sturrock, P.~A. \& Scargle, J.~D. \ 2010, \apj, 718, 527
%
\bibitem[Tan \& Cheng(2012)]{tan12} Tan, B., \& Cheng, Z.\ 2012, \apss, 382
%
\bibitem[Thompson et al.(2003)]{thompson03} Thompson, M.~J., Christensen-Dalsgaard, J., Miesch, M.~S. \& Toomre, J. \ 2003, \araa, 2003, 599
%
\bibitem[Tull et al.(1998)]{tull98} Tull, R.~G. et al. \ 1998, Proc. SPIE, 3355, 387
%
%
\bibitem[Vogt et al.(2010)]{vogt10} Vogt, S.~S., Butler, R.~P., Rivera, E.~J. et al. \ 2010, \apj, 723, 954
%
\bibitem[Vogt et al.(2012)]{vogt12} Vogt, S.~S., Butler, R.~P., \& Haghighipour, N. \ 2012, arXiv:1207.4515
%
\bibitem[Walkowicz et al.(2004)]{walkowicz04} Walkowicz, L.~M., Hawley, S.~L., \& West, A.~A. \ 2004, \pasp, 116, 1105
%
\bibitem[West et al.(2004)]{west04} West, A.~A., Hawley, S.~L., Walkowicz, L.~M. et al. \ 2004, \aj, 128, 426
%
\bibitem[West et al.(2008)]{west08} West, A.~A., Hawley, S.~L., Bochanski, J.~J. et al. \ 2008, \aj, 135, 785
%
\bibitem[West et al.(2009)]{west09} West, A.~A., Hawley, S.~L., Bochanski, J.~J., Covey, K.~R., \& Burgasser, A.~J. \ 2009, Proc. IAU, 258, 327
%
\bibitem[Wright(2004)]{wright04} Wright, J.~T. \ 2004, \aj, 128, 1273
%
\bibitem[Wright et al.(2011)]{wright11} Wright, N.~J.,Drake, J.~J., Mamajek, E.~E. \& Henry, G.~W. \ 2011, \aj, 743, 48
%
\bibitem[Zechmeister \& K\"{u}rster(2009)]{zk09} Zechmeister, M. \& K\"{u}rster, M. \ 2009, \aap, 496, 577
%
\bibitem[Zechmeister et al.(2009)]{zechmeister09} Zechmeister, M., K\"{u}rster, M., \& Endl, M. \ 2009, \aap, 505, 859
%
\end{thebibliography}
\end{document}